\documentclass[10pt,graphicx,subfigure,axodraw]{article}
\usepackage{bm}
\usepackage{tabularx}
\usepackage{multirow}
\usepackage{cancel}
\usepackage{amsfonts}
\usepackage{amssymb}
\usepackage{epsfig}
\usepackage{float}
\usepackage{amsmath}
\usepackage{amsthm}
\usepackage{slashed}
\usepackage[title]{appendix}
\restylefloat{table}
\usepackage{tabulary}
\usepackage{hyperref}
\usepackage{cleveref}
\usepackage[usenames,dvipsnames]{color}
\usepackage[parfill]{parskip}
\usepackage[numbers,sort&compress]{natbib}
\usepackage{kbordermatrix}
\usepackage{braket}

\def\non{\nonumber\\}

\newcommand{\beqn}{\begin{eqnarray}}
\newcommand{\eeqn}{\end{eqnarray}}

\newcommand{\bl}[1]{{\color{blue}{#1}}}

\renewcommand{\kbldelim}{(}
\renewcommand{\kbrdelim}{)}
\usepackage{tikz}

\begin{document}

\newcommand{\Nu}{\mbox{\boldmath\large$\nu$}}
\newcommand{\Tau}{\mbox{\boldmath\Large$\tau$}}
\newcommand{\bb}{\mbox{\normalsize$\boldsymbol{b}$}}
\newcommand{\bt}{\mbox{\normalsize$\boldsymbol{t}$}}
\newcommand{\bU}{\mbox{\normalsize$\mathbf{U}$}}
\newcommand{\bD}{\mbox{\normalsize$\mathbf{D}$}}
\newcommand{\bE}{\mbox{\normalsize$\mathbf{E}$}}
\newcommand{\mTau}{\mbox{$\tau$}}
\def\non{\nonumber\\}

{\small

\begin{titlepage}

\author{Jinzheng Li$^a$\footnote{Email:li.jinzh@northeastern.edu},
 ~Pran Nath$^a$\footnote{Email: p.nath@northeastern.edu} ~and Raza M. Syed$^{a,b}$\footnote{
Email: rsyed@aus.edu}\\~\\
$^{a}$\textit{\normalsize Department of Physics, Northeastern University, Boston, MA 02115-5000, USA} \\
$^{b}$\textit{\normalsize Department of Physics, American University of Sharjah, P.O. Box 26666, Sharjah, UAE\footnote{Permanent address}} \\}

\title{A natural MSSM from a novel  $\mathsf{SO(10)}$, Yukawa unification, 
light sparticles,  and  SUSY implications at LHC}

\maketitle
\begin{center}
\end{center}

\begin{abstract}
The $\mathsf{SO(10)}$ model with a heavy Higgs spectrum consisting of
$\mathsf{560+\overline{560}}$ and a light Higgs spectrum consisting of $2\mathsf{\times 10+320}$ plet  representations of $\mathsf{SO(10)}$ is unique among $\mathsf{SO(10)}$ models.
It has the remarkable property that VEVs of $\mathsf{560}$ and $\mathsf{\overline{560}}$
can simultaneously reduce the rank of the gauge group and further
reduce the remaining symmetry down to the standard model gauge group.
 Additionally, on mixing with the light fields all the Higgs fields
become heavy except for one pair of light Higgs doublets just as in MSSM. This  model has not been fully
explored thus far because of the technical difficulty of computing~
the couplings of the heavy and the light Higgs sectors, specifically the
interaction $(\mathsf{560\times 560)\cdot 320}$ involving the coupling of tensor-spinors
with a third rank ~{mixed} tensor $\mathsf{320}$. An explicit analysis of such couplings is
given in this paper. 
Spontaneous symmetry breaking of the
$\mathsf{SO(10)}$ symmetry is carried out  reducing the gauge group
to $\mathsf{SU(3)}_c\times \mathsf{SU(2)}_L\times \mathsf{U(1)}_Y$ with just one  pair of light Higgs.
 Thus a natural deduction of MSSM arises from the 
 $\mathsf{SO(10)}$ model with no fine tuning needed.
Further, it is shown that the light Higgs doublet of the
model is a linear combination of the Higgs doublet fields of the $2\mathsf{\times 10}$
and the $\mathsf{320}$ Higgs fields. 
{It is shown that in this class of $\mathsf{SO(10)}$ models $b-t-\tau$ unification
can be achieved with $\tan\beta$ as low as 5-10. An analysis of the sparticle
spectrum within ${\tilde g}$SUGRA  renormalization group evolution is given which
leads to a bi-modal sparticle spectrum consisting of a compressed low mass 
spectrum for sleptons and weakinos and a high mass spectrum of gluino, squarks,
and heavy Higgs.}
While the LSP is the light neutralino, the  NLSP is found to be the light stau lying close to the LSP, while the remaining leptons, and the weakinos
 are also in close proximity to the LPS with masses in the few hundred GeV range. 
 The cross section for slepton production and weakino production are estimated and appear promising for SUSY at the LHC. However, a more dedicated analysis
is needed to predict the size of the supersymmetric signatures at the LHC.  

\end{abstract}

\end{titlepage}
\tableofcontents

\section{Introduction}
Among grand unified theories
 $\mathsf{SO(10)}$ unification is the most appealing
since aside from unifying the standard model gauge group
$\mathsf{SU(3)_C\times SU(2)_L\times U(1)_Y}$ it also accommodates a full generation
of quarks and leptons in one $\mathsf{16}$-plet spinor representation of $\mathsf{SO(10)}$
 \cite{Georgi:1974my, Fritzsch:1974nn} (for a review of grand unified theories see, e.g.\cite{Nath:2006ut}).
There is a further feature of $\mathsf{SO(10)}$ unification
which sets it apart from the minimal  unifying gauge group $\mathsf{SU(5)}$.
 Thus in $\mathsf{SU(5)}$, just one $\mathsf{24}$-plet of Higgs which breaks the 
 GUT symmetry
down to  the Standard Model gauge group. For $\mathsf{SO(10)}$ there exist
 a variety of possibilities with both small as well large representations (see, e.g.,  ~\cite{Clark:1982ai,Aulakh:1982sw,Babu:1992ia,Aulakh:2003kg,Bajc:2004xe,Aulakh:2004hm,Babu:2006nf,Aulakh:2008sn}  and for applications see, e.g.,
\cite{Harvey:1980je,Ananthanarayan:1992cd,Chen:2000fp,Blazek:2002ta,Fukuyama:2004ti,Dermisek:2006dc,Nath:2015kaa}).
  Additionally unlike $\mathsf{SU(5)}$
 we need to reduce the rank of the gauge group. To check which Higgs representation can produce rank reduction we look at the $\mathsf{SU(5)\times U(1)}$ quantum numbers of some of some possible candidate tensor representations, such as $\mathsf{45}$, $\mathsf{54}$, $\mathsf{210}$. Their $\mathsf{SU(5)\times U(1)}$
 decomposition are: $\mathsf{45}=\mathsf{1}(0)+ \mathsf{10}(4)+ \mathsf{\overline{10}}(-4) + \mathsf{24}(0)$,
 $\mathsf{54}=\mathsf{15}(4)+\mathsf{\overline{15}}(-4) + \mathsf{24}(0)$ and $\mathsf{210}=\mathsf{1}(0) + \mathsf{5}(-8)+ \mathsf{\overline{5}}(8)
 + \mathsf{10}(4)+ \mathsf{\overline{10}}(-4) + \mathsf{24}(0) + \mathsf{40}(-4) + \mathsf{\overline{40}}(4) + \mathsf{75}(0)$.
  The $\mathsf{SU(5)}$ representations whose VEVs preserve the standard model gauge group are $\mathsf{1}(0)$, $\mathsf{24}(0)$ and $\mathsf{75}(0)$ but all of these have a zero $\mathsf{U(1)}$ quantum number and thus cannot reduce the rank. A possible Higgs representation
  which can reduce the rank is the spinor representation $\mathsf{16}$ which under $\mathsf{SU(5)\times U(1)}$ has the decomposition $\mathsf{16}=\mathsf{1}(-5)+ \mathsf{\overline{5}}(-3) + \mathsf{10}(-1)$
  where the singlet has a non-vanishing $\mathsf{U(1)}$ quantum number.  Thus in this
  case the minimal Higgs representation which can break $\mathsf{SO(10)}$ down to
  standard model gauge group is $\mathsf{16+\overline{16}+ 45}$. Another possibility for
  reducing the rank is to use $\mathsf{126}$-plet of Higgs which under $\mathsf{SU(5)\times U(1)}$ has the decomposition $\mathsf{126}= \mathsf{1}(-10)+ \mathsf{\overline{5}}(-2) + \mathsf{10}(-6)
  + \mathsf{\overline{15}}(6) +\mathsf{45}(2) +\mathsf{\overline{50}}(-2)$. Here the singlet field
  has a non-vanishing $\mathsf{U(1)}$ quantum number and can reduce the rank.
  Thus in  this case using $\mathsf{45+126+\overline{126}}$ will break $\mathsf{SO(10)}$
  to the Standard Model gauge group. However, we further need to break
  the symmetry down to $\mathsf{SU(3)_C\times U(1)_{em}}$ which requires a $\mathsf{10}$-plet
  of Higgs which under $\mathsf{SU(5)\times U(1)}$ has the decomposition
  $\mathsf{10}=\mathsf{5}(2)+ \mathsf{\overline{5}}(-2)$ and has Higgs doublets.
  The discussion above illustrates that the
  breaking of $\mathsf{SO(10)}$ to $\mathsf{SU(3)_C\times U(1)_{em}}$ requires three
  sets of Higgs representations, i.e., one to reduce the rank, the second
  to break the rest of the symmetry down to the standard model gauge group
  and the third to break it further to the residual symmetry $\mathsf{SU(3)_C\times U(1)_{em}}$. A while back another proposal was made where the combination
presentations
 $\mathsf{144+\overline{144}}$  was used to break the $\mathsf{SO(10)}$ GUT symmetry \cite{Babu:2005gx,Babu:2006rp}. Under $\mathsf{SU(5)\times U(1)}$
 decomposition \cite{Slansky:1981yr} $\mathsf{144}= \mathsf{\overline{5}}(3)+ \mathsf{5}(7) + \mathsf{10}(-1)+  \mathsf{15}(-1)+ \mathsf{24}(-5) + \mathsf{40}(-1) +\mathsf{\overline{45}}(3)$. Here since $\mathsf{24}(-5)$ has a non-vanishing
 $\mathsf{U(1)}$ quantum number,  a VEV of  $\mathsf{24}(-5)$ will reduce the rank and
  then break the rest of the symmetry of $\mathsf{SO(10)}$ down to the Standard Model gauge group just in one step.

 The $\mathsf{144+\overline{144}}$ model discussed above, however,
 suffers from the problem typical of most grand unified models.
    Thus, as well known, typically in grand unified models the Higgs doublets would turn out to be the same size in mass as the Higgs triplets which are required to be heavy to suppress rapid proton decay (see, e.g.,\cite{Nath:2006ut}).
     There are a variety of
    proposals on how to produce a light Higgs doublet needed for electroweak
    symmetry breaking. One such proposal is to use fine tuning to force one Higgs  doublet pair to be   light. However, such a procedure is viewed
    unnatural since the fine tuning involved is extreme.
     There are various other possibilities where a light pair of doublets
    could arise from a symmetry
    principle, or a group theoretic constraint or a more fundamental aspect of
    the theory itself.  Such models
    may be viewed as natural.
     Models of this type exist in strings, see, e.g.,
    \cite{Braun:2005nv,Bouchard:2006dn,Anderson:2009mh,Bouchard:2005ag}.
      One of the early ways to get light Higgs doublets while keeping the Higgs 
      color triplets heavy is to arrange the vacuum expectation value (VEV)
      so that the spontaneous breaking of $\mathsf{SO(10)}$ occurs only along the $B-L$
      preserving direction
      (see, e.g.,\cite{Dimopoulos:1981xm,Babu:1994dq,Lee:1993jw}).      
    Thus, if the $\mathsf{SO(10)}$ symmetry breaking occurs along the
    $B-L$ preserving direction, then the Higgs color triplets will get masses and
    the Higgs doublets will remain light.
    
    There is yet another mechanism for naturally generating a doublet-triplet
    splitting. It is the well known missing partner mechanism which does not work for 
    $\mathsf{SU(5)}$ but is feasible  in $\mathsf{SO(10)}$ model building.  This mechanism
    involves two Higgs sectors: one heavy and the other light.
  Here the choice of heavy vs light
    Higgs representations is such that there is an exact match between
    the number of heavy Higgs color triplets/anti-triplets and light Higgs color triplets/anti-triplets
    but there is a mismatch in the number of heavy Higgs doublet pairs $n_{h}$
    and the number of light Higgs doublet pairs $n_{l}$  such that
    $n_{l}=n_{h}+1$. In this case in the presence of mixing of light and heavy
    Higgs fields, all Higgs triplets/anti-triplets will become heavy while only
     $2n_h$ number of Higgs doublet pairs  will become heavy leaving one
    Higgs doublet pair light. A key element of this construction is the requirement
    that there exist a heavy Higgs representation where the number of Higgs
    doublet pairs in one less than the number of Higgs triplet-anti-triplet pairs.
    An example of such  representations in
     $\mathsf{SU(5)}$ is the combination $\mathsf{50+\overline{50}}$-plets of heavy Higgs which
    has one pair of heavy Higgs triplets/anti-triplets and there are no Higgs doublets.
    In this case if one chooses the heavy sector to be $\mathsf{50+\overline{50} + 75}$
    and the light sector to be $\mathsf{5+\overline{5}}$, the $\mathsf{75}$-plet induces a mixing
    between the heavy $\mathsf{50+\overline{50}}$ Higgs and $\mathsf{5+\overline{5}}$ light
    Higgs  generating
    heavy masses for all the Higgs triplets/anti-triplets and leaving one Higgs
    doublet pair light \cite{Masiero:1982fe,Grinstein:1982um}.
       An example of such a model in $\mathsf{SO(10)}$ was was discussed in~\cite{Babu:2006nf} where the heavy sector consists of $\mathsf{126+\overline{126}
       +210}$ and the light sector consists of $2\mathsf{\times 10+120}$,
       and the $\mathsf{126+\overline{126}}$ contain $\mathsf{50+\overline{50}}$ providing the
       key element for the missing partner mechanism to work in $\mathsf{SO(10)}$.
       A more exhaustive analysis of the missing partner mechanism in $\mathsf{SO(10)}$ was given in \cite{Babu:2011tw} where several other models producing a light Higgs doublet pair were discussed.

      Of specific interest for the work in this paper is the model
      discussed in \cite{Babu:2011tw} consisting of
          \begin{align}
      \text{Heavy} ~\{\mathsf{560+ \overline{560}}\} + \text {Light} ~\{2\mathsf{\times 10+ 320}\}
      \end{align}
    The $\mathsf{560}$-plet is a  second rank tensor spinor $\Theta^{\mu\nu}$
    where $\mu,\nu =1\cdots 10$ are $\mathsf{SO(10)}$ indices. It
    is anti-symmetric in the tensor indices, and traceless which leads
    to $\mathsf{16\times 45-160=560}$ components. It has the $\mathsf{SU(5)}$ decomposition
    \begin{eqnarray}
\label{560}
\mathsf{560} &=& \mathsf{1}(-5)+\mathsf{\overline{5}}(3)+\mathsf{\overline{10}}(-9)+\mathsf{10_1}(-1)+\mathsf{10_2}(-1)+\mathsf{24}(-5)+40(-1)\nonumber \nonumber\\
~~~&&+~\mathsf{45}(7)+\mathsf{\overline{45}}(3)+\mathsf{\overline{50}}(3)+\mathsf{\overline{70}}(3)+\mathsf{75}(-5)+\mathsf{175}(-1)~.
\label{560}
\end{eqnarray}
It is remarkable in the following way: It has features of $\mathsf{144}$ as well as of
$\mathsf{126}$ in the following way. Thus from Eq.(\ref{560}) we see that in $\mathsf{SU(5)}$
decomposition it has
 three representations in it which carry non-vanishing $\mathsf{U(1)}$
quantum numbers, i.e., $\mathsf{1}(-5),~\mathsf{24}(-5),~ \mathsf{75}(-5)$ and which can reduce the rank
and break the GUT symmetry down to the standard model gauge group
all at once just like the $\mathsf{144+\overline{144}}$ does.

What makes it unique is that it also has $\mathsf{\overline{50}}(3)$ which
as we know enters in the missing partner mechanism.
Thus $\mathsf{560}$
 contains one color triplet and four color anti--triplets,
 one up-type Higgs doublet, and three down-type doublets. Combined with
 $\mathsf{\overline{560}}$ one finds that $\mathsf{560+\overline{560}}$ contains five pairs of Higgs
 color triplets/anti-triplets and four pairs of Higgs doublet pairs. If one allows a light Higgs sector
 consisting of $2\mathsf{\times 10 + 320}$ of Higgs fields, it brings in five pair of color triplets/anti-triplets and and five pairs of light Higgs doublets. A mixing of the heavy and light sectors will then make all five color triplet pairs of the light sector and four Higgs doublets pairs of the light sector heavy leaving us
 with one pair of Higgs doublets from the light sector which would be the Higgs doublet pair of MSSM.

   While some preliminary analysis of the model with $\mathsf{560+\overline{560}}$
   multiplet  was given in reference \cite{Babu:2011tw} the details of this model have never been
   worked out.  The main reason for it is the explicit computation of the
   interaction   $\mathsf{560\cdot\overline{560}\cdot320}$  is difficult as it involves a special
   technique to handle spinor-tensor coupling. The technique to handle
   spinor-vector couplings involving $\mathsf{144}-$plet was developed in \cite{Babu:2011tw,Nath:2005bx} based on previous works \cite{Nath:2001uw,Nath:2001yj,Nath:2003rc}
   using oscillator technique \cite{Mohapatra:1979nn,Wilczek:1981iz}
    and here we extend it to the coupling involving
  spinor-tensor. Thus one of the
   contributions of this work is that  we carry out
   an explicit computation of these couplings and give a concrete analysis of
   the symmetry breaking chain. We identify the MSSM Higgs that emerges
   from the analysis as a linear combination of five Higgs doublet pairs from
    the light Higgs sector where two Higgs doublet pairs arise from $\mathsf{10_1}$ and $\mathsf{10_2}$ and three Higgs doublets arise from the $\mathsf{320}-$plet.

The outline of this paper is as follows:  In Section \ref{sec:2} we give
a description of the model. In Section \ref{sec:3} we give
an analysis of  breaking of ${\mathsf{SO(10)}}$ to
$\mathsf{SU(3)_C\times SU(2)_L \times U(1)_Y}$ with a natural doublet-triplet splitting. An analysis of the Higgs doublet
mass matrix is given in section \ref{sec:4}. ~{In section \ref{sec:5} 
we discuss the spontaneous breaking of $\mathsf{SO(10)}$ down to the standard model gauge group symmetry and  generation of the light Higgs doublet.
In section \ref{pheno}, phenomenological aspects of the model are discussed. 
Here it is shown that $b-t-\tau$ unification can be achieved at low $\tan\beta$ with
$\tan\beta$ as low as 5-10. The analysis is carried out in $\tilde g$SUGRA framework where 
the radiative breaking of the electroweak symmetry is accomplished by a heavy gluino, which 
leads to a split sparticle mass spectrum and also a split Higgs boson mass spectrum where the
light spectrum lies in the few hundred GeV region and the heavy spectrum lies in the several 
TeV region consistent with the current experimental constraints on the Higgs boson 
mass, on the relic density of dark matter and on the muon anomaly. 
 Further, it is shown that while the lightest sparticle (LSP) is the neutralino, the next to
 lightest particle (NLSP) is the light stau. An analysis of the production cross section for some for the sleptons and for the light weakinos is given and shown to be substantial indicating the possibility of detection of 
some of the light sparticles at 14 TeV  high luminosity LHC (HL-LHC), and at 27 TeV high energy LHC 
(HE-LHC). Conculsions are given in Section \ref{sec:7}. 

A more elaborate discussion of the mathematical
details of the analysis are given in appendices \ref{sec:A} - \ref{sec:G}. Thus in
Appendix \ref{sec:A} we give an analysis of $\mathsf{560 + \overline{560}}$  tensor-spinor in its $\mathsf{SU(5)}$ oscillator modes.
Further in this appendix we compute the normalization of kinetic energies of irreducible $\mathsf{SU(5)}$ multiplets arising in the decomposition of
$\mathsf{560 + \overline{560}}$ multiplets.  In Appendix \ref{sec:B}, we give a computation of the couplings
of the $\mathsf{SU(5)}$ components fields $\mathsf{1},~\mathsf{24},~ \mathsf{75}$ that appear in $\mathsf{560}$ and
$\mathsf{\overline{560}}$ multiplets. Such couplings determine the spontaneous
breaking of $\mathsf{SO(10)}$ to the Standard Model gauge group when they
develop VEVs. {
{In Appendix  \ref{sec:C} we exhibit extraction and normalization of all $\mathsf{SU(2)_L}$ doublet fields
appearing in the model}. In Appendix \ref{sec:D}  we give a detailed discussion
of how the  symmetric and anti-symmetric $\mathsf{320}$ three index tensors
are constructed. In Appendix \ref{sec:E}, we discuss the all the possible cubic couplings of the $\mathsf{320}-$plet with the
 $\mathsf{560}-$plet and their associated generational symmetries. In Appendix \ref{sec:F}, we give extensive details for
constructing normalized $\mathsf{70}-$, $\mathsf{45}-$ and $\mathsf{5}-$plets with diagonal kinetic energy terms contained in $\Lambda^{\mathsf{320}_s}_{(\mu\nu)\lambda}$ of $\mathsf{SO(10)}$.
The couplings of $\mathsf{320}$ and $\mathsf{10}$ to
$\mathsf{560}$ and $\mathsf{\overline{560}}$ multiplets are given in Appendix \ref{sec:G}. These
couplings play a central role in doublet-triplet splitting to produce
a light Higgs that enters in electroweak symmetry breaking.

}

\section{The model \label{sec:2}}
The model makes use of a single $\mathsf{560}+\overline{\mathsf{560}}$ multiplet, a single $\mathsf{320}$ multiplet and two $\mathsf{10}$ multiplets of $\mathsf{SO(10)}$.
The superpotential of the model is given by
\begin{align}
W&= W_{\textsc{gut}}+ W_{\textsc{dt}}\nonumber\\
W_{\textsc{gut}}&= \mathsf{45}\cdot\mathsf{45}+\mathsf{560}\cdot
\mathsf{\overline{560}}\cdot\mathsf{45}+ \mathsf{560}\cdot\mathsf{\overline{560}},\non
W_{\textsc{dt}}&\sim \mathsf{560}\cdot\mathsf{\overline{560}}~+~\mathsf{{560}}\cdot\mathsf{{560}}\cdot \mathsf{10}_1
~+~\mathsf{\overline{560}}\cdot\mathsf{\overline{560}}\cdot\left(\mathsf{10}_1+\mathsf{10}_2\right)~+~\mathsf{{560}}\cdot\mathsf{{560}}\cdot \mathsf{320}
~+~\mathsf{\overline{560}}\cdot\mathsf{\overline{560}}\cdot \mathsf{320}.
\end{align}
where $W_{\textsc{gut}}$ breaks the GUT symmetry down to the Standard
Model gauge group and $W_{\textsc{dt}}$ generates the doublet-triplet splitting
and leads to just one light Higgs doublet pair. The content of Higgs
triplets/anti-triplets and of Higgs doublet pairs for $\mathsf{560+\overline{560}}$,
for $\mathsf{320}$ and for $\mathsf{10_1+10_2}$ is shown in Table 1. In Eq.(\ref{560})
 we exhibited the decomposition of $\mathsf{560}$ under $\mathsf{SU(5)\times U(1)}$.
 The $\mathsf{SU(5)\times U(1)}$ decomposition of $\mathsf{10}$-plets is simple, i.e.,
 $\mathsf{10}=\mathsf{5}(2)+ \mathsf{\overline{5}}(-2)$ as noted earlier. For the $\mathsf{320}$ multiplet we have the
 decomposition
 \begin{align}
 \mathsf{320}&= \mathsf{5}(2) + \overline{\mathsf{5}}(-2) + \mathsf{40}(-6) + \overline{\mathsf{40}}(6) + 45(2) +\overline{\mathsf{45}}(-2) + \mathsf{70}(2) + \overline{\mathsf{70}}(-2).
 \end{align}
 Explicit tensorial structure of various terms in the superpotential will be
 discussed in sections below where we also give details of the computation
 of the couplings.
{{\tiny
\begin{table}[htb]
\begin{center}
\begin{tabular}{|c|c|c|c|c|}
\hline
&&&&\\
 Heavy Fields & Light Fields &Pairs of D and T & Pairs of D and T  &  Residual Set \\
    &   & in Heavy Fields   & in Light Fields  & of Light Modes   \\

$\mathsf{560}+\overline{\mathsf{560}}$&$ \mathsf{320} + 2 \times \mathsf{10}$&4D+5T& (3D+3T)+ (2D+2T) &1D \\
&&&&\\
\hline
\end{tabular}\\~\\
\label{table1}
\caption{
Exhibition of Higgs doublet pairs (D) consisting of up--type and down--type Higgs doublets,
and Higgs triplet/anti-triplet (T) pairs in the $\mathsf{SO(10)}$ missing partner model discussed in this work.}
\end{center}
\label{tab:1}
\end{table}
}}

Before proceeding further we exhibit the
symmetric ($s$) and the antisymmetric ($a$) tensor products of
$\mathsf{560}$ with itself.
\begin{eqnarray}
\left(\mathsf{560} \times \mathsf{560}\right)_{s} &=& \mathsf{10} + \mathsf{126} + \overline{\mathsf{126_1}}+\overline{\mathsf{126_2}}  + \mathsf{210'}+ \mathsf{320}+
 \mathsf{1728_1} + \mathsf{1728_2} + \mathsf{2970_1}+\mathsf{2970_2} \nonumber\\
&&+~\mathsf{3696^{\prime}}+\mathsf{4410}+\mathsf{4950}+\overline{\mathsf{4950}}+ \mathsf{6930'} + \mathsf{10560}+ \mathsf{27720} + \mathsf{36750}  + \overline{\mathsf{46800}},\\
\left(\mathsf{560} \times \mathsf{560}\right)_{a} &=& \mathsf{120_1} + \mathsf{120_2} + \mathsf{320} + \mathsf{1728_1} + \mathsf{1728_2} + \mathsf{2970} + \overline{\mathsf{3696'_1}}+ \overline{\mathsf{3696'_2}}+ \overline{\mathsf{3696'_3}}  \nonumber \\
&&+~ \mathsf{4312_1} + \mathsf{4312_2} + \mathsf{10560}+ \mathsf{34398} + \mathsf{36750}  + \overline{\mathsf{48114}}.
\label{decomp}
\end{eqnarray}
Here we find that both the symmetric and the anti-symmetric tensor
products of $\mathsf{560}$ with itself contain a $\mathsf{320}$. Correspondingly there
are two $\mathsf{320}$ multiplets which is symmetric in two of the three tensor
indices with the totally symmetric part and the trace on two indices
removed and the other which is anti-symmetric in two of its indices
with the totally anti-symmetric part and the trace removed. In this
analysis we consider the coupling of the $\mathsf{320}_s$ to $\mathsf{560\times 560}$
and to $\mathsf{\overline{560}\times \overline{560}}$.
 The irreducible tensor-spinor $\ket{\Theta_{\mu\nu}^{(\mathsf{560})}}$ can be obtained from the $\mathsf{720}~(=\mathsf{16}\times\mathsf{45})$ dimensional reducible tensor-spinor $\ket{\Theta_{\mu\nu}^{\mathsf{720}}}$   by removing  $\mathsf{160}~(=\mathsf{16}\times\mathsf{10})$ components, given by $\Gamma_{\mu}\ket{\Theta_{\mu\nu}^{(\mathsf{720})}}$, out of the $\mathsf{720}$ components of the unconstrained tensor-spinor. Thus we define the $\mathsf{560}$
  multiplet so that so that it is anti-symmetric in its two tensor indices and
  traceless given by
     \begin{subequations}
		\begin{align}
     \textnormal{Traceless:}&~~~~~\Gamma_{\mu}\ket{\Theta_{\mu\nu}^{\breve{s}~(\mathsf{560})}}=0,\label{560 Constraint 1}\\
     \textnormal{Antisymmetric:}&~~~~~ \ket{\Theta_{\mu\nu}^{\breve{s}~(\mathsf{560})}}=-\ket{\Theta_{\nu\mu}^{\breve{s}~(\mathsf{560})}}\label{560 Constraint 2}.
     \end{align}
          \label{def560}
          	\end{subequations}
Here $\breve{s}=1,\dots,16$ is the spinor index and  the Greek letters $\mu,\nu,\dots=1,2,\dots, 10$ are the $\mathsf{SO(10)}$ indices and
$\Gamma_{\mu}$ satisfy the rank$-10$ Clifford algebra, that is,
$\{\Gamma_{\mu}, \Gamma_{\nu}\} = 2\delta_{\mu\nu}$.
Further details of the construction of $\mathsf{560}$ and of $\mathsf{\overline{560}}$ are
discussed in Appendix \ref{sec:A}
where we also give an explicit decomposition
of them in terms of  irreducible representations of $\mathsf{SU(5)\times U(1)}$.

Regarding $\mathsf{320}$ multiplet there are two: one of which is symmetric in two
indices while the other is anti-symmetric. We can construct them from
the 1000 dimensional multiplet $T_{\mu\nu\lambda}$. Thus the
 anti-symmetric one is given
by {
\begin{align}
\Lambda^{(\mathsf{320}_a)}_{[\mu\nu]\lambda}&=\frac{1}{6} \Big[2 \big(T_{\mu\nu\lambda}- T_{\nu\mu\lambda}\big)+ T_{\lambda\nu\mu}+T_{\mu\lambda\nu}- T_{\nu\lambda\mu}-T_{\lambda\mu\nu}\Big]-\frac{1}{18}\Big(\delta_{\nu\lambda}\delta_{\mu\beta}
-\delta_{\mu\lambda}\delta_{\nu\beta}\Big)\big(T_{\beta\alpha\alpha}-T_{\alpha\beta\alpha}\big)
\label{320a} 
\end{align}
}
while the symmetric $\mathsf{320}$-plet is given by
{
\begin{align}
\Lambda^{(\mathsf{320}_s)}_{(\mu\nu)\lambda}&= \frac{1}{6} \Big[2 \big(T_{\mu\nu\lambda}+ T_{\nu\mu\lambda}\big)-\big( T_{\lambda\nu\mu}+T_{\mu\lambda\nu}+ T_{\nu\lambda\mu}+T_{\lambda\mu\nu}\big)\Big]\nonumber\\
&~~~+\frac{1}{54}\big(\delta_{\nu\lambda}\delta_{\mu\beta}
+\delta_{\mu\lambda}\delta_{\nu\beta}-2\delta_{\mu\nu}\delta_{\lambda\beta}\big)
\big(-T_{\beta\alpha\alpha}+2T_{\alpha\alpha\beta}-T_{\alpha\beta\alpha}\big)
\label{320s}
\end{align}
}
The details of their construction are given in Appendix \ref{sec:D}.

\section{One step breaking of $\boldsymbol{\mathsf{SO(10)}}$ to
$\boldsymbol{\mathsf{SU(3)_C}\times\mathsf{SU(2)_L} \times\mathsf{U(1)_Y}}$ and natural doublet-triplet splitting \label{sec:3}}

The form of the superpotential responsible for a one-scale breaking of $\mathsf{SO(10)}$ symmetry is
\begin{eqnarray}
W_{\textsc{gut}}&\sim &\mathsf{45}\cdot\mathsf{45}+\mathsf{560}\cdot
\mathsf{\overline{560}}\cdot\mathsf{45}+ \mathsf{560}\cdot\mathsf{\overline{560}}.
\end{eqnarray}
In the computation of these couplings we will use the method developed in
 \cite{Nath:2001uw, Nath:2001yj, Nath:2005bx} using oscillator
 techniques~\cite{Mohapatra:1979nn, Wilczek:1981iz}.
Exhibiting its tensorial structure, the superpotential can be written as
\begin{eqnarray}\label{GUTsymmetry1}
W_{\textsc{gut}}&=&\frac{1}{2!}M_{45}\Phi^{(45)}_{\mu\nu}\Phi^{(45)}_{\mu\nu}+\lambda_{45}\braket{\Theta_{\mu\nu}^{(\mathsf{560})*}|B|
\overline{\Theta}_{\nu\sigma}^{(\overline{\mathsf{560}})}}\Phi^{(45)}_{\mu\sigma}+ M_{560}\braket{\Theta_{\mu\nu}^{(\mathsf{560})*}|B|
\overline{\Theta}_{\mu\nu}^{(\overline{\mathsf{560}})}},
\end{eqnarray}
where $B$ is the $\mathsf{SO(10)}$ charge conjugation operator  given by $B=\prod_{\mu=\textnormal{odd}}\Gamma_{\mu}=-i\prod_{k=1}^5\left(b_k-b_k^{\dagger}\right)$. Integrating out $\Phi^{45}_{\mu\nu}$ in Eq.(\ref{GUTsymmetry1}), expanding out as in Eqs.(\ref{GUTsymmetryB2} - \ref{GUTsymmetryB4}) and finally substituting Eqs.(\ref{GUTsymmetryB5} - \ref{GUTsymmetryB7})we get
{
  \begin{eqnarray}\label{GUTsymmetryvevs}
  W_{\textsc{gut}}
&=&\frac{\lambda_{45}^2}{4M_{45}}\left[-\left(\frac{7}{432}\right){{\mathbf S}}_{75}^2{\overline{\mathbf S}}_{75}^2
-\left({\frac{\sqrt{5}}{216}}\right){\mathbf S}_{75}^2{\overline{\mathbf S}}_{75}{\overline{\mathbf S}}_{24}-\left({\frac{\sqrt{5}}{216}}\right){\mathbf S}_{75}{\overline{\mathbf S}}_{75}^2{\mathbf S}_{24}\right.\nonumber\\
&&~~~~~~~~~\left.-\left({\frac{349}{8640}}\right){\mathbf S}_{75}{\overline{\mathbf S}}_{75}{\mathbf S}_{24}{\overline{\mathbf S}}_{24}
+\left({\frac{1}{180}}\right){\mathbf S}_{75}{\overline{\mathbf S}}_{75}{\mathbf S}_{24}{\overline{\mathbf S}}_{1}+\left({\frac{1}{180}}\right){\mathbf S}_{75}{\overline{\mathbf S}}_{75}{\overline{\mathbf S}}_{24}{\mathbf S}_{1}\right.\nonumber\\
&&~~~~~~~~~\left.-\left(\frac{1}{90}\right){\mathbf S}_{75}{\overline{\mathbf S}}_{75}{\mathbf S}_{1}{\overline{\mathbf S}}_{1}
-\left({\frac{25}{3456}}\right){\overline{\mathbf S}}_{75}^2{\mathbf S}_{24}^2-\left({\frac{25}{3456}}\right){\mathbf S}_{75}^2{\overline{\mathbf S}}_{24}^2\right.\nonumber\\
&&~~~~~~~~~\left.+\left({\frac{\sqrt{5}}{288}}\right){\mathbf S}_{75}{\overline{\mathbf S}}_{24}{\mathbf S}_{24}{\overline{\mathbf S}}_{1}
+\left({\frac{\sqrt{5}}{288}}\right){\mathbf S}_{75}{\overline{\mathbf S}}_{24}^2{\mathbf S}_{1}+\left({\frac{\sqrt{5}}{288}}\right){\overline{\mathbf S}}_{75}{\mathbf S}_{24}^2{\overline{\mathbf S}}_{1}\right.\nonumber\\
&&~~~~~~~~~\left.+\left({\frac{\sqrt{5}}{288}}\right){\overline{\mathbf S}}_{75}{\mathbf S}_{24}{\overline{\mathbf S}}_{24}{\mathbf S}_{1}
-\left({\frac{49}{8640\sqrt{5}}}\right){\mathbf S}_{75}{\overline{\mathbf S}}_{24}^2{\mathbf S}_{24}-\left({\frac{49}{8640\sqrt{5}}}\right){\overline{\mathbf S}}_{75}{\mathbf S}_{24}^2{\overline{\mathbf S}}_{24}\right.\nonumber\\
&&~~~~~~~~~\left.~{-}\left({\frac{{217}}{17280}}\right){\overline{\mathbf S}}_{24}^2{\mathbf S}_{24}^2
+\left({\frac{1}{1440}}\right){\mathbf S}_{24}^2{\overline{\mathbf S}}_{24}{\overline{\mathbf S}}_{1}+\left({\frac{1}{1440}}\right){\overline{\mathbf S}}_{24}^2{\mathbf S}_{24}{\mathbf S}_{1}\right.\nonumber\\
&&~~~~~~~~~\left.-\left({\frac{1}{480}}\right){\mathbf S}_{24}^2{\overline{\mathbf S}}_{1}^2
-\left({\frac{1}{480}}\right){\overline{\mathbf S}}_{24}^2{\mathbf S}_{1}^2+\left({\frac{1}{144}}\right){\overline{\mathbf S}}_{24}{\mathbf S}_{24}{\mathbf S}_{1}{\overline{\mathbf S}}_{1}~{-}\left(\frac{1}{~{405}}\right){\overline{\mathbf S}}_{1}^2{\mathbf S}_{1}^2\right]\nonumber\\
&&~~~~~~~~~+i M_{560}\left[\frac{1}{4}{\overline{\mathbf S}}_{75}
{\mathbf S}_{75}+\left({1}\right){\overline{\mathbf S}}_{24}
{\mathbf S}_{24}+\left(1\right){\overline{\mathbf S}}_{1}
{\mathbf S}_{1}\right].
\end{eqnarray}
Where $\mathbf S$ and $\overline{\mathbf S}$ represent the $\mathsf{SU(3)_C\times SU(2)_l\times U(1)_Y}$ singlets contained in $SU(5)$'s $\mathsf{1}+, +\mathsf{24}+\mathsf{24}$ and $\mathsf{\overline{1}}+\mathsf{\overline{24}}+\mathsf{\overline{75}}$ multiplets.} Further details of the analysis are given in the Appendix  \ref{sec:B}.
 
Vanishing of the $F-$terms require that we look for solutions of Eq.(\ref{GUTsymmetryvevs}) for which
\begin{eqnarray}\label{F-terms}
\frac{\partial W_{\textsc{gut}}}{\partial{\mathbf S}_{75}}=0,~~~\frac{\partial W_{\textsc{gut}}}{\partial {\overline{\mathbf S}}_{75}}=0,~~~\frac{\partial W_{\textsc{gut}}}{\partial{\mathbf S}_{24}}=0,~~~\frac{\partial W_{\textsc{gut}}}{\partial{\overline{\mathbf S}}_{24}}=0,~~~\frac{\partial W_{\textsc{gut}}}{\partial{\mathbf S}_{1}}=0,~~~\frac{\partial W_{\textsc{gut}}}{\partial{\overline{\mathbf S}}_{1}}=0,
\end{eqnarray}
are satisfied simultaneously.
The solutions to the Eqs.(\ref{F-terms}) are carried out numerically on Mathematica. The number of allowed solutions is quite large.  
We display solutions for the special case when  the VEVs of $\mathsf{1}(-5)$ and $\overline {\mathsf{1}}(5)$, $\mathsf{24}(-5)$ and $\overline{\mathsf{24}}(5)$,  $\mathsf{75}(-5)$ and $\overline{\mathsf{75}}(5)$ are equal. We choose to exhibit a total  of 10 solutions (see Table \ref{table-singlets}), two for each  value of 
{$\displaystyle M^2_{\text{eff}}\equiv \frac{M_{45}\cdot M_{560}}{\lambda_{45}^2}$
where we identify $M_{\text{eff}}$ as the GUY scale $M_{\text{GUT}}$.}

   In this model the doublet-triplet splitting is automatic because of the missing
   partner constraint.
The superpotential that triggers doublet-triplet splitting in symbolic form is given by
\begin{eqnarray}\label{Doublet-Triplet Splitting 1}
W_{\textsc{dt}}&\sim& \mathsf{560}\cdot\mathsf{\overline{560}}~+~\mathsf{{560}}\cdot\mathsf{{560}}\cdot \mathsf{10}_1
~+~\mathsf{\overline{560}}\cdot\mathsf{\overline{560}}\cdot\left(\mathsf{10}_1+\mathsf{10}_2\right)~+~\mathsf{{560}}\cdot\mathsf{{560}}\cdot \mathsf{320}
~+~\mathsf{\overline{560}}\cdot\mathsf{\overline{560}}\cdot \mathsf{320}.
\end{eqnarray}
With full tensorial structure incorporated, Eq.(\ref{Doublet-Triplet Splitting 1}) takes the form
\begin{eqnarray}\label{Doublet-Triplet Splitting 2}
W_{\textsc{dt}}&=&  M_{560}~\braket{\Theta_{\mu\nu}^{(\mathsf{560})*}|B|\overline{\Theta}_{\mu\nu}^{(\overline{\mathsf{560}})}}
+\alpha~\braket{\Theta_{\mu\nu}^{(\mathsf{560})*}|B\Gamma_{\alpha}|\Theta_{\mu\nu}^{(\mathsf{560})}}\Omega^{(\mathsf{10}_1)}_{\alpha}
+\sum_{r=1}^2 {\overline {\alpha}}_r~\braket{\overline{\Theta}_{\mu\nu}^{(\overline{\mathsf{560}})*}|B\Gamma_{\alpha}|\overline{\Theta}_{\mu\nu}^{(\overline{\mathsf{560}})}}
\Omega^{(\mathsf{10_r})}_{\alpha}\nonumber\\
&&+\beta~\braket{\Theta_{\mu\sigma}^{(\mathsf{560})*}|B\Gamma_{\lambda}|\Theta_{\nu\sigma}^{(\mathsf{560})}}\Lambda^{(\mathsf{320_s})}
_{(\mu\nu)\lambda}
+\overline{\beta}~\braket{\overline{\Theta}_{\mu\sigma}^{(\overline{\mathsf{560}})*}|B\Gamma_{\lambda}|
\overline{\Theta}_{\nu\sigma}^{(\overline{\mathsf{560}})}}\Lambda^{(\mathsf{320_s})}_{(\mu\nu)\lambda}.
\end{eqnarray}
Further details of the analysis are given in Appendix \ref{sec:G}.
\section{The Higgs doublet mass matrix \label{sec:4}}
As shown in Table 1
the Higgs boson doublets arise from the
heavy fields $\mathsf{560 +\overline{560}}$ and the light fields which consist of
two $\mathsf{10}-$plets and a $\mathsf{320}-$plet of $\mathsf{SO(10)}$.  We list them below.
\begin{align}
\{{\cal D}^i\} &=
(^{(5_{10_1})}{\mathbf D}^a, ~^{(5_{10_2})}{\mathbf D}^a, ~^{(5_{320})}{\mathbf D}^a,
 ~^{(5_{\overline{560}})}{\mathbf D}^a
 ~^{(45_{320})}{\mathbf D}^a,
   ~^{(45_{560})}{\mathbf D}^a,
   ~^{(45_{\overline{560}})}{\mathbf D}^a,
   ~^{(70_{320})}{\mathbf D}^a,
  ~^{(70_{\overline{560}})}{\mathbf D}^a).
\end{align}
Similarly we have the anti-doublets  given by
\begin{align}
\{\overline{{\cal D}}_j\}&=(^{(\bar{5}_{10_1})}{\mathbf D}_a, ~^{(\bar{5}_{10_2})}{\mathbf D}_a, ~^{(\bar{5}_{320})}{\mathbf D}_a,
 ~^{(\bar{5}_{{560}})}{\mathbf D}_a
 ~^{(\overline{45}_{320})}{\mathbf D}_a,
   ~^{(\overline{45}_{\overline{560}})}{\mathbf D}_a,
   ~^{(\overline{45}_{{560}})}{\mathbf D}_a,
   ~^{(\overline{70}_{320})}{\mathbf D}_a,
  ~^{(\overline{70}_{{560}})}{\mathbf D}_a).
\end{align}
{The above leads to a $9\times 9$ dimensional Higgs doublet  mass matrix.}
  The mass matrix is constructed from the superpotential
  of Eq.(\ref{Doublet-Triplet Splitting 2}) after spontaneous breaking and one has the following
  result:\\

  {\small
\begin{eqnarray}
\label{doublet mass matrix}
 M_{\text{doublet}}=\kbordermatrix{
&{}^{(\overline{5}_{10_1})}\!{{\mathbf D}}_{a} &
{}^{(\overline{5}_{10_2})}\!{\mathbf D}_{a} &
{}^{(\overline{5}_{320})}\!{\mathbf D}_{a}&
 {}^{(\overline{5}_{{560}})}\!{\mathbf D}_{a} &
 {}^{(\overline{45}_{{320}})}\!{\mathbf D}_{a} &
 {}^{(\overline{45}_{\overline{560}})}\!{\mathbf D}_{a} &
{}^{(\overline{45}_{{560}})}\!{\mathbf D}_{a}
&{}^{(\overline{70}_{{320}})}\!{\mathbf D}_{a}&
{}^{(\overline{70}_{{560}})}\!{\mathbf D}_{a}
\cr\cr
{}^{({5}_{10_1})}\!{\mathbf D}^{a} & 0& 0& 0&d_3 &0 &\bar d_2 &d_4&0&d_5\cr\cr
{}^{({5}_{10_2})}\!{\mathbf D}^{a} & 0& 0& 0&0 &0 &\frac{\bar d_2 \bar \alpha_2}{\bar \alpha_1} &0&0&0 \cr\cr
{}^{({5}_{320})}\!{\mathbf D}^{a}& 0& 0& 0&d_7 &0 &\bar d_6 &d_8&0&d_{15} \cr\cr
{}^{({5}_{\overline{560}})}\!{\mathbf D}^{a}& \bar d_3& \frac{\bar d_3 \bar \alpha_2}{\bar \alpha_1}&\bar d_7&d_1 &\bar d_{13} &0 &0&\bar d_{16}&0\cr\cr
{}^{({45}_{{320}})}\!{\mathbf D}^{a}& 0& 0& 0&d_{13} &0 &\bar d_{14} &d_{12}&0&0\cr\cr
{}^{({45}_{560})}\!{\mathbf D}^{a} & d_2& 0& d_6&0 &d_{14} &-\frac{d_1}{2} &0&d_{11}&0 \cr\cr
{}^{({45}_{\overline{560}})}\!{\mathbf D}^{a}& \bar d_4& \frac{\bar d_4 \bar \alpha_2}{\bar \alpha_1}& \bar d_8&0 &\bar d_{12} &0 &\frac{d_1}{2} &\bar d_9&0\cr\cr
{}^{({70}_{{320}})}\!{\mathbf D}^{a}& 0& 0& 0&d_{16} &0 &\bar d_{11} &d_9&0&d_{10}\cr\cr
{}^{({70}_{\overline{560}})}\!{\mathbf D}^{a}& \bar d_5& \frac{d_5 \bar \alpha_2}{\bar \alpha_1}& \bar d_{15}&0 &0 &0 &0&\bar d_{10}&\frac{d_1}{2}\cr\cr
},
\end{eqnarray}
}
  where $d_i$ and $\bar{d}_j$ are defined as follows
  \begin{align}
  d_1&= i M_{560}; &d_2= {\frac{-i\alpha}{2\sqrt{6}} \left(\sqrt{5}{\mathcal V}_{24} + 4
  {\mathcal V}_{75}\right)},\non
  d_3&= {i\alpha\sqrt{\frac{2}{10815}} \left(97{\mathcal V}_{1} +6
  {\mathcal V}_{24}\right)};
    &d_4={\frac{i\alpha}{5\sqrt{87}} \left(25 {\mathcal V}_{24} -
     \sqrt{5}{\mathcal V}_{75}\right)},  \non
     d_5&= {-\frac{1}{4}\sqrt{\frac{15}{2}}} i\alpha {\mathcal V}_{24};  &
     d_6= {\frac{i\beta}{144 \sqrt 2}  \left(23 \sqrt{5} {\mathcal V}_{24} - 16 {\mathcal V}_{75}\right)},  \non
       d_7&=
       {\frac{i\beta}{ 504 \sqrt{7210}} \left(13480{\mathcal V}_{1} + 7761 {\mathcal V}_{24}\right)};  & d_8=
       {\frac{i\beta}{1008 \sqrt{29}} \left(2695{\mathcal V}_{24} + 212\sqrt{5}  {\mathcal V}_{75}\right)}, \non
     d_9&=
       {\frac{i\beta}{120\sqrt{174}} \left(65{\mathcal V}_{24} + 184 \sqrt{5} {\mathcal V}_{75}\right)};
   & d_{10}=
      {\frac{i\beta}{40\sqrt{3}} \left(-8\sqrt{5} {\mathcal V}_1-\sqrt{5}{\mathcal V}_{24}
       +20{\mathcal V}_{75}\right)},\non
d_{11}&=
       {\frac{i\beta}{48\sqrt{3}} \left(-5\sqrt{5}{\mathcal V}_{24} + 16 {\mathcal V}_{75}\right)};
   &d_{12}=
       {\frac{i\beta}{60\sqrt{29}} \left(196{\mathcal V}_1-129{\mathcal V}_{24} + 2\sqrt{5} {\mathcal V}_{75}\right)},   \non
        d_{13}&=
       {\frac{i\beta}{24\sqrt{1442}} \left(217\sqrt{5}{\mathcal V}_{24} - 8{\mathcal V}_{75}\right)};
    &d_{14}=
       {\frac{i\beta}{48\sqrt{10}} \left(-16{\mathcal V}_1+33{\mathcal V}_{24}\right)},  \non
   d_{15}&= {\frac{37 i\beta}{168}\sqrt{\frac{5}{2}} {\mathcal V}_{24}};  &d_{16}= {\frac{37 i\beta}{8}\sqrt{\frac{5}{2163}} {\mathcal V}_{24}},\non
   \bar d_2&=  {\frac{-i\bar\alpha}{2\sqrt{6}} \left(\sqrt{5}\overline{\mathcal V}_{24} + 4
  \overline{\mathcal V}_{75}\right)};
  &\bar d_3={i\bar\alpha\sqrt{\frac{2}{10815}} \left(97\overline{\mathcal V}_{1} +6
  \overline{\mathcal V}_{24}\right)},\non
    \bar d_4&={\frac{i\bar\alpha}{5\sqrt{87}} \left(25 \overline{\mathcal V}_{24} -
     \sqrt{5}\overline{\mathcal V}_{75}\right)};  
     &\bar d_5= {-\frac{1}{4}\sqrt{\frac{15}{2}}} i\bar\alpha  \overline{\mathcal V}_{24},\non
   \bar d_6&= {\frac{i\bar\beta}{144\sqrt{2}} (23\sqrt{5}  \overline{\mathcal V}_{24} -16\overline{\mathcal V}_{75})};
    & \bar d_7={\frac{i\bar\beta}{504\sqrt{7210}} (13480\overline{\mathcal V}_{1} +7761  \overline{\mathcal V}_{24})},\non
      \bar d_8&= {\frac{i\bar\beta}{1008\sqrt{29}} (2695 \overline{\mathcal V}_{24} +
     212\sqrt{5} \overline{\mathcal V}_{75})};
    & \bar d_9={\frac{i\bar\beta}{120\sqrt{174}} (65\overline{\mathcal V}_{24} +84\sqrt{5}\overline{\mathcal V}_{75})},
    \non
       \bar d_{10}&= {\frac{i\bar\beta}{40\sqrt{3}} (-8\sqrt{5} \overline{\mathcal V}_{1}-\sqrt{5}\overline{\mathcal V}_{24}
  + 20 \overline{\mathcal V}_{75})};
    & \bar d_{11}={\frac{i\bar\beta}{48\sqrt{3}} (-5\sqrt{5}\overline{\mathcal V}_{24}+ 16\overline{\mathcal V}_{75})},\non
       \bar d_{12}&= {\frac{i\bar\beta}{60\sqrt{29}} (196\overline{\mathcal V}_1 -129\overline{\mathcal V}_{24} + 2\sqrt 5\overline{\mathcal V}_{75})};
    & \bar d_{13}={\frac{i\bar\beta}{24\sqrt{1442}} (217\sqrt{5}\overline{\mathcal V}_{24}-8\overline{\mathcal V}_{75})},\non
      \bar d_{14}&= {\frac{i\bar\beta}{48\sqrt{10}} (
      - 16\overline{\mathcal V}_1 +33  \overline{\mathcal V}_{24})};
      &\bar d_{15}= {\frac{37i\bar\beta}{168}\sqrt{\frac{5}{2}}\overline{\mathcal V}_{24}},\non
     \bar d_{16}&={\frac{37i\bar\beta}{8}\sqrt{\frac{5}{2163}}\overline{\mathcal V}_{24}}.
     \label{higgs-matrix-elements}
\end{align}

The Higgs doublet mass matrix is diagonalized by two unitary matrices $U$ and $V$
such that
\begin{align}
{U_d^\dagger M_{\text{Higgs-doub}} V_d=  M_{\text{Higgs-doub}}^{\text{diag}} =(m_{d_1},~ m_{d_2},\cdots,~ m_{d_8},0),}
\label{UV}
\end{align}
whose relevant elements are displayed in Eq. (\ref{doublet mass eigenstates}),
\begin{eqnarray}\label{doublet mass eigenstates}
  ~~\begin{pmatrix}{}^{({\overline{5}}_{10_1})}\!{\mathbf D}_{a}\cr {}^{({\overline{5}}_{10_2})}\!{\mathbf D}_{a}\cr{}^{({\overline{5}}_{320})}\!{\mathbf D}_{a}\cr{}^{({\overline{5}}_{\overline{560}})}\!{\mathbf D}_{a} \cr{}^{(\overline{45}_{{320}})}\!{\mathbf D}_{a} \cr{}^{(\overline{45}_{\overline{560}})}\!{\mathbf D}_{a} \cr{}^{(\overline{45}_{{560}})}\!{\mathbf D}_{a}
 \cr{}^{({\overline{70}}_{{320}})}\!{\mathbf D}_{a}
 \cr{}^{({\overline{70}}_{{560}})}\!{\mathbf D}_{a}
\end{pmatrix}= 
{\begin{pmatrix}
\cdots & &  \cdots & & V_{d_{19}}&  \\
 \cdots  &  &  \cdots& & V_{d_{29}}& \\
   \cdots  &  & \cdots& &V_{d_{39}} &  \\
  \cdots   &  & \cdots& & 0 &  \\
   \cdots&  & \cdots &  &V_{d_{59}} & \\
  \cdots  &  & \cdots && 0 &  \\
   \cdots &  &  \cdots&& 0 & \\
  \cdots & &  \cdots&&  V_{d_{89}} &\\
  \cdots & & \cdots &&  0 &\\
 \end{pmatrix}}
 {\begin{pmatrix}
 {}^{1}\!{\mathbf D}^{\prime }_a\cr {}^{2}\!{\mathbf D}^{\prime}_a\cr{}^{3}\!{\mathbf D}^{\prime}_a\cr{}^{4}\!{\mathbf D}^{\prime}_a \cr{}^{5}\!{\mathbf D}^{\prime}_a \cr{}^{6}\!{\mathbf D}^{\prime}_a \cr{}^{7}\!{\mathbf D}^{\prime}_a
  \cr{}^{8}\!{\mathbf D}^{\prime}_a  \cr {\mathbf{H_d}}_a
 \end{pmatrix}}
  ;~~\begin{pmatrix}{}^{({5}_{10_1})}\!{\mathbf D}^{a}\cr {}^{({5}_{10_2})}\!{\mathbf D}^{a}\cr{}^{({5}_{320})}\!{\mathbf D}^{a}\cr{}^{({5}_{\overline{560}})}\!{\mathbf D}^{a} \cr{}^{({45}_{{320}})}\!{\mathbf D}^{a} \cr{}^{({45}_{560})}\!{\mathbf D}^{a} \cr{}^{({45}_{\overline{560}})}\!{\mathbf D}^{a}
 \cr{}^{({70}_{{320}})}\!{\mathbf D}^{a}
 \cr{}^{({70}_{\overline{560}})}\!{\mathbf D}^{a}
\end{pmatrix}= 
{\begin{pmatrix}
 \cdots & &  \cdots & & U_{d_{19}}&  \\
 \cdots  &  &  \cdots& & U_{d_{29}}& \\
   \cdots  &  & \cdots& &U_{d_{39}} &  \\
  \cdots   &  & \cdots& & 0 &  \\
   \cdots&  & \cdots &  &U_{d_{59}} & \\
  \cdots  &  & \cdots && 0 &  \\
   \cdots &  &  \cdots&& 0 & \\
  \cdots & &  \cdots&&  U_{d_{89}} &\\
  \cdots & & \cdots &&  0 &\\
 \end{pmatrix}}
 {\begin{pmatrix}{}^{1}\!{\mathbf D}^{\prime a}\cr {}^{2}\!{\mathbf D}^{\prime a}\cr{}^{3}\!{\mathbf D}^{\prime a}\cr{}^{4}\!{\mathbf D}^{\prime a} \cr{}^{5}\!{\mathbf D}^{\prime a} \cr{}^{6}\!{\mathbf D}^{\prime a} \cr{}^{7}\!{\mathbf D}^{\prime a}
  \cr{}^{8}\!{\mathbf D}^{\prime a}  \cr {\mathbf{H_u}}^a
 \end{pmatrix}},
\end{eqnarray}
 where $\mathbf D$'s and ${\mathbf D}^{\prime}$'s represent the normalized kinetic energy basis and normalized kinetic and mass eigenbasis, respectively of the doublet mass matrix of Eq. (\ref{doublet mass matrix}). The pair of doublets $({\mathbf{H_d}}_a,{\mathbf{H_u}}^a)$ are identified to be light and are the
 normalized   electroweak Higgs doublets of the minimal supersymmetric standard model (MSSM).
 Numerical values of the non-zero matrix elements of  $U$ and $V$ are displayed in Tables \ref{table-U} and \ref{table-V} for benchmarks of Table \ref{table-singlets}.

The doublet mass matrix has a zero mode which gives us the electroweak
doublet pair  of MSSM, i.e., ${\mathbf{H_u}}^a$ which couples to the up-quarks and
${\mathbf{H_d}}_a$ which couples to the down quarks and the leptons. The electroweak
doublets do not have any components in the doublets arising from
$\mathsf{560}$ and $\mathsf{\overline{560}}$ since they are superheavy and are linear
combinations only of the doublets arising from $\mathsf{10}_1$ and $\mathsf{10}_2$ of Higgs
and the doublets arising from the $\mathsf{320}-$plet of Higgs. Thus one may write
$~{{\mathbf{H_u}}^a}$ and $~{{\mathbf{H_d}}_a}$ as the following combination of fields 

\begin{align}
~{{\mathbf{H_u}}^a}=\sum_{k=1}^{5} c_{uk} D^a_k, ~~ ~{{\mathbf{H_d}}_a}=\sum_{k=1}^{5}c_{dk}D_{ak}
 \label{five-Hu}
 \end{align}
where
\begin{align}
\left\{{D}^a_{k=1,\dots,5}\right\} &=~{
\left({}^{({5}_{10_1})}\!{\mathbf D}^{a}, ~{}^{({5}_{10_2})}\!{\mathbf D}^{a}, ~{}^{({5}_{320})}\!{\mathbf D}^{a},
~{}^{({45}_{320})}\!{\mathbf D}^{a},
   ~{}^{({70}_{320})}\!{\mathbf D}^{a}\right)}.
   \label{five-Du}
\end{align}
Similarly we have the anti-doublets  which we display as
\begin{align}
\left\{{D}_{ak=1,\dots,5}\right\} &=~{\left({}^{(\overline{5}_{10_1})}\!{\mathbf D}_{a}, ~{}^{(\overline{5}_{10_2})}\!{\mathbf D}_{a}, ~{}^{(\overline{5}_{320})}\!{\mathbf D}_{a},
~{}^{(\overline{45}_{320})}\!{\mathbf D}_{a},
   ~{}^{(\overline{70}_{320})}\!{\mathbf D}_{a}\right)}.
\end{align}
where $c_{uk}$ and $c_{dk}$ are determined by diagonalizing Higgs doublet
mass matrix of Eq. (\ref{doublet mass matrix}).
The $\mathsf{320}$ has no cubic couplings with quarks and leptons and thus only the components of
the two Higgs doublets arising from  $\mathsf{10}_1$ and $\mathsf{10}_2$ (and not all five components 
arising from Eqs. (\ref{five-Hu}) and (\ref{five-Du})) 
enter in the couplings of the Higgs bosons to quarks and leptons.


\section{Numerical analysis of breaking of $\boldsymbol{\mathsf{SO(10)}}$ to $\boldsymbol{\mathsf{SU(3)_C\times SU(2)_L\times U(1)_Y}}$
and generation of light Higgs doublets}
\label{sec:5}}

 {We discuss now the numerical analysis of the
  $\mathsf{560+\overline{560}+ 320+}2\mathsf{\times 10}$ model to break the $\mathsf{SO(10)}$
  symmetry to the standard model gauge group,  and analyze the Higgs doublet structure to exhibit explicitly a pair of light Higgs doublets that enter in the electroweak symmetry breaking and would generate quark 
  and lepton masses.  First we will
 numerically solve the spontaneous symmetry breaking equations in Section \ref{sec:3}
 and determine the allowed values of VEVs ${\mathcal V}_{1}, {\mathcal V}_{24}, {\mathcal V}_{75}$.  The mass scale that enters in the determination of these VEVs
 is $M_{eff}$ where
  $M_{eff}^2=M_{45}\cdot M_{560}/\lambda^2_{45}$. To determine the
 VEVs we solve the spontaneous symmetry breaking equations Eqs. (\ref{F-terms}). We list in Table \ref{table-singlets},
 the computations of ${\mathcal V}_{1}, {\mathcal V}_{24}, {\mathcal V}_{75}$ for a range of 
   $M_{eff}$ from $10^{15}$ GeV to $10^{17}$ GeV. We note that all of the VEVs, i.e.,
   ${\mathcal V}_{1}$, ${\mathcal V}_{24}$, and ${\mathcal V}_{75}$ are comparable in 
   size and thus enter essentially with equal strength in breaking the gauge group $\mathsf{SO(10)}$ 
   down to the Standard Model  gauge group.   
 The analysis  of Table \ref{table-singlets} in then used to   
 numerically compute the Higgs doublet mass matrix of Eq. (\ref{doublet mass matrix}) where the elements of 
 this mass matrix consisting of $d_1\cdots d_{16}$, $\bar d_6,\cdots ,\bar d_{16}$ 
 given by Eq. (\ref{higgs-matrix-elements}) are determined in terms of the heavy VEVs   ${\mathcal V}_{1}$, ${\mathcal V}_{24}$, and ${\mathcal V}_{75}$.  Diagonalization of the mass matrix of Eq. (\ref{doublet mass matrix}) by 
 the biunitary transformation of Eq. (\ref{UV})  allows us to identify the  massless modes, i.e., the electroweak doublet. The exact composition of the electroweak doublet is given by the 
 transformations of Eq. (\ref{doublet mass eigenstates}) which connect the primitive field doublets labeled by ${\mathbf D}_a,
 {\mathbf D}^a$  to the ones in mass diagonal basis ${\mathbf D}^{\prime }_a, {\mathbf D}^{\prime a}$.   
 The composition of the massless doublets is given in terms of the diagonalizing matrices 
 $U$ and $V$ that appear in Eq. (\ref{doublet mass eigenstates}). The numerical composition of the massless up-Higgs 
 is given in Table \ref{table-U} for a range of $M_{eff}$ values as given in Table \ref{table-singlets}. Similarly  the  numerical composition of the massless down-Higgs  is given in Table \ref{table-V} for same range of $M_{eff}$ values. 
  The analysis above allows a computation of the quark and lepton masses using the
  VEVs of the electroweak doublets.}\\
{
\begin{table}[H]

\begin{center}
\resizebox{\linewidth}{!}{
\begin{tabular}{|c|c|c|c|}
\multicolumn{4}{c}{${\mathcal V}_{75}={\overline{\mathcal V}}_{75}$, ~${\mathcal V}_{24}={\overline{\mathcal V}}_{24}$, ~${\mathcal V}_{1}={\overline{\mathcal V}}_{1}$}\cr \hline \hline
 $\displaystyle M_{eff}^2\equiv \frac{M_{45}\cdot M_{560}}{\lambda_{45}^2}~({\rm GeV}^2)$&${\mathcal V}_{75}~({\rm GeV})$& ${\mathcal V}_{24}~({\rm GeV})$&
 ${\mathcal V}_{1}~({\rm GeV})$
 \\
 \hline\hline
\multirow{2}{*}{$10^{30}$}& {\small{$(-1.082+\imath 2.927)\times10^{16}$}} & {\small{$(-2.844-\imath 6.446)\times 10^{15}$}}&
{\small{$(-4.716-\imath 2.297)\times 10^{16}$}}\\
\cline{2-4}
&{\small{$(7.762+\imath 8.288)\times10^{15}$}} & {\small{$(-7.933+\imath 6.054)\times 10^{15}$}}&
{\small{$(1.156+\imath 0.980)\times 10^{16}$}} \\
\hline\hline
\multirow{2}{*}{$10^{31}$}& {\small{$(7.205+\imath 7.205)\times10^{14}$}} & {\small{$(-3.356-\imath 3.356)\times 10^{16}$}}&
{\small{$(6.813+\imath 6.813)\times 10^{16}$}} \\
\cline{2-4}
&{\small{$(-1.661-\imath 2.608)\times10^{16}$}} & {\small{$(2.207-\imath 0.100)\times 10^{16}$}}&
{\small{$(3.578-\imath 5.238)\times 10^{15}$}}\\
\hline\hline
\multirow{2}{*}{$10^{32}$}& {\small{$(-0.588-\imath 1.223)\times10^{17}$}} & {\small{$(1.342-\imath 0.251)\times 10^{17}$}}&
{\small{$(1.731+\imath 1.577)\times 10^{17}$}} \\
\cline{2-4}
&{\small{$(-3.691-\imath 3.691)\times10^{16}$}} & {\small{$(-4.226-\imath 4.226)\times 10^{15}$}}&
{\small{$(2.287+\imath2.287)\times 10^{14}$}} \\
\hline\hline
\multirow{2}{*}{$10^{33}$}& {\small{$(-9.258+\imath 3.422)\times10^{17}$}} & {\small{$(2.038+\imath 0.899)\times 10^{17}$}}&
{\small{$(0.726+\imath 1.491)\times 10^{18}$}}\\
\cline{2-4}
&{\small{$(2.621+\imath 2.455)\times10^{17}$}} & {\small{$(1.915-\imath2.509)\times 10^{17}$}}&
{\small{$(3.098-\imath3.655)\times 10^{17}$}} \\
\hline\hline
\multirow{2}{*}{$10^{34}$}& {\small{$(-8.248-\imath 5.251)\times10^{17}$}} & {\small{$(-0.318+\imath6.978 )\times 10^{17}$}}&
{\small{$(-1.657+\imath 1.132)\times 10^{17}$}}\\
\cline{2-4}
&{\small{$(-0.588-\imath 1.223)\times10^{18}$}} & {\small{$(1.342-\imath0.251)\times 10^{18}$}}&
{\small{$(1.731+\imath1.577)\times 10^{18}$}}\\
\hline
\hline
\end{tabular}}
\caption{Numerical estimates of the VEVs of the  singlet, the $\mathsf{24-}$plet and the $\mathsf{75-}$plet fields in $\mathsf{560}+\overline{\mathsf{560}}$ multiplets
in the spontaneous breaking of the $\mathsf{SO(10)}$ gauge symmetry {to $\mathsf{SU(5)\times U(1)}$}
at the  GUT scale.}
\label{table-singlets}
\end{center}
\end{table}
 {
\begin{table}[H]

\begin{center}
\resizebox{\linewidth}{!}{
\begin{tabular}{|c|c|c|c|c|c|c|c|}
\multicolumn{8}{c}
{~{\Large$M_{560}=1.0\times 10^{18}~{\rm GeV}$,  $\alpha=\overline{\alpha}_1=\overline{\alpha}_2=1.0$, $\beta=\overline{\beta}=1.0$}}\\
\hline\hline
\Large{${\mathcal V}_{75}~({\rm GeV})$}&\Large{${\mathcal V}_{24}~({\rm GeV})$}&\Large{${\mathcal V}_{1}~({\rm GeV})$}&\Large{$U_{d{_{19}}}$}&\Large{$U_{d{_{29}}}$}&\Large{${U_{d{_{39}}}}$}&\Large{${U_{d{_{59}}}}$}&\Large{${U_{d{_{89}}}}$} \\
\hline
 {\small{$(-1.082+\imath 2.927)\times10^{16}$}} & {\small{$(-2.844-\imath 6.446)\times 10^{15}$}}&
{\small{$(-4.716-\imath 2.297)\times 10^{16}$}} &{\small{$0.230 +\imath0.000$}}  & {\small{$-0.169+\imath0.074$}} & {\small{$-0.926-\imath0.058$}} & {\small{$0.059+\imath0.097$}} &{\small{ $-0.196+\imath0.020$}}\\
\hline
 {\small{$(7.762+\imath 8.288)\times10^{15}$}} & {\small{$(-7.933+\imath 6.054)\times 10^{15}$}}&
{\small{$(1.156+\imath 0.980)\times 10^{16}$}} &{\small{$-0.019 +\imath0.000$}}  & {\small{$0.289+\imath0.239$}} & {\small{$-0.052+\imath0.147$}} & {\small{$-0.232+\imath0.170$}} &{\small{ $0.846-\imath0.193$ }}\\
\hline
{\small{$(7.205+\imath 7.205)\times10^{14}$}} & {\small{$(-3.356-\imath 3.356)\times 10^{16}$}}&
{\small{$(6.813+\imath 6.813)\times 10^{16}$}} &{\small{$0.112 
+\imath0.000$}}  & {\small{$-0.672+\imath0.000$}} & {\small{$-0.504+\imath0.000$}} & {\small{$-0.105+\imath0.000$}} &{\small{ $0.520+\imath0.000$}}\\
\hline
{\small{$(-1.661-\imath 2.608)\times10^{16}$}} & {\small{$(2.207-\imath 0.100)\times 10^{16}$}}&
{\small{$(3.578-\imath 5.238)\times 10^{15}$}} &{\small{$-0.196
+\imath0.000$}}  & {\small{$0.189+\imath0.230$}} & {\small{$0.370-\imath0.314$}} & {\small{$-0.361+\imath0.073$}} &{\small{ $0.648+\imath0.286$ }}\\
\hline
{\small{$(-0.588-\imath 1.223)\times10^{17}$}} & {\small{$(1.342-\imath 0.251)\times 10^{17}$}}&
{\small{$(1.731+\imath 1.577)\times 10^{17}$}} &{\small{$0.235
+\imath0.000$}}  & {\small{$-0.084+\imath0.108$}} & {\small{$-0.715+\imath0.257$}} & {\small{$-0.003+\imath0.145$}} &{\small{ $-0.390-\imath0.418$ }}\\
\hline
{\small{$(-3.691-\imath 3.691)\times10^{16}$}} & {\small{$(-4.226-\imath 4.226)\times 10^{15}$}}&
{\small{$(2.287+\imath2.287)\times 10^{14}$}} &{\small{$-0.630
+\imath0.000$}}  & {\small{$0.552+\imath0.000$}} & {\small{$0.489+\imath0.000$}} & {\small{$-0.045+\imath0.000$}} &{\small{ $-0.240+\imath0.000$ }}\\
\hline
{\small{$(-9.258+\imath 3.422)\times10^{17}$}} & {\small{$(2.038+\imath 0.899)\times 10^{17}$}}&
{\small{$(0.726+\imath 1.491)\times 10^{18}$}} &{\small{$-0.230
+\imath0.000$}}  & {\small{$0.169+\imath0.074$}} & {\small{$0.926-\imath0.058$}} & {\small{$-0.059+\imath0.097$}} &{\small{ $0.196+\imath0.020$ }}\\
\hline
{\small{$(2.621+\imath 2.455)\times10^{17}$}} & {\small{$(1.915-\imath2.509)\times 10^{17}$}}&
{\small{$(3.098+\imath3.655)\times 10^{17}$}}  &{\small{$-0.019
+\imath0.000$}}  & {\small{$0.289-\imath0.239$}} & {\small{$-0.052-\imath0.147$}} & {\small{$-0.232-\imath0.170$}} &{\small{ $0.846+\imath0.193$}}\\
\hline
{\small{$(-8.248-\imath 5.252)\times10^{17}$}} & {\small{$(-0.318+\imath6.978 )\times 10^{17}$}}&
{\small{$(-1.657+\imath 1.132)\times 10^{17}$}} &{\small{$0.196
+\imath0.000$}}  & {\small{$-0.189+\imath0.230$}} & {\small{$-0.370-\imath0.314$}} & {\small{$0.361+\imath0.073$}} &{\small{ $-0.648+\imath0.286$ }}\\
\hline
{\small{$(-0.588-\imath 1.223)\times10^{18}$}} & {\small{$(1.342-\imath0.251)\times 10^{18}$}}&
{\small{$(1.731+\imath1.577)\times 10^{18}$}} &{\small{$-0.235
+\imath0.000$}}  & {\small{$0.084-\imath0.108$}} & {\small{$0.715-\imath0.257$}} & {\small{$0.003-\imath0.145$}} &{\small{ $0.390+\imath0.418$}}\\
\hline\hline
\end{tabular}}
\caption{{\small A numerical estimate of the elements of the \emph{up} Higgs zero mode eigenvector using the analysis of Table \ref{table-singlets} and the Higgs doublet mass matrix, $M_d$, displayed in Eq. (\ref{doublet mass matrix}) {and $U$ and $V$ defined in Eq.(\ref{UV})}}}
\label{table-U}
\end{center}
\end{table}
\begin{table}[H]

\begin{center}
\resizebox{\linewidth}{!}{
\begin{tabular}{|c|c|c|c|c|c|c|c|}
\multicolumn{8}{c}
{~{\Large $M_{560}=1.0\times 10^{18}~{\rm GeV}$,  $\alpha=\overline{\alpha}_1=\overline{\alpha}_2=1.0$, $\beta=\overline{\beta}=1.0$}}\\
\hline\hline
\Large{${\mathcal V}_{75}~({\rm GeV})$}&\Large{${\mathcal V}_{24}~({\rm GeV})$}&\Large{${\mathcal V}_{1}~({\rm GeV})$}&\Large{$V_{d{_{19}}}$}&\Large{$V_{d{_{29}}}$}&\Large{${V_{d{_{39}}}}$}&\Large{${V_{d{_{59}}}}$}&\Large{${V_{d{_{89}}}}$} \\
\hline\rule{0pt}{3ex}
 {\small{$(-1.082+\imath 2.927)\times10^{16}$}} & {\small{$(-2.844-\imath 6.446)\times 10^{15}$}}&
{\small{$(-4.716-\imath 2.297)\times 10^{16}$}} &{\small{$0.086 +\imath0.049$}}  & {\small{$-0.044+\imath0.183$}} & {\small{$-0.223-\imath0.923$}} & {\small{$0.109+\imath0.043$}} &{\small{ $-0.014-\imath0.201$ }}\\
\hline
{\small{$(7.762+\imath 8.288)\times10^{15}$}} & {\small{$(-7.933+\imath 6.054)\times 10^{15}$}}&
{\small{$(1.156+\imath 0.980)\times 10^{16}$}} &{\small{$-0.296 +\imath0.166$}}  & {\small{$0.313-\imath0.162$}} & {\small{$0.018+\imath0.145$}} & {\small{$0.180+\imath0.202$}} &{\small{ $-0.740-\imath0.345$ }}\\
\hline
{\small{$(7.205+\imath 7.205)\times10^{14}$}} & {\small{$(-3.356-\imath 3.356)\times 10^{16}$}}&
{\small{$(6.813+\imath 6.813)\times 10^{16}$}} &{\small{$0.408 
-\imath0.272$}}  & {\small{$-0.490+\imath0.327$}} & {\small{$0.368-\imath0.246$}} & {\small{$0.076-\imath0.051$}} &{\small{ $-0.379+\imath0.253$ }}\\
\hline
{\small{$(-1.661-\imath 2.608)\times10^{16}$}} & {\small{$(2.207-\imath 0.100)\times 10^{16}$}}&
{\small{$(3.578-\imath 5.238)\times 10^{15}$}} &{\small{$-0.115
-\imath0.197$}}  & {\small{$-0.057+\imath0.290$}} & {\small{$0.472+\imath0.100$}} & {\small{$-0.350+\imath0.106$}} &{\small{ $0.432-\imath0.554$ }}\\
\hline
{\small{$(-0.588-\imath 1.223)\times10^{17}$}} & {\small{$(1.342-\imath 0.251)\times 10^{17}$}}&
{\small{$(1.731+\imath 1.577)\times 10^{17}$}} &{\small{$-0.022
+\imath0.186$}}  & {\small{$-0.138-\imath0.011$}} & {\small{$0.680-\imath0.358$}} & {\small{$0.110+\imath0.096$}} &{\small{$-0.046-\imath0.576$ }}\\
\hline
{\small{$(-3.691-\imath 3.691)\times10^{16}$}} & {\small{$(-4.226-\imath 4.226)\times 10^{15}$}}&
{\small{$(2.287+\imath2.287)\times 10^{14}$}}  &{\small{$0.036
-\imath0.094$}}  & {\small{$0.250-\imath0.661$}} & {\small{$-0.222+\imath0.586$}} & {\small{$0.020-\imath0.054$}} &{\small{ $0.109-\imath0.288$}}\\
\hline
{\small{$(-9.258+\imath 3.422)\times10^{17}$}} & {\small{$(2.038+\imath 0.899)\times 10^{17}$}}&
{\small{$(0.726+\imath 1.491)\times 10^{18}$}} &{\small{$-0.075
-\imath0.064$}}  & {\small{$-0.157+\imath0.103$}} & {\small{$0.944-\imath0.098$}} & {\small{$-0.076-\imath0.088$}} &{\small{ $0.194-\imath0.053$ }}\\
\hline
{\small{$(2.621+\imath 2.455)\times10^{17}$}} & {\small{$(1.915-\imath2.509)\times 10^{17}$}}&
{\small{$(3.098-\imath3.655)\times 10^{17}$}} &{\small{$-0.088
-\imath0.328$}}  & {\small{$0.103-\imath0.338$}} & {\small{$0.117-\imath0.088$}} & {\small{$0.270-\imath0.012$}} &{\small{ $-0.764-\imath0.288$ }}\\
\hline
{\small{$(-8.248-\imath 5.252)\times10^{17}$}} & {\small{$(-0.318+\imath6.978 )\times 10^{17}$}}&
{\small{$(-1.657+\imath 1.132)\times 10^{17}$}} &{\small{$-0.047
+\imath0.224$}}  & {\small{$-0.146-\imath0.257$}} & {\small{$0.415-\imath0.245$}} & {\small{$-0.365+\imath0.011$}} &{\small{ $0.586+\imath0.389$ }}\\
\hline
{\small{$(-0.588-\imath 1.223)\times10^{18}$}} & {\small{$(1.342-\imath0.251)\times 10^{18}$}}&
{\small{$(1.731+\imath1.577)\times 10^{18}$}} &{\small{$0.185
-\imath0.029$}}  & {\small{$0.027+\imath0.136$}} & {\small{$-0.528-\imath0.558$}} & {\small{$0.063-\imath0.132$}} &{\small{ $-0.542+\imath0.200$ }}\\
\hline\hline
\end{tabular}}
\caption{{\small A numerical estimate of the elements of the \emph{down} Higgs zero mode eigenvector using the analysis of Table \ref{table-singlets} and the Higgs doublet mass matrix, $M_d$, displayed in Eq. (\ref{doublet mass matrix})  and $U$ and $V$ defined in Eq.(\ref{UV}).
     }}
\label{table-V}
\end{center}
\end{table}
}

\section{
 Yukawa unification, light sparticles, and  SUSY implications at LHC
\label{pheno}}

In this section, we discuss several phenomenological aspects of the model
within supergravity grand unification\cite{Chamseddine:1982jx,Hall:1983iz,Arnowitt:1992aq}. These include 
Yukawa unification for the third generation of quarks and leptons, the muon anomaly,
a split sparticle spectrum with a compressed low mass sparticle spectrum lying in the 
few hundred GeV  region  and a high mass spectrum with masses ranging over several TeV. 
The compressed low mass spectrum looks promising for SUSY discovery at the LHC.
Regarding $b-t-\tau$ unification, in typical $\mathsf{SO(10)}$ models
such a unification comes about for a large $\tan\beta$, as large as $\tan\beta=50$\cite{Ananthanarayan:1991xp}. However, for the case of  $\mathsf{SO(10)}$  under consideration here, one finds that a splitting of the $b-t$ Yukawa couplings occurs at the GUT scale which leads to a $b-t-\tau$ unification with a low $\tan\beta$ which can be as low as $\tan\beta=5-10$.
We give now some further details of the Yukawa unification analysis. We note that
since there is no cubic interaction of matter fermions with the $\mathsf{320}$ and only $\mathsf{10_1},\mathsf{10_2}$ enters in the Yukawa couplings, so the cubic superpotential of fermions with the Higgs fields takes the form
\begin{align}
    W_3 = \sum_{r=1}^2f^{10_r}\left<\Psi^*_{(+)}|B\Gamma_\mu|\Psi_{(+)}\right>^r\Omega_\mu
\end{align}
Thus the third generation Yukawas are given by
\begin{align}
    h_{\tau}^0 = i2\sqrt{2}\sum_{r=1}^2f^{10_r}V_{d_{r9}},\quad h_{b}^0 = -i2\sqrt{2}\sum_{r=1}^2f^{10_r}V_{d_{r9}},\quad h_{t}^0 = -i2\sqrt{2}\sum_{r=1}^2f^{10_r}U_{d_{r9}}.\label{eq:yukawa0}
\end{align}
The numerical analysis of the Yukawa couplings at the electroweak scale
which generate the observe quark lepton masses,
is done within the framework of supergravity grand unified models using non-universities of gaugino masses. In order to do $b-t-\tau$ unification, we need to run renormalization group equations and calculate mass spectrum with \textbf{SPheno-4.0.4} \cite{Porod:2003um,Porod:2011nf}. 
After the RG evolution, the quantities of interest at the electroweak scale are
{$h_t,h_b,h_{\tau}, m_{h^0}, \Delta a_\mu,\Omega h^2$, where $m_{h^0}$ is Higgs mass, $\Delta a_\mu$ is muon anomaly defined so that $\Delta a_{\mu} =\Delta (g_\mu -2)/2$
 and $\Omega h^2$ is the dark matter  relic density calculated by \textbf{MicrOmega-6.1.15} \cite{Alguero:2023zol}. For the Higgs mass and the
 relic density we take the experiment constraints so that 
\begin{align}
    m_{h^0} &= 125\pm2 ~\text{GeV} \label{h0constraint}\\
    \Omega h^2 &<0.12 \label{relicconstraint}.
\end{align}

Regarding $\Delta a_\mu$,
the recent Fermilab experimental measurement\cite{Muong-2:2023cdq}
 finds
a deviation from the standard model result of $\Delta a_\mu = (24.9\pm 4.8)\times10^{-10}$\cite{Muong-2:2023cdq,Chakraborti:2023pis}.
 An important contribution to 
the muon moment comes from the hadronic vacuum polarization (HVP). 
The Fermilab analysis uses $e^+e^-$ cross section 
to hadrons in its evaluation of HVP.   The analyses 
of HVP as given by lattice gauge calculations\cite{Borsanyi:2020mff,Ce:2022kxy,ExtendedTwistedMass:2022jpw}
 indicate a somewhat smaller
deviation from the standard model prediction. However, the situation on the 
lattice gauge calculations is not fully settled. For that reason we will use the Fermilab
$\Delta a_\mu$ constraint but with a two $\sigma$ corridor.  
As pointed out early on\cite{Yuan:1984ww} 
the light sleptons and light weakinos can contribute
to the muon anomalous moment and produce a sizable electroweak correction and thus there exists a connection between 
the Fermilab experiment and the search for sparticles at the LHC\cite{Everett:2001tq,Chattopadhyay:2001vx,Aboubrahim:2021xfi}.}

The analysis of $b-t-\tau$ unification
is done within the renormalization group analysis of
$\tilde g$SUGRA~\cite{Akula:2013ioa} with gluino mass driven radiative
breaking of the electroweak symmetry.}
To identify the allowed $\tilde g$SUGRA parameters, we  scan the parameter space and find models that (1) satisfy all experimental constraints and (2) yield third generation fermion masses arising from 
Eq. (\ref{eq:yukawa0}) consistent with data. 
{A direct Monte Carlo (MC) scan of the parameter space is done in the following range 
\begin{align}
    m_0 &\in (0,1000)~\textrm{GeV}\quad A_0 \in (-5000,5000)~\textrm{GeV}\\
    m_1 &\in (0,1500)~\textrm{GeV}\quad m_2 \in (0,1500)~\textrm{GeV}\quad m_3 \in (0,10000)~\textrm{GeV}\\
    \tan{\beta} &\in (0,30)\quad m_t \in (172.2,173.0)~\textrm{GeV}\quad \bar{m}_b(\bar{m}_b)\in (4.15,4.22)~\textrm{GeV}
\end{align}
With values of \( U_{d_{r9}} \) and \( V_{d_{r9}} \) provided in Tables \ref{table-U} and \ref{table-V} and input parameter $f^{10_r}$ in Table(\ref{tab:yukawa}), we find the result of Yukawa couplings of the top, the bottom and the tau using  Eq.(\ref{eq:yukawa0}).}  
In Table(\ref{tab:SUGRA}) we exhibit  SUGRA parameter sets used in RG analysis 
consistent with the experimental data on the top and the  bottom quark masses and for the tau lepton mass.
In Table(\ref{tab:constraints})
we display the light sparticle mass spectrum consistent with the relic density constraint and with 
the experimental $g_\mu-2$ constraint. Here it is seen that the lightest sparticles are the neutralino, 
the stau, the tau-sneutrino, the first two generation sneutrinos, and the first two generation sleptons. 
In Fig.(\ref{fig:RG}) we exhibit the RG evolution of the  sparticles from the GUT scale to the electroweak 
scale in $\tilde g$SUGRA model for the input parameters of the  SUGRA model (a)  of 
Table(\ref{tab:SUGRA}). Here in the left panel, we find that while the sleptons and squarks all start with the same universal 
 $m_0$ for the scalar mass, the RG evolution splits their masses widely at the electroweak scale. Thus
the slepton masses remain in the few hundred GeV region, while the squark masses are driven to high
values in the several TeV region due to their color interactions driven by the high gluino mass.
The right panel exhibits the RG evolution of the Yukawa couplings from the GUT scale to the electroweak
scale. Quite remarkably one finds that there is splitting of the $b-t$ Yukawas as the GUT scale due to 
the bi-unitary transformations needed to diagonalize the Higgs boson mass as given by Eq.(\ref{UV}).
This splitting $b-t$ is significant enough to achieve a $b-t-\tau$ unification for values of $\tan\beta$
between $\sim (5-10)$ as illustrated in Fig.(\ref{fig:ybytm0A0}). Further, as noted already the $\tilde g$SUGRA model evolution splits the sparticle spectrum into two distinct categories: a light sparticle spectrum consisting of sleptons and weakinos and a heavy sparticle spectrum consisting of heavy gluino, heavy squarks and 
heavy Higgses $H^0$, $A^0$ and $H^\pm$. This is illustrated in Fig. (3).
Here we illustrate the light spectrum vs heavy spectrum split for the case of two models (a) and (c) of
Table(\ref{tab:SUGRA}) where the upper panels are for case (a) and the lower panels are for case (c).
 The left panel in each case exhibits the scale of splitting of the light and heavy mass spectrum, 
 while the right panel focuses on the mass hierarchy of the light mass spectrum. The right panels are
 of special interest for the possible discovery of sparticles at the LHC because of their relatively low 
 masses and there is the possibility that one or two of these sparticles may be accessible at the LHC in
 future runs. \\

  As noted above because of the split sparticle spectrum in the  $\mathsf{SO(10)}$ model with $\tilde g$SUGRA 
  RG running,  possibility exists for the discovery of light sparticles at the LHC. To make it more concrete
  we estimate now the production cross sections for some of the light sparticles at the 
  LHC. Thus the production cross sections for the first and the second generation sleptons in $pp$ collisions, i.e.,
  $\sigma(pp\to \tilde e_L \tilde e_L)$, $\sigma(pp\to \tilde e_R \tilde e_R)$,  
$\sigma(pp\to \tilde \mu_L \tilde \mu_L)$, $\sigma(pp\to \tilde \mu_R \tilde \mu_R)$,
and $\sigma(pp\to \tilde \nu_{\mu} \tilde \nu_{\mu})$  (while $\sigma(pp\to \tilde \nu_{e} \tilde \nu_{e})$
is similar size as for the $\mu$ case)   at $\sqrt{s} = 14$ TeV and $\sqrt{s}=27$ TeV, are given in 
Table (\ref{tab:crosssection}).  
In Table (\ref{tab:crosssection-tau}) we give the production cross-sections for the staus, and for the
stau-neutrinos, i.e., for 
$\sigma(pp\rightarrow\tilde{\tau_1}\tilde{\tau_1})$,$\sigma(pp\rightarrow\tilde{\tau_2}\tilde{\tau_2})$,
$\sigma(pp\rightarrow\tilde{\tau_1}\tilde{\tau_2})$, 
$\sigma(pp\rightarrow\tilde{\nu}_{\tau}\tilde{\nu}_{\tau})$,  and 
$\sigma(pp\rightarrow\tilde{\tau}_1\tilde{\nu}_{\tau})$, at $\sqrt{s} = 14$ TeV and $\sqrt{s}=27$ TeV.  
 Table (\ref{stau-BR}) gives a display of the branching ratios of $\tilde \tau_1$, $\tilde \tau_2$ and 
  $\tilde \nu_\tau$ exhibiting the fact that the stau-production and their subsequent decays produce a 
  rich array of signatures.  
 (For recent works on stau phenomenology 
  see\cite{Aboubrahim:2021phn,Chakraborti:2023pis,Baer:2024hgq} 
  and for related earlier works see \cite{Ellis:1998kh,Arnowitt:2006jq,Aboubrahim:2017aen}. 
    One of the most interesting signatures in supersymmetry is the 
  trileptonic signature arising from the final state $\chi^0_2\chi_1^{\pm}$
\cite{Weinberg:1982tp,Arnowitt:1983ah,Chamseddine:1983eg,Nath:1983iz,Dicus:1983cb,Baer:1986dv,Nath:1987sw}. 
  Here the decay of the
  $\chi_2^0\to \ell^+\ell^-\chi^0_1$ and $\chi_1^{\pm}\to \ell^{\pm}+ \nu(\bar\nu)+ \chi^0_1$  will lead to 
  the trileptonic signature $\chi^0_2\chi_1^{\pm}  \to \ell_1^+\ell_2^+\ell_2^- + \text{missing ~energy}$.
  Typically this signal has a low background from the standard model processes. In the analysis of the low
 scale mass spectrum one finds that the masses of $\chi^0_2\chi_1^{\pm}$ lie in the few hundred GeV
 region. It is of interest to estimate the size of the production cross section $ \sigma(pp\to \chi^0_2\chi_1^{\pm})$.   
 The estimate of the sizes for the chosen benchmarks are given in  Table (\ref{tab:crosssection-chi2}). 
 From the table one finds the $ \sigma(pp\to \chi^0_2\chi_1^{\pm})$ cross section can be as large as
 $\sim 20$ fb for some entries which implies that HL-LHC with an integrated luminosity of 3000 fb$^{-1}$
 will produce $\sim 6\times 10^4$ events. There is then a good chance that some of these event may
 survive the stringent cuts at the LHC and be detectable. A more detailed analysis of signal events 
  is outside the scope of this work but is of interest for further analysis.

\begin{table}[h]
    \centering
    \scriptsize
    \begin{tabular}{|c|c|ccc|}
    \hline
        Model &   $f^{(10_r)}$ & $h_{t}^0$ & $h_{b}^0$ & $h_{\tau}^0$\\
    \hline
         (a) &  $(0.707-0.283i,-0.954-0.019i)$& 0.484 & 0.051 & 0.051\\
         (b) &  (0.567,0.046)& 0.498 & 0.033 & 0.033\\
         (c) &  (0.475,0.103)& 0.486 & 0.046 & 0.046\\
         (d) &  $(1.09+0.918i,-0.013+0.936i)$& 0.503 & 0.050 & 0.050\\
         (e) &  $(0.570+0.379i,1.12-0.362i)$& 0.486 & 0.062 & 0.062\\
         (f) &  $(0.218,0.213)$& 0.487 & 0.056 & 0.056\\
         (g) &  $(0.595+0.196i,0.781-0.063i)$& 0.485 & 0.044 & 0.044\\
         (h) &  $(0.459,0.056)$& 0.489 & 0.043 & 0.043\\
         (i) &  $(1.08+0.901i,-0.0062+0.927i)$& 0.485 & 0.043 & 0.043\\
         (j) &  $(0.566+0.378i,1.12-0.360i)$& 0.484 & 0.055 & 0.055\\
    
    \hline
    \end{tabular}
    \caption{Input parameter $f^{10_r}$ and {the corresponding} 
 Yukawa couplings of top, bottom and tau  given by Eq.(\ref{eq:yukawa0}). }
    \label{tab:yukawa}
\end{table}

\begin{table}[h]
    \small
    \centering
    \begin{tabular}{|c|cccccc|ccc|ccc|}
    \hline
        Model &  $m_0$ & $A_0$ & $m_1$ & $m_2$ & $m_3$ & $\tan\beta$ & $m_t$ & $\bar{m_b}(\bar{m_b})$ & $m_\tau$\\
    \hline
         (a)  &  533.0& -2502 & 814 & 815 &  7945& 9.7 & 172.4 & 4.15& 1.77682\\
         (b)  &  363.7& -473 & 740 & 766 &  6722& 5.9 & 172.9 & 4.20& 1.77682 \\
         (c)  &  464.2& -3681 & 561 & 676 &  6616& 8.6 & 172.6 & 4.15& 1.77682 \\
         (d)  &  398.8& -1759 & 859 & 1111 &  8566& 9.2 & 172.7 & 4.21& 1.77682 \\
         (e)  &  486.4& 1233 & 882 & 876 &  7514& 11.6 & 172.6 & 4.20& 1.77682 \\
         (f)  &  568.0& 831 & 896 & 733 &  7987& 10.4 & 172.8 & 4.17& 1.77682  \\
         (g)  &  373.4& -3413 & 674 & 1013 & 7732& 8.1 & 172.2 & 4.19& 1.77682  \\
         (h)  &  497.9& 2901 & 894 & 759 &  7589& 7.9 & 172.6 & 4.22& 1.77682  \\
         (i)  &  433.7& 930 & 762 & 833 &  7653& 8.1 & 172.2 & 4.16& 1.77682  \\
         (j)  &  624.8& 4216 & 764 & 742 &  8836& 10.2 & 172.4 & 4.19& 1.77682  \\
         
    \hline
    \end{tabular}
    \caption{The SUGRA parameters sets used for renormalization group analysis of masses of the top, the bottom and the $\tau$ lepton, using their Yukawa couplings at the GUT scale as 
    input as given in Table (\ref{tab:yukawa}).}
    \label{tab:SUGRA}
\end{table}

\begin{table}[!h]
    \centering
    \small
    \begin{tabular}{|c|cccccccccccc|}
    \hline
    Model & $h_0$ & $\Tilde{\chi}_1^0$ & $\Tilde{\tau}_1$ & $\Tilde{\nu_\tau}$ & $\Tilde{\nu_\mu}$ & $\Tilde{\mu}$ & $\Tilde{\tau}_2$ & $\Tilde{\chi}_1^\pm$ & $\Tilde{\chi}_2^0$ & $\Tilde{\chi}_2^\pm$ & $\Omega h^2$ & $\Delta a_\mu (\times10^{-9})$\\
    \hline
    (a) & 124.4 & 313 & 331 & 410 & 449 & 456 & 628 & 575 & 575 & 7993 & 0.0643 & 2.15\\  
    (b) & 123.3 & 283 & 284 & 325 & 334 & 343 & 490 & 543 & 543 & 6772 & 0.0183 & 3.18\\
    (c) & 123.7 & 206 & 229 & 326 & 379 & 387 & 519 & 473 & 473 & 7072 & 0.1220 & 2.81\\
    (d) & 124.2 & 330 & 339 & 466 & 495 & 500 & 590 & 824 & 824 & 8418 & 0.0146 & 2.65\\
    (e) & 124.3 & 344 & 366 & 460 & 483 & 489 & 627 & 630 & 630 & 7116 & 0.0937 & 1.92\\
    (f) & 124.8 & 347 & 361 & 419 & 440 & 447 & 673 & 498 & 497 & 7602 & 0.0538 & 2.05\\
    (g) & 123.1 & 251 & 273 & 416 & 455 & 460 & 526 & 752 & 752 & 7965 & 0.0771 & 3.07\\
    (h) & 124.0 & 346 & 347 & 386 & 409 & 417 & 596 & 521 & 521 & 7099 & 0.0240 & 1.93\\
    (i) & 123.7 & 288 & 296 & 358 & 373 & 381 & 551 & 588 & 588 & 7303 & 0.0281 & 3.72\\
    (j) & 124.1 & 280 & 303 & 370 & 436 & 443 & 642 & 487 & 487 & 8016 & 0.1210 & 2.86\\
    
    \hline
    \end{tabular}
    \caption{Low scale SUSY mass spectrum showing the Higgs boson $h^0$ mass consistent with 
    the constraint of Eq.(\ref{h0constraint}), and the masses of     
    $\Tilde{\chi}_1^0$, $\Tilde{\tau}_1$, $\Tilde{\nu_\tau}$, $\Tilde{\nu_\mu}$, $\Tilde{\mu}$, 
    $\Tilde{\tau}_2$, $\Tilde{\chi}_1^\pm$, $\Tilde{\chi}_2^0$ and $\Tilde{\chi}_2^\pm$
    for the set of benchmarks (a)-(j) of Table (\ref{tab:SUGRA}).    
    The corresponding  relic density $\Omega h^2$ consistent with the constraint of Eq.(\ref{relicconstraint})
    and the muon anomaly  
     $\Delta a_\mu (\times10^{-9})$
    consistent with the muon anomaly constraint is also exhibited.}    
    \label{tab:constraints}
\end{table}
\begin{table}[h]
    \centering
    \footnotesize
    \begin{tabular}{|c|cc|cc|cc|cc|cc|cc|}
        \hline
    Model & \multicolumn{2}{c|}{$\sigma(pp\rightarrow\tilde{e}_L\tilde{e}_L)$} & \multicolumn{2}{c|}{$\sigma(pp\rightarrow\tilde{\mu}_L\tilde{\mu}_L)$}  & \multicolumn{2}{c|}{$\sigma(pp\rightarrow\tilde{e}_R\tilde{e}_R)$} & \multicolumn{2}{c|}{$\sigma(pp\rightarrow\tilde{\mu}_R\tilde{\mu}_R)$}  & \multicolumn{2}{c|}{$\sigma(pp\rightarrow\tilde{\nu}_e\tilde{\nu}_e)$} & \multicolumn{2}{c|}{$\sigma(pp\rightarrow\tilde{\nu}_\mu\tilde{\nu}_\mu)$}   \\
        \hline
          & 14TeV & 27TeV &  14TeV & 27TeV & 14TeV & 27TeV &  14TeV & 27TeV& 14TeV & 27TeV &  14TeV & 27TeV\\
    \hline
    (a) & 0.936 & 3.619& 0.940 & 3.632 & 0.077 & 0.382 & 0.078 & 0.382 & 0.932 & 3.621 & 0.933 & 3.625 \\
    (b) & 3.208 & 10.577& 3.216 & 10.603 & 0.317 & 1.236 & 0.316& 1.235& 3.358 & 1.110 & 3.359 & 1.111\\
    (c) & 1.920 & 6.738& 1.931 & 6.771 & 0.179 & 0.762 & 0.179 & 0.763& 1.964 & 6.912 & 1.968 &6.931\\
    (d) & 0.572 & 2.379& 0.570 & 2.372 & 0.343 & 1.323 & 0.345 & 1.330& 0.560& 2.344 & 5.598 & 2.344\\
    (e) & 0.677 & 2.745& 0.680 & 2.757 & 0.106 & 0.491 & 0.105 & 0.491& 0.668& 2.722 & 0.668 & 2.724\\
    (f) & 1.022 & 3.903& 1.025 & 3.913 & 0.060 & 0.309 & 0.060 & 0.309& 1.020& 3.916 & 1.021 & 3.918\\
    (g) & 0.883 & 3.442& 0.899 & 3.498 & 0.278 & 1.106 & 0.274 & 1.093& 0.877& 3.436& 0.878 & 3.440\\
    (h) & 1.396 & 5.105& 1.399 & 5.117 & 0.101 & 0.472 & 0.101 & 0.472& 1.408& 5.176& 1.410 & 5.180\\
    (i) & 2.049 & 7.131& 2.056 & 7.154 & 0.185 & 0.782 & 0.184 & 0.782& 2.102& 7.345& 2.103 & 7.349\\
    (j) & 1.066 & 4.046& 1.070 & 4.061 & 0.051 & 0.271 & 0.051 & 0.271& 1.065& 4.065& 1.068 & 4.073\\
    \hline
    \end{tabular}
 \caption{The NLO+NNLL pair production cross-sections, in fb, of  first and second generation sleptions at $\sqrt{s} = 14$TeV and 27TeV for all benchmark models.} 
    \label{tab:crosssection}
\end{table}
\begin{table}[h]
    \centering
    \footnotesize
    \begin{tabular}{|c|cc|cc|cc|cc|cc|}
        \hline
    Model & \multicolumn{2}{c|}{$\sigma(pp\rightarrow\tilde{\tau_1}\tilde{\tau_1})$} & \multicolumn{2}{c|}{$\sigma(pp\rightarrow\tilde{\tau_2}\tilde{\tau_2})$}  & \multicolumn{2}{c|}{$\sigma(pp\rightarrow\tilde{\tau}_1\tilde{\tau}_2)$}  & \multicolumn{2}{c|}{$\sigma(pp\rightarrow\tilde{\nu}_{\tau}\tilde{\nu}_\tau)$} & \multicolumn{2}{c|}{$\sigma(pp\rightarrow\tilde{\tau}_{1}\tilde{\nu}_\tau)$}    \\
        \hline
          & 14TeV & 27TeV &  14TeV & 27TeV & 14TeV & 27TeV & 14TeV & 27TeV & 14TeV & 27TeV\\
    \hline
    (a) & 2.584 & 8.335 & 0.076 & 0.369 & 1.398 & 5.144 & 0.112 & 0.459 & 0.634 & 2.427\\
    (b) & 5.065 & 15.29 & 0.246 & 0.981 & 3.745 & 12.238& 0.266 & 0.950 & 1.281 & 4.429\\
    (c) & 10.171& 28.29 & 0.195 & 0.810 & 3.713 & 12.144& 0.376 & 1.325 & 2.684 & 8.903\\
    (d) & 0.893 & 3.011 & 0.242 & 1.052 & 0.637 & 2.616 & 0.186 & 0.733 & 0.951 & 3.984\\
    (e) & 1.460 & 4.935 & 0.080 & 0.390 & 0.841 & 3.316 & 0.122 & 0.509 & 0.551 & 2.253\\
    (f) & 2.008 & 6.769 & 0.053 & 0.275 & 1.267 & 4.722 & 0.060 & 0.260 & 0.354 & 1.397\\
    (g) & 3.621 & 10.649& 0.232 & 0.975 & 1.311 & 4.865 & 0.366 & 1.333 & 1.970 & 7.257\\
    (h) & 2.416 & 7.993 & 0.098 & 0.454 & 1.821 & 6.477 & 0.088 & 0.354 & 0.430 & 1.637\\
    (i) & 4.142 & 12.715& 0.144 & 0.625 & 2.509 & 8.585 & 0.191 & 0.721 & 1.049 & 3.755\\
    (j) & 4.134 & 12.835& 0.068 & 0.335 & 2.184 & 7.596 & 0.090 & 0.364 & 0.690 & 2.505\\
    \hline
    \end{tabular}
       \caption{The stau and stau-neutrino pair production cross-sections at NLO+NNLL order in fb, at $\sqrt{s} = 14$TeV and 27TeV for all benchmark models. The largest cross sections are for the production of $\tilde \tau_1 \tilde\tau_1$ followed by the
   production cross section for $\tilde\tau_1 \tilde\tau_2$.}
    \label{tab:crosssection-tau}     
\end{table}

\begin{table}[!h]
    \centering
    \begin{tabular}{|cccccccc|}
        \hline
    Model & $\tilde{\tau_1}\rightarrow \tau \tilde{\chi}_1^0$ &$\tilde{\tau_2}\rightarrow \tau\tilde{\chi}_1^0$ & $\tilde{\tau}_2\rightarrow \tilde{\nu}_{\tau} W^-$ &  $\tilde{\tau}_2\rightarrow \tilde{\tau}_1 Z[h^0]$& $\tilde{\nu}_\tau\rightarrow \tau \tilde{\chi}_1^0$ &$\tilde{\nu}_\tau\rightarrow \tilde{\tau}_1 W^+$ &$\tilde{\nu}_\tau\rightarrow \tilde{\tau}_1 q \bar{q}$ \\
    \hline
    (a) & 100\% & 12\% & 41\% & 35\%[12\%]  & 91\% & $<$1\% & 9\%  \\
    (b) & 100\% & 23\% & 37\% & 28\%[12\%]  & 99\% & $<$1\% & 1\%   \\
    (c) & 100\% & 99\% & $<$1\% & $<$1\%[$<$1\%]  & 31\% & 69\% & $<$1\%\\
    (d) & 100\% & 15\% & 22\% & 62\%[$<$1\%]  & 9\% & 91\% &$<$1\% \\
    (e) & 100\% & 14\% & 36\% & 44\%[6\%]  & 32\% & 68\% & $<$1\% \\
    (f) & 100\% & 12\% & 42\%& 29\% [15\%] & 98\% & $<$1\% & 2\% \\
    (g) & 100\% & 19\% & 14\% & 67\%[$<$1\%] & 9\% & 91\% &$<$1\% \\
    (h) & 100\% & 98\% & $<$1\% & $<$1\%[$<$1\%]  & 99\% & $<$1\% & $<$1\%  \\
    (i) & 100\% & 16\% & 39\% & 32\%[12\%]  & 98\% & $<$1\% &2\% \\
    (j) & 100\% & 98\% & $<$1\% & $<$1\% [$<$1\%]  & 97\% & $<$1\%  & 3\% \\
    \hline
    \end{tabular}
    \caption{While the decay branching ratio of $\tilde \tau_1\to \tau \tilde \chi_1^0$ is 100\%, the decay  
    branching ratios of $\tilde \tau_2$ and $\tilde \nu_\tau$ are  diverse and show a variety of model dependent signatures  providing differentiation  among cases (a)-(j) for this class of $\mathsf{SO(10)}$ models.}
 \label{stau-BR}
\end{table}
\begin{table}[h]
    \centering
    \footnotesize
    \begin{tabular}{|c|cc|}
        \hline
    Model & \multicolumn{2}{c|}{$\sigma(pp\rightarrow\tilde{\chi}_2^0\tilde{\chi}_1^\pm)$}   \\
        \hline
          & 14TeV & 27TeV \\
    \hline
    (a) & 8.773 & 40.105 \\
    (b) & 11.446 & 50.159\\
    (c) & 21.770 & 87.013\\
    (d) & 0.731 & 5.284\\
    (e) & 5.564 & 27.362\\
    (f) & 17.400 & 71.840\\
    (g) & 2.216 & 12.828\\
    (h) & 13.977 & 59.533\\
    (i) & 7.838 & 36.461\\
    (j) & 19.198 & 78.215\\
    \hline
    \end{tabular}
    \caption{The production cross-section for $\tilde{\chi}_2^0\tilde{\chi}_1^\pm$     
    in fb at $\sqrt{s} = 14$ TeV and 27 TeV for all models. As discussed in text the process
    $pp\rightarrow\tilde{\chi}_2^0\tilde{\chi}_1^\pm$ is of significant interest and it leads to the well-known     tri-leptonic signal for supersymmetry.} 
    \label{tab:crosssection-chi2}
\end{table}
\begin{figure}[h]
    \centering
    \includegraphics[width=0.45\linewidth]{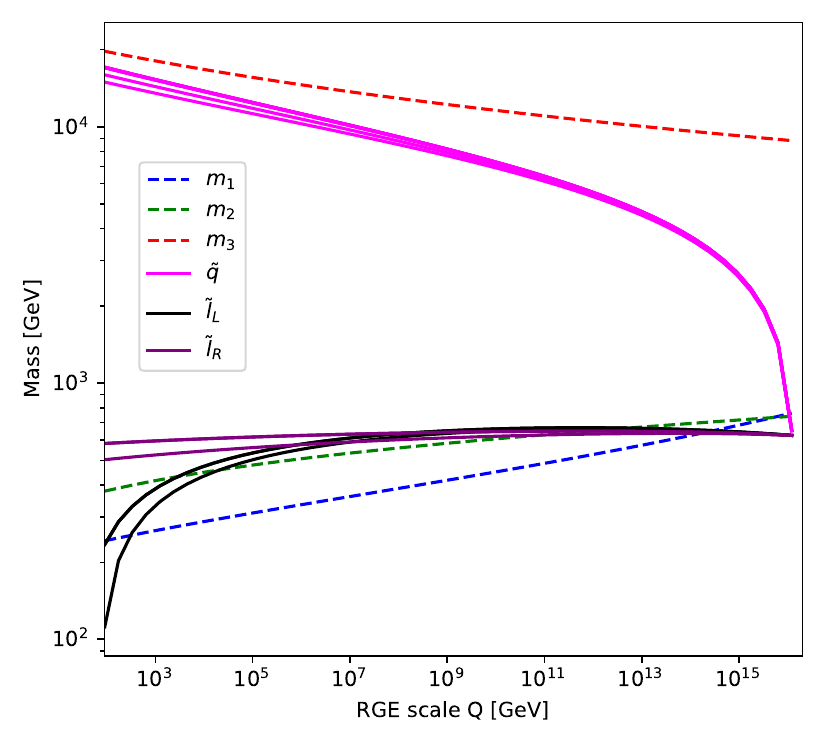}
    \includegraphics[width=0.45\linewidth]{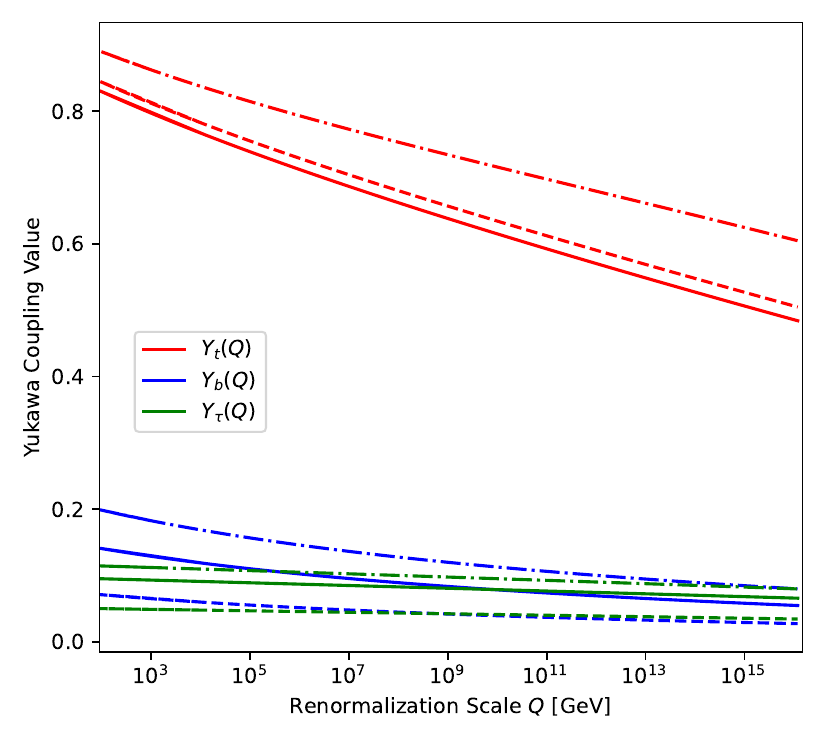}
   \caption{Left panel: Renormalization group evolution of sleptons, squarks and of the gaugino masses in $\tilde g$SUGRA for model (a). 
    Right panel:  Renormalization group evolution of Yukawa couplings of $b$ and $t$ quarks
    and of the  $\tau$ lepton.
     The dash line,  solid line and dash dot line correspond to $\tan{\beta}$ = 5, 9.8 and 14.}
    \label{fig:RG}
\end{figure}
\begin{figure}[h]
    \centering
    \includegraphics[width=0.45\linewidth]{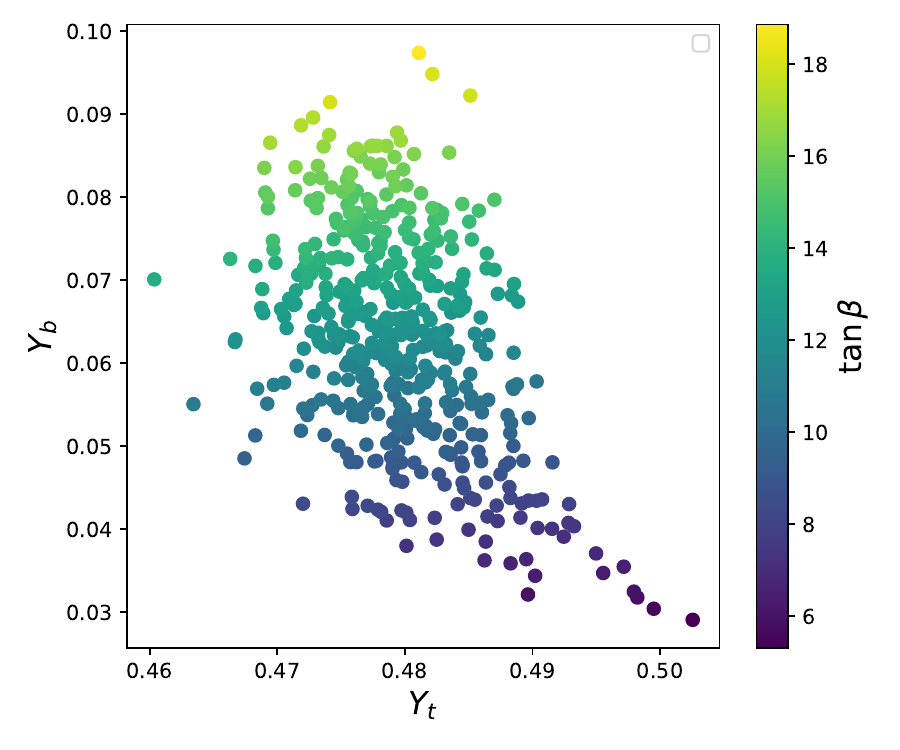}
    \includegraphics[width=0.45\linewidth]{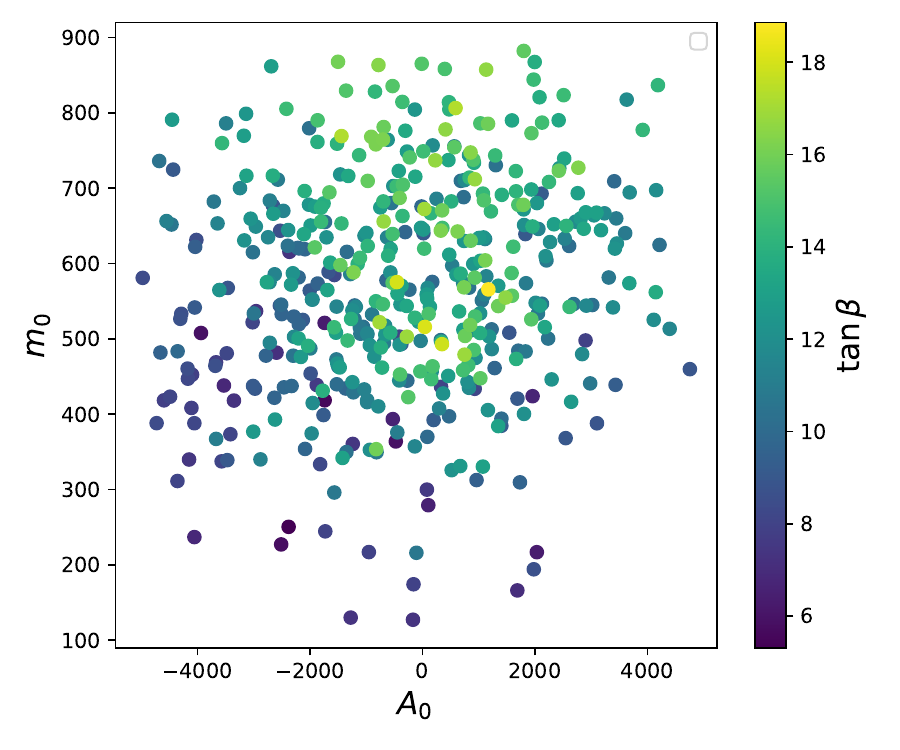}
    \caption{
  Left panel:  An MC scan of the top and the bottom Yukawa couplings at the GUT scale with color map giving 
    $\tan{\beta}$. Right panel: An MC scan of $m_0$ v.s. $A_0$ with color map giving $\tan{\beta}$. All the scatter points satisfy the Higgs boson mass, the relic density 
    and the muon anomaly constraint.    }
    \label{fig:ybytm0A0}
\end{figure}

\begin{figure}[htb]
   \label{fig:massspectrum}
    \begin{minipage}[t]{.45\textwidth}
        \centering
        \begin{tikzpicture}[thick,scale=0.35, every node/.style={scale=0.8}]
    \tiny
  \draw (-3.000000,0.000000) rectangle (19.000000,15.400000);
  \draw (-3.000000,0.000000) node[left] {0};
  \draw (-3.000000,1.466667) node[left] {1600};
  \draw (-2.700000,1.466667) -- (-3.000000,1.466667);
  \draw (-3.000000,2.933333) node[left] {3200};
  \draw (-2.700000,2.933333) -- (-3.000000,2.933333);
  \draw (-3.000000,4.400000) node[left] {4800};
  \draw (-2.700000,4.400000) -- (-3.000000,4.400000);
  \draw (-3.000000,5.866667) node[left] {6400};
  \draw (-2.700000,5.866667) -- (-3.000000,5.866667);
  \draw (-3.000000,7.333333) node[left] {8000};
  \draw (-2.700000,7.333333) -- (-3.000000,7.333333);
  \draw (-3.000000,8.800000) node[left] {9600};
  \draw (-2.700000,8.800000) -- (-3.000000,8.800000);
  \draw (-3.000000,10.266667) node[left] {11200};
  \draw (-2.700000,10.266667) -- (-3.000000,10.266667);
  \draw (-3.000000,11.733333) node[left] {12800};
  \draw (-2.700000,11.733333) -- (-3.000000,11.733333);
  \draw (-3.000000,13.200000) node[left] {14400};
  \draw (-2.700000,13.200000) -- (-3.000000,13.200000);
  \draw (-3.000000,14.666667) node[left] {16000};
  \draw (-2.700000,14.666667) -- (-3.000000,14.666667);
  \draw (-5.200000,15.400000) node[left,rotate=90] {Mass / GeV};

  \draw[-stealth,line width=0.00pt,dashed,color=black!10.0586837983] (0.800000,7.311117) -- (6.200000,0.303628);
  \draw[-stealth,line width=0.00pt,dashed,color=black!10.02012791152] (0.800000,7.311117) -- (10.800000,0.286685);
  \draw[-stealth,line width=0.00pt,dashed,color=black!10.02012791152] (0.800000,7.311117) -- (10.800000,0.526727);
  \draw[-stealth,line width=0.00pt,dashed,color=black!10.05911449342] (0.800000,7.311117) -- (11.200000,0.526857);
  \draw[-stealth,line width=0.00pt,dashed,color=black!10.0586837983] (0.800000,7.311117) -- (6.200000,0.575362);
  \draw[-stealth,line width=0.00pt,dashed,color=black!10.1938014844] (0.800000,7.311085) -- (6.200000,0.303628);
  \draw[-stealth,line width=0.00pt,dashed,color=black!10.02173223124] (0.800000,7.311085) -- (10.800000,0.286685);
  \draw[-stealth,line width=0.00pt,dashed,color=black!10.02173223124] (0.800000,7.311085) -- (10.800000,0.526727);
  \draw[-stealth,line width=0.00pt,dashed,color=black!10.0648295032] (0.800000,7.311085) -- (11.200000,0.526857);
  \draw[-stealth,line width=0.00pt,dashed,color=black!10.1938014844] (0.800000,7.311085) -- (6.200000,0.575362);
  \draw[-stealth,line width=0.00pt,dashed,color=black!10.1145131902] (1.200000,7.311343) -- (6.200000,0.375896);
  \draw[-stealth,line width=0.00pt,dashed,color=black!10.04712582202] (1.200000,7.311343) -- (10.800000,0.286685);
  \draw[-stealth,line width=0.00pt,dashed,color=black!10.04712582202] (1.200000,7.311343) -- (11.200000,0.526857);
  \draw[-stealth,line width=0.01pt,dashed,color=black!10.742136076] (15.900000,11.679861) -- (10.800000,0.286685);
  \draw[-stealth,line width=0.33pt,dashed,color=black!29.7506311] (15.900000,11.679861) -- (10.800000,0.526727);
  \draw[-stealth,line width=0.09pt,dashed,color=black!15.103633330000001] (16.200000,10.882672) -- (16.200000,10.100881);
  \draw[-stealth,line width=0.01pt,dashed,color=black!10.5260650402] (16.200000,10.882672) -- (10.800000,0.286685);
  \draw[-stealth,line width=0.23pt,dashed,color=black!23.984592040000003] (16.200000,10.882672) -- (10.800000,0.526727);
  \draw[-stealth,line width=0.00pt,dashed,color=black!10.1288519608] (16.200000,10.882672) -- (10.800000,7.327266);
  \draw[-stealth,line width=0.00pt,dashed,color=black!10.1289759802] (16.200000,10.882672) -- (10.800000,7.327480);
  \draw[-stealth,line width=0.00pt,dashed,color=black!10.01585191042] (16.200000,10.100881) -- (10.800000,0.286685);
  \draw[-stealth,line width=0.00pt,dashed,color=black!10.212588799] (16.200000,10.100881) -- (10.800000,0.526727);
  \draw[-stealth,line width=0.01pt,dashed,color=black!10.4253568378] (16.200000,10.100881) -- (11.200000,0.526857);
  \draw[-stealth,line width=0.17pt,dashed,color=black!20.2946335] (16.200000,10.100881) -- (10.800000,7.327266);
  \draw[-stealth,line width=0.17pt,dashed,color=black!20.11360636] (16.200000,10.100881) -- (10.800000,7.327480);
  \draw[-stealth,line width=0.34pt,dashed,color=black!30.3162169] (16.200000,10.100881) -- (11.200000,7.327505);
  \draw[-stealth,line width=1.00pt,dashed,color=black!69.99999274] (5.800000,0.418019) -- (10.800000,0.286685);
  \draw[-stealth,line width=1.00pt,dashed,color=black!69.99998452] (5.800000,0.411787) -- (10.800000,0.286685);
  \draw[-stealth,line width=1.00pt,dashed,color=black!70.0] (6.200000,0.303628) -- (10.800000,0.286685);
  \draw[-stealth,line width=0.04pt,dashed,color=black!12.554216566000001] (6.200000,0.375896) -- (6.200000,0.303628);
  \draw[-stealth,line width=0.91pt,dashed,color=black!64.88672896] (6.200000,0.375896) -- (10.800000,0.286685);
  \draw[-stealth,line width=0.03pt,dashed,color=black!12.021146806] (15.700000,13.998858) -- (15.900000,11.679861);
  \draw[-stealth,line width=0.06pt,dashed,color=black!13.426422898] (15.700000,13.998858) -- (16.200000,10.882672);
  \draw[-stealth,line width=0.08pt,dashed,color=black!14.989340824] (15.700000,13.998858) -- (16.200000,10.100881);
  \draw[-stealth,line width=0.03pt,dashed,color=black!11.993481574] (15.700000,13.998858) -- (15.900000,11.697354);
  \draw[-stealth,line width=0.03pt,dashed,color=black!12.057262876] (15.700000,13.998858) -- (16.200000,11.657202);
  \draw[-stealth,line width=0.06pt,dashed,color=black!13.44430522] (15.700000,13.998858) -- (16.200000,10.887762);
  \draw[-stealth,line width=0.06pt,dashed,color=black!13.531799626] (10.800000,0.526727) -- (5.800000,0.418019);
  \draw[-stealth,line width=0.06pt,dashed,color=black!13.892863162] (10.800000,0.526727) -- (5.800000,0.411787);
  \draw[-stealth,line width=0.15pt,dashed,color=black!18.92469436] (10.800000,0.526727) -- (6.200000,0.303628);
  \draw[-stealth,line width=0.10pt,dashed,color=black!16.20070822] (10.800000,0.526727) -- (6.200000,0.375896);
  \draw[-stealth,line width=0.13pt,dashed,color=black!17.79340624] (11.200000,0.526857) -- (5.800000,0.411787);
  \draw[-stealth,line width=0.21pt,dashed,color=black!22.40722174] (11.200000,0.526857) -- (6.200000,0.375896);
  \draw[-stealth,line width=0.04pt,dashed,color=black!12.186043324] (10.800000,7.327266) -- (0.800000,0.114026);
  \draw[-stealth,line width=0.01pt,dashed,color=black!10.3026165832] (10.800000,7.327266) -- (6.200000,0.303628);
  \draw[-stealth,line width=0.04pt,dashed,color=black!12.186043324] (10.800000,7.327266) -- (10.800000,0.286685);
  \draw[-stealth,line width=0.10pt,dashed,color=black!16.03836586] (10.800000,7.327266) -- (10.800000,0.526727);
  \draw[-stealth,line width=0.02pt,dashed,color=black!10.91583196] (10.800000,7.327266) -- (11.200000,0.526857);
  \draw[-stealth,line width=0.01pt,dashed,color=black!10.3300813432] (10.800000,7.327266) -- (6.200000,0.575362);
  \draw[-stealth,line width=0.06pt,dashed,color=black!13.839208462] (10.800000,7.327480) -- (0.800000,0.114026);
  \draw[-stealth,line width=0.00pt,dashed,color=black!10.2949478542] (10.800000,7.327480) -- (6.200000,0.303628);
  \draw[-stealth,line width=0.03pt,dashed,color=black!12.05703211] (10.800000,7.327480) -- (10.800000,0.286685);
  \draw[-stealth,line width=0.10pt,dashed,color=black!15.713313094] (10.800000,7.327480) -- (10.800000,0.526727);
  \draw[-stealth,line width=0.02pt,dashed,color=black!10.908834814] (10.800000,7.327480) -- (11.200000,0.526857);
  \draw[-stealth,line width=0.01pt,dashed,color=black!10.3335451306] (10.800000,7.327480) -- (6.200000,0.575362);
  \draw[-stealth,line width=0.30pt,dashed,color=black!28.147922440000002] (11.200000,7.327505) -- (0.800000,0.114026);
  \draw[-stealth,line width=0.01pt,dashed,color=black!10.631303824] (11.200000,7.327505) -- (6.200000,0.375896);
  \draw[-stealth,line width=0.01pt,dashed,color=black!10.327281496] (11.200000,7.327505) -- (10.800000,0.286685);
  \draw[-stealth,line width=0.02pt,dashed,color=black!10.931006242] (11.200000,7.327505) -- (10.800000,0.526727);
  \draw[-stealth,line width=0.00pt,dashed,color=black!10.0344822949] (11.200000,7.327505) -- (11.200000,0.526857);
  \draw[-stealth,line width=1.00pt,dashed,color=black!69.99907311999999] (15.900000,11.697354) -- (10.800000,0.286685);
  \draw[-stealth,line width=0.01pt,dashed,color=black!10.635710524] (16.200000,11.657202) -- (0.800000,0.114026);
  \draw[-stealth,line width=0.01pt,dashed,color=black!10.61819962] (16.200000,11.657202) -- (16.200000,10.882672);
  \draw[-stealth,line width=0.00pt,dashed,color=black!10.05597725728] (16.200000,11.657202) -- (16.200000,10.100881);
  \draw[-stealth,line width=0.74pt,dashed,color=black!54.14606602] (16.200000,11.657202) -- (10.800000,0.286685);
  \draw[-stealth,line width=0.00pt,dashed,color=black!10.01202177256] (16.200000,11.657202) -- (10.800000,0.526727);
  \draw[-stealth,line width=0.06pt,dashed,color=black!13.324583452] (16.200000,11.657202) -- (10.800000,7.327266);
  \draw[-stealth,line width=0.06pt,dashed,color=black!13.325496622] (16.200000,11.657202) -- (10.800000,7.327480);
  \draw[-stealth,line width=0.02pt,dashed,color=black!11.166408832] (16.200000,11.657202) -- (16.200000,10.887762);
  \draw[-stealth,line width=0.04pt,dashed,color=black!12.475996894] (16.200000,10.887762) -- (0.800000,0.114026);
  \draw[-stealth,line width=0.04pt,dashed,color=black!12.475996894] (16.200000,10.887762) -- (16.200000,10.100881);
  \draw[-stealth,line width=0.01pt,dashed,color=black!10.5705204808] (16.200000,10.887762) -- (10.800000,0.286685);
  \draw[-stealth,line width=0.23pt,dashed,color=black!23.9089549] (16.200000,10.887762) -- (10.800000,0.526727);
  \draw[-stealth,line width=0.46pt,dashed,color=black!37.82808886] (16.200000,10.887762) -- (11.200000,0.526857);
  \draw[-stealth,line width=0.10pt,dashed,color=black!16.06040482] (16.200000,10.887762) -- (10.800000,7.327266);
  \draw[-stealth,line width=0.10pt,dashed,color=black!16.14442786] (16.200000,10.887762) -- (10.800000,7.327480);
  \draw[-stealth,line width=0.01pt,dashed,color=black!10.3382770108] (16.200000,10.887762) -- (11.200000,7.327505);
  \draw[-stealth,line width=1.00pt,dashed,color=black!69.9999727] (5.800000,0.580309) -- (10.800000,0.286685);
  \draw[-stealth,line width=0.12pt,dashed,color=black!17.1223522] (6.200000,0.575362) -- (0.800000,0.114026);
  \draw[-stealth,line width=0.12pt,dashed,color=black!17.1223522] (6.200000,0.575362) -- (6.200000,0.303628);
  \draw[-stealth,line width=0.41pt,dashed,color=black!34.348729899999995] (6.200000,0.575362) -- (6.200000,0.375896);
  \draw[-stealth,line width=0.12pt,dashed,color=black!17.46007114] (6.200000,0.575362) -- (10.800000,0.286685);
  \draw[-stealth,line width=0.00pt,dashed,color=black!10.0693430296] (6.200000,0.575362) -- (10.800000,0.526727);

  \draw[color=blue,thick] (-0.200000,0.114026) -- (1.800000,0.114026);
  \draw[color=blue,thick] (-0.200000,7.311117) -- (1.800000,7.311117);
  \draw[color=blue,thick] (-0.200000,7.311085) -- (1.800000,7.311085);
  \draw[color=red,thick] (0.200000,7.311343) -- (2.200000,7.311343);
  \draw[color=blue,thick] (14.900000,11.679861) -- (16.900000,11.679861);
  \draw[color=black!50!blue!30!green,thick] (15.200000,10.882672) -- (17.200000,10.882672);
  \draw[color=red,thick] (15.200000,10.100881) -- (17.200000,10.100881);
  \draw[color=blue,thick] (4.800000,0.418019) -- (6.800000,0.418019);
  \draw[color=blue,thick] (4.800000,0.411787) -- (6.800000,0.411787);
  \draw[color=red,thick] (5.200000,0.303628) -- (7.200000,0.303628);
  \draw[color=red,thick] (5.200000,0.375896) -- (7.200000,0.375896);
  \draw[color=black!50!blue!30!green,thick] (14.700000,13.998858) -- (16.700000,13.998858);
  \draw[color=blue,thick] (9.800000,0.286685) -- (11.800000,0.286685);
  \draw[color=blue,thick] (9.800000,0.526727) -- (11.800000,0.526727);
  \draw[color=red,thick] (10.200000,0.526857) -- (12.200000,0.526857);
  \draw[color=blue,thick] (9.800000,7.327266) -- (11.800000,7.327266);
  \draw[color=blue,thick] (9.800000,7.327480) -- (11.800000,7.327480);
  \draw[color=red,thick] (10.200000,7.327505) -- (12.200000,7.327505);
  \draw[color=blue,thick] (14.900000,11.697354) -- (16.900000,11.697354);
  \draw[color=black!50!blue!30!green,thick] (15.200000,11.657202) -- (17.200000,11.657202);
  \draw[color=red,thick] (15.200000,10.887762) -- (17.200000,10.887762);
  \draw[color=blue,thick] (4.800000,0.580309) -- (6.800000,0.580309);
  \draw[color=red,thick] (5.200000,0.575362) -- (7.200000,0.575362);

  \draw (-0.400000,0.381787) node[left] {\small $h^0$};
  \draw (-0.400000,6.992945) node[left] {\small $A^0$};
  \draw (-0.400000,7.629257) node[left] {\small $H^0$};
  \draw (2.400000,7.311343) node[right] {\small $H^\pm$};
  \draw (14.700000,11.370452) node[left] {\small $\tilde{q}_\text{L}$};
  \draw (14.700000,12.006763) node[left] {\small $\tilde{q}_\text{R}$};
  \draw (17.400000,10.567061) node[right] {\small $\tilde{b}_1$};
  \draw (17.400000,11.203373) node[right] {\small $\tilde{t}_2$};
  \draw (17.400000,10.100881) node[right] {\small $\tilde{t}_1$};
  \draw (4.600000,-0.140264) node[left] {\small $\tilde{\nu}_\text{L}$};
  \draw (4.600000,0.418019) node[left] {\small $\tilde{\ell}_\text{L}$};
  \draw (4.600000,1.132360) node[left] {\small $\tilde{\ell}_\text{R}$};
  \draw (7.400000,-0.157737) node[right] {\small $\tilde{\tau}_1$};
  \draw (7.400000,0.381787) node[right] {\small $\tilde{\nu}_\tau$};
  \draw (7.400000,1.114886) node[right] {\small $\tilde{\tau}_2$};
  \draw (14.500000,13.998858) node[left] {\small $\tilde{g}$};
  \draw (9.600000,0.136101) node[left] {\small $\tilde{\chi}_1^0$};
  \draw (9.600000,0.772413) node[left] {\small $\tilde{\chi}_2^0$};
  \draw (12.400000,0.526857) node[right] {\small $\tilde{\chi}_1^\pm$};
  \draw (9.600000,7.009217) node[left] {\small $\tilde{\chi}_3^0$};
  \draw (9.600000,7.645529) node[left] {\small $\tilde{\chi}_4^0$};
  \draw (12.400000,7.327505) node[right] {\small $\tilde{\chi}_2^\pm$};
  \draw (17.400000,11.657202) node[right] {\small $\tilde{b}_2$};
\end{tikzpicture}
    \end{minipage}
    \hfill
    \begin{minipage}[t]{.45\textwidth}
        \centering
        \tiny
    \begin{tikzpicture}[thick,scale=0.35, every node/.style={scale=0.8}]

  \draw (-3.000000,0.000000) rectangle (19.000000,15.400000);
  \draw (-3.000000,0.000000) node[left] {0};
  \draw (-3.000000,2.200000) node[left] {100};
  \draw (-2.700000,2.200000) -- (-3.000000,2.200000);
  \draw (-3.000000,4.400000) node[left] {200};
  \draw (-2.700000,4.400000) -- (-3.000000,4.400000);
  \draw (-3.000000,6.600000) node[left] {300};
  \draw (-2.700000,6.600000) -- (-3.000000,6.600000);
  \draw (-3.000000,8.800000) node[left] {400};
  \draw (-2.700000,8.800000) -- (-3.000000,8.800000);
  \draw (-3.000000,11.000000) node[left] {500};
  \draw (-2.700000,11.000000) -- (-3.000000,11.000000);
  \draw (-3.000000,13.200000) node[left] {600};
  \draw (-2.700000,13.200000) -- (-3.000000,13.200000);
  \draw (-3.000000,15.400000) node[left] {700};
  \draw (-5.200000,15.400000) node[left,rotate=90] {Mass / GeV};

  \draw[-stealth,line width=1.00pt,dashed,color=black!69.99999274] (5.800000,10.032455) -- (10.800000,6.880442);
  \draw[-stealth,line width=1.00pt,dashed,color=black!69.99998452] (5.800000,9.882890) -- (10.800000,6.880442);
  \draw[-stealth,line width=1.00pt,dashed,color=black!70.0] (6.200000,7.287069) -- (10.800000,6.880442);
  \draw[-stealth,line width=0.04pt,dashed,color=black!12.554216566000001] (6.200000,9.021501) -- (6.200000,7.287069);
  \draw[-stealth,line width=0.91pt,dashed,color=black!64.88672896] (6.200000,9.021501) -- (10.800000,6.880442);
  \draw[-stealth,line width=0.06pt,dashed,color=black!13.531799626] (10.800000,12.641437) -- (5.800000,10.032455);
  \draw[-stealth,line width=0.06pt,dashed,color=black!13.892863162] (10.800000,12.641437) -- (5.800000,9.882890);
  \draw[-stealth,line width=0.15pt,dashed,color=black!18.92469436] (10.800000,12.641437) -- (6.200000,7.287069);
  \draw[-stealth,line width=0.10pt,dashed,color=black!16.20070822] (10.800000,12.641437) -- (6.200000,9.021501);
  \draw[-stealth,line width=0.13pt,dashed,color=black!17.79340624] (11.200000,12.644559) -- (5.800000,9.882890);
  \draw[-stealth,line width=0.21pt,dashed,color=black!22.40722174] (11.200000,12.644559) -- (6.200000,9.021501);
  \draw[-stealth,line width=1.00pt,dashed,color=black!69.9999727] (5.800000,13.927405) -- (10.800000,6.880442);
  \draw[-stealth,line width=0.12pt,dashed,color=black!17.1223522] (6.200000,13.808693) -- (0.800000,2.736626);
  \draw[-stealth,line width=0.12pt,dashed,color=black!17.1223522] (6.200000,13.808693) -- (6.200000,7.287069);
  \draw[-stealth,line width=0.41pt,dashed,color=black!34.348729899999995] (6.200000,13.808693) -- (6.200000,9.021501);
  \draw[-stealth,line width=0.12pt,dashed,color=black!17.46007114] (6.200000,13.808693) -- (10.800000,6.880442);
  \draw[-stealth,line width=0.00pt,dashed,color=black!10.0693430296] (6.200000,13.808693) -- (10.800000,12.641437);

  \draw[color=blue,thick] (-0.200000,2.736626) -- (1.800000,2.736626);
  \draw[color=blue,thick] (4.800000,10.032455) -- (6.800000,10.032455);
  \draw[color=blue,thick] (4.800000,9.882890) -- (6.800000,9.882890);
  \draw[color=red,thick] (5.200000,7.287069) -- (7.200000,7.287069);
  \draw[color=red,thick] (5.200000,9.021501) -- (7.200000,9.021501);
  \draw[color=blue,thick] (9.800000,6.880442) -- (11.800000,6.880442);
  \draw[color=blue,thick] (9.800000,12.641437) -- (11.800000,12.641437);
  \draw[color=red,thick] (10.200000,12.644559) -- (12.200000,12.644559);
  \draw[color=blue,thick] (4.800000,13.927405) -- (6.800000,13.927405);
  \draw[color=red,thick] (5.200000,13.808693) -- (7.200000,13.808693);

  \draw (-0.400000,2.736626) node[left] {\small $h^0$};
  \draw (4.600000,9.641141) node[left] {\small $\tilde{\nu}_\text{L}$};
  \draw (4.600000,10.274205) node[left] {\small $\tilde{\ell}_\text{L}$};
  \draw (7.400000,7.287069) node[right] {\small $\tilde{\tau}_1$};
  \draw (7.400000,9.021501) node[right] {\small $\tilde{\nu}_\tau$};
  \draw (9.600000,6.880442) node[left] {\small $\tilde{\chi}_1^0$};
  \draw (9.600000,12.641437) node[left] {\small $\tilde{\chi}_2^0$};
  \draw (12.400000,12.644559) node[right] {\small $\tilde{\chi}_1^\pm$};
  \draw (4.600000,13.927405) node[left] {\small $\tilde{\ell}_\text{R}$};
  \draw (7.400000,13.808693) node[right] {\small $\tilde{\tau}_2$};
\end{tikzpicture}
    \end{minipage}

\begin{minipage}[t]{.45\textwidth}
        \centering
        \tiny
        \begin{tikzpicture}[thick,scale=0.35, every node/.style={scale=0.8}]

  \draw (-3.000000,0.000000) rectangle (19.000000,15.400000);
  \draw (-3.000000,0.000000) node[left] {0};
  \draw (-3.000000,1.735211) node[left] {1600};
  \draw (-2.700000,1.735211) -- (-3.000000,1.735211);
  \draw (-3.000000,3.470423) node[left] {3200};
  \draw (-2.700000,3.470423) -- (-3.000000,3.470423);
  \draw (-3.000000,5.205634) node[left] {4800};
  \draw (-2.700000,5.205634) -- (-3.000000,5.205634);
  \draw (-3.000000,6.940845) node[left] {6400};
  \draw (-2.700000,6.940845) -- (-3.000000,6.940845);
  \draw (-3.000000,8.676056) node[left] {8000};
  \draw (-2.700000,8.676056) -- (-3.000000,8.676056);
  \draw (-3.000000,10.411268) node[left] {9600};
  \draw (-2.700000,10.411268) -- (-3.000000,10.411268);
  \draw (-3.000000,12.146479) node[left] {11200};
  \draw (-2.700000,12.146479) -- (-3.000000,12.146479);
  \draw (-3.000000,13.881690) node[left] {12800};
  \draw (-2.700000,13.881690) -- (-3.000000,13.881690);
  \draw (-5.200000,15.400000) node[left,rotate=90] {Mass / GeV};

  \draw[-stealth,line width=0.00pt,dashed,color=black!10.0978864252] (0.800000,7.679036) -- (6.200000,0.248153);
  \draw[-stealth,line width=0.00pt,dashed,color=black!10.02660783634] (0.800000,7.679036) -- (10.800000,0.223030);
  \draw[-stealth,line width=0.00pt,dashed,color=black!10.02660783634] (0.800000,7.679036) -- (10.800000,0.513298);
  \draw[-stealth,line width=0.00pt,dashed,color=black!10.0783223686] (0.800000,7.679036) -- (11.200000,0.513492);
  \draw[-stealth,line width=0.00pt,dashed,color=black!10.0978864252] (0.800000,7.679036) -- (6.200000,0.562493);
  \draw[-stealth,line width=0.01pt,dashed,color=black!10.4929764076] (0.800000,7.678994) -- (6.200000,0.248153);
  \draw[-stealth,line width=0.00pt,dashed,color=black!10.0289461426] (0.800000,7.678994) -- (10.800000,0.223030);
  \draw[-stealth,line width=0.00pt,dashed,color=black!10.0289461426] (0.800000,7.678994) -- (10.800000,0.513298);
  \draw[-stealth,line width=0.00pt,dashed,color=black!10.0867700344] (0.800000,7.678994) -- (11.200000,0.513492);
  \draw[-stealth,line width=0.01pt,dashed,color=black!10.4929764076] (0.800000,7.678994) -- (6.200000,0.562493);
  \draw[-stealth,line width=0.01pt,dashed,color=black!10.3447992946] (1.200000,7.679338) -- (6.200000,0.353696);
  \draw[-stealth,line width=0.00pt,dashed,color=black!10.062608653] (1.200000,7.679338) -- (10.800000,0.223030);
  \draw[-stealth,line width=0.00pt,dashed,color=black!10.062608653] (1.200000,7.679338) -- (11.200000,0.513492);
  \draw[-stealth,line width=0.01pt,dashed,color=black!10.739260396] (15.900000,11.679720) -- (10.800000,0.223030);
  \draw[-stealth,line width=0.33pt,dashed,color=black!29.751424] (15.900000,11.679720) -- (10.800000,0.513298);
  \draw[-stealth,line width=0.11pt,dashed,color=black!16.82004874] (16.200000,10.808477) -- (16.200000,9.940972);
  \draw[-stealth,line width=0.01pt,dashed,color=black!10.5276203266] (16.200000,10.808477) -- (10.800000,0.223030);
  \draw[-stealth,line width=0.23pt,dashed,color=black!24.07911052] (16.200000,10.808477) -- (10.800000,0.513298);
  \draw[-stealth,line width=0.00pt,dashed,color=black!10.08595243] (16.200000,10.808477) -- (10.800000,7.668895);
  \draw[-stealth,line width=0.00pt,dashed,color=black!10.0861010356] (16.200000,10.808477) -- (10.800000,7.669193);
  \draw[-stealth,line width=0.00pt,dashed,color=black!10.0170914065] (16.200000,9.940972) -- (10.800000,0.223030);
  \draw[-stealth,line width=0.01pt,dashed,color=black!10.353592099] (16.200000,9.940972) -- (10.800000,0.513298);
  \draw[-stealth,line width=0.01pt,dashed,color=black!10.707603622] (16.200000,9.940972) -- (11.200000,0.513492);
  \draw[-stealth,line width=0.15pt,dashed,color=black!19.28709916] (16.200000,9.940972) -- (10.800000,7.668895);
  \draw[-stealth,line width=0.15pt,dashed,color=black!19.012507900000003] (16.200000,9.940972) -- (10.800000,7.669193);
  \draw[-stealth,line width=0.30pt,dashed,color=black!28.20817702] (16.200000,9.940972) -- (11.200000,7.669221);
  \draw[-stealth,line width=1.00pt,dashed,color=black!69.99998962000001] (5.800000,0.419937) -- (10.800000,0.223030);
  \draw[-stealth,line width=1.00pt,dashed,color=black!69.9999097] (5.800000,0.411315) -- (10.800000,0.223030);
  \draw[-stealth,line width=1.00pt,dashed,color=black!70.0] (6.200000,0.248153) -- (10.800000,0.223030);
  \draw[-stealth,line width=0.69pt,dashed,color=black!51.43378696] (6.200000,0.353696) -- (6.200000,0.248153);
  \draw[-stealth,line width=0.31pt,dashed,color=black!28.56621304] (6.200000,0.353696) -- (10.800000,0.223030);
  \draw[-stealth,line width=0.03pt,dashed,color=black!11.967480244] (15.700000,13.944793) -- (15.900000,11.679720);
  \draw[-stealth,line width=0.06pt,dashed,color=black!13.519867108] (15.700000,13.944793) -- (16.200000,10.808477);
  \draw[-stealth,line width=0.09pt,dashed,color=black!15.28334812] (15.700000,13.944793) -- (16.200000,9.940972);
  \draw[-stealth,line width=0.03pt,dashed,color=black!11.939725288] (15.700000,13.944793) -- (15.900000,11.697295);
  \draw[-stealth,line width=0.03pt,dashed,color=black!11.997854428] (15.700000,13.944793) -- (16.200000,11.660646);
  \draw[-stealth,line width=0.06pt,dashed,color=black!13.543960263999999] (15.700000,13.944793) -- (16.200000,10.816137);
  \draw[-stealth,line width=0.05pt,dashed,color=black!12.787397746] (10.800000,0.513298) -- (5.800000,0.419937);
  \draw[-stealth,line width=0.05pt,dashed,color=black!13.26132244] (10.800000,0.513298) -- (5.800000,0.411315);
  \draw[-stealth,line width=0.18pt,dashed,color=black!20.8433365] (10.800000,0.513298) -- (6.200000,0.248153);
  \draw[-stealth,line width=0.12pt,dashed,color=black!17.02303324] (10.800000,0.513298) -- (6.200000,0.353696);
  \draw[-stealth,line width=0.11pt,dashed,color=black!16.53433888] (11.200000,0.513492) -- (5.800000,0.411315);
  \draw[-stealth,line width=0.23pt,dashed,color=black!24.05244682] (11.200000,0.513492) -- (6.200000,0.353696);
  \draw[-stealth,line width=0.04pt,dashed,color=black!12.122112934] (10.800000,7.668895) -- (0.800000,0.134151);
  \draw[-stealth,line width=0.00pt,dashed,color=black!10.2364100014] (10.800000,7.668895) -- (6.200000,0.248153);
  \draw[-stealth,line width=0.00pt,dashed,color=black!10.249950187] (10.800000,7.668895) -- (10.800000,0.223030);
  \draw[-stealth,line width=0.01pt,dashed,color=black!10.760642206] (10.800000,7.668895) -- (10.800000,0.513298);
  \draw[-stealth,line width=0.01pt,dashed,color=black!10.86579727] (10.800000,7.668895) -- (11.200000,0.513492);
  \draw[-stealth,line width=0.00pt,dashed,color=black!10.2663830092] (10.800000,7.668895) -- (6.200000,0.562493);
  \draw[-stealth,line width=0.06pt,dashed,color=black!13.775200911999999] (10.800000,7.669193) -- (0.800000,0.134151);
  \draw[-stealth,line width=0.00pt,dashed,color=black!10.2285164896] (10.800000,7.669193) -- (6.200000,0.248153);
  \draw[-stealth,line width=0.00pt,dashed,color=black!10.2489002386] (10.800000,7.669193) -- (10.800000,0.223030);
  \draw[-stealth,line width=0.01pt,dashed,color=black!10.75828474] (10.800000,7.669193) -- (10.800000,0.513298);
  \draw[-stealth,line width=0.01pt,dashed,color=black!10.85861212] (10.800000,7.669193) -- (11.200000,0.513492);
  \draw[-stealth,line width=0.00pt,dashed,color=black!10.2710021674] (10.800000,7.669193) -- (6.200000,0.562493);
  \draw[-stealth,line width=0.30pt,dashed,color=black!27.762000519999997] (11.200000,7.669221) -- (0.800000,0.134151);
  \draw[-stealth,line width=0.01pt,dashed,color=black!10.4969866234] (11.200000,7.669221) -- (6.200000,0.353696);
  \draw[-stealth,line width=0.01pt,dashed,color=black!10.3099917154] (11.200000,7.669221) -- (10.800000,0.223030);
  \draw[-stealth,line width=0.01pt,dashed,color=black!10.875439744] (11.200000,7.669221) -- (10.800000,0.513298);
  \draw[-stealth,line width=0.03pt,dashed,color=black!11.593759198] (11.200000,7.669221) -- (11.200000,0.513492);
  \draw[-stealth,line width=1.00pt,dashed,color=black!69.99893098] (15.900000,11.697295) -- (10.800000,0.223030);
  \draw[-stealth,line width=0.01pt,dashed,color=black!10.69858759] (16.200000,11.660646) -- (0.800000,0.134151);
  \draw[-stealth,line width=0.01pt,dashed,color=black!10.677766] (16.200000,11.660646) -- (16.200000,10.808477);
  \draw[-stealth,line width=0.00pt,dashed,color=black!10.0844925856] (16.200000,11.660646) -- (16.200000,9.940972);
  \draw[-stealth,line width=0.79pt,dashed,color=black!57.1086727] (16.200000,11.660646) -- (10.800000,0.223030);
  \draw[-stealth,line width=0.00pt,dashed,color=black!10.01438549554] (16.200000,11.660646) -- (10.800000,0.513298);
  \draw[-stealth,line width=0.04pt,dashed,color=black!12.517828612] (16.200000,11.660646) -- (10.800000,7.668895);
  \draw[-stealth,line width=0.04pt,dashed,color=black!12.518798002] (16.200000,11.660646) -- (10.800000,7.669193);
  \draw[-stealth,line width=0.02pt,dashed,color=black!11.269069042] (16.200000,11.660646) -- (16.200000,10.816137);
  \draw[-stealth,line width=0.06pt,dashed,color=black!13.307880688000001] (16.200000,10.816137) -- (0.800000,0.134151);
  \draw[-stealth,line width=0.06pt,dashed,color=black!13.307880688000001] (16.200000,10.816137) -- (16.200000,9.940972);
  \draw[-stealth,line width=0.01pt,dashed,color=black!10.5898151488] (16.200000,10.816137) -- (10.800000,0.223030);
  \draw[-stealth,line width=0.23pt,dashed,color=black!23.976361660000002] (16.200000,10.816137) -- (10.800000,0.513298);
  \draw[-stealth,line width=0.47pt,dashed,color=black!37.96787362] (16.200000,10.816137) -- (11.200000,0.513492);
  \draw[-stealth,line width=0.08pt,dashed,color=black!15.09892318] (16.200000,10.816137) -- (10.800000,7.668895);
  \draw[-stealth,line width=0.09pt,dashed,color=black!15.209838598000001] (16.200000,10.816137) -- (10.800000,7.669193);
  \draw[-stealth,line width=0.00pt,dashed,color=black!10.2656788954] (16.200000,10.816137) -- (11.200000,7.669221);
  \draw[-stealth,line width=1.00pt,dashed,color=black!69.99998253999999] (5.800000,0.578382) -- (10.800000,0.223030);
  \draw[-stealth,line width=0.09pt,dashed,color=black!15.11089006] (6.200000,0.562493) -- (0.800000,0.134151);
  \draw[-stealth,line width=0.09pt,dashed,color=black!15.11089006] (6.200000,0.562493) -- (6.200000,0.248153);
  \draw[-stealth,line width=0.38pt,dashed,color=black!32.78377588] (6.200000,0.562493) -- (6.200000,0.353696);
  \draw[-stealth,line width=0.15pt,dashed,color=black!19.04938678] (6.200000,0.562493) -- (10.800000,0.223030);
  \draw[-stealth,line width=0.00pt,dashed,color=black!10.092894889] (6.200000,0.562493) -- (10.800000,0.513298);

  \draw[color=blue,thick] (-0.200000,0.134151) -- (1.800000,0.134151);
  \draw[color=blue,thick] (-0.200000,7.679036) -- (1.800000,7.679036);
  \draw[color=blue,thick] (-0.200000,7.678994) -- (1.800000,7.678994);
  \draw[color=red,thick] (0.200000,7.679338) -- (2.200000,7.679338);
  \draw[color=blue,thick] (14.900000,11.679720) -- (16.900000,11.679720);
  \draw[color=black!50!blue!30!green,thick] (15.200000,10.808477) -- (17.200000,10.808477);
  \draw[color=red,thick] (15.200000,9.940972) -- (17.200000,9.940972);
  \draw[color=blue,thick] (4.800000,0.419937) -- (6.800000,0.419937);
  \draw[color=blue,thick] (4.800000,0.411315) -- (6.800000,0.411315);
  \draw[color=red,thick] (5.200000,0.248153) -- (7.200000,0.248153);
  \draw[color=red,thick] (5.200000,0.353696) -- (7.200000,0.353696);
  \draw[color=black!50!blue!30!green,thick] (14.700000,13.944793) -- (16.700000,13.944793);
  \draw[color=blue,thick] (9.800000,0.223030) -- (11.800000,0.223030);
  \draw[color=blue,thick] (9.800000,0.513298) -- (11.800000,0.513298);
  \draw[color=red,thick] (10.200000,0.513492) -- (12.200000,0.513492);
  \draw[color=blue,thick] (9.800000,7.668895) -- (11.800000,7.668895);
  \draw[color=blue,thick] (9.800000,7.669193) -- (11.800000,7.669193);
  \draw[color=red,thick] (10.200000,7.669221) -- (12.200000,7.669221);
  \draw[color=blue,thick] (14.900000,11.697295) -- (16.900000,11.697295);
  \draw[color=black!50!blue!30!green,thick] (15.200000,11.660646) -- (17.200000,11.660646);
  \draw[color=red,thick] (15.200000,10.816137) -- (17.200000,10.816137);
  \draw[color=blue,thick] (4.800000,0.578382) -- (6.800000,0.578382);
  \draw[color=red,thick] (5.200000,0.562493) -- (7.200000,0.562493);

  \draw (-0.400000,0.380313) node[left] {\small $h^0$};
  \draw (-0.400000,7.362088) node[left] {\small $A^0$};
  \draw (-0.400000,7.995942) node[left] {\small $H^0$};
  \draw (2.400000,7.679338) node[right] {\small $H^\pm$};
  \draw (14.700000,11.371580) node[left] {\small $\tilde{q}_\text{L}$};
  \draw (14.700000,12.005435) node[left] {\small $\tilde{q}_\text{R}$};
  \draw (17.400000,10.495380) node[right] {\small $\tilde{b}_1$};
  \draw (17.400000,11.129234) node[right] {\small $\tilde{t}_2$};
  \draw (17.400000,9.940972) node[right] {\small $\tilde{t}_1$};
  \draw (4.600000,-0.139006) node[left] {\small $\tilde{\nu}_\text{L}$};
  \draw (4.600000,0.419937) node[left] {\small $\tilde{\ell}_\text{L}$};
  \draw (4.600000,1.128703) node[left] {\small $\tilde{\ell}_\text{R}$};
  \draw (7.400000,-0.162452) node[right] {\small $\tilde{\tau}_1$};
  \draw (7.400000,0.380313) node[right] {\small $\tilde{\nu}_\tau$};
  \draw (7.400000,1.105257) node[right] {\small $\tilde{\tau}_2$};
  \draw (14.500000,13.944793) node[left] {\small $\tilde{g}$};
  \draw (9.600000,0.129878) node[left] {\small $\tilde{\chi}_1^0$};
  \draw (9.600000,0.763732) node[left] {\small $\tilde{\chi}_2^0$};
  \draw (12.400000,0.513492) node[right] {\small $\tilde{\chi}_1^\pm$};
  \draw (9.600000,7.352117) node[left] {\small $\tilde{\chi}_3^0$};
  \draw (9.600000,7.985971) node[left] {\small $\tilde{\chi}_4^0$};
  \draw (12.400000,7.669221) node[right] {\small $\tilde{\chi}_2^\pm$};
  \draw (17.400000,11.660646) node[right] {\small $\tilde{b}_2$};
\end{tikzpicture}
    \end{minipage}
    \hfill
    \begin{minipage}[t]{.45\textwidth}
        \centering
        \tiny
    \begin{tikzpicture}[thick,scale=0.35, every node/.style={scale=0.8}]

  \draw (-3.000000,0.000000) rectangle (19.000000,15.400000);
  \draw (-3.000000,0.000000) node[left] {0};
  \draw (-3.000000,2.566667) node[left] {100};
  \draw (-2.700000,2.566667) -- (-3.000000,2.566667);
  \draw (-3.000000,5.133333) node[left] {200};
  \draw (-2.700000,5.133333) -- (-3.000000,5.133333);
  \draw (-3.000000,7.700000) node[left] {300};
  \draw (-2.700000,7.700000) -- (-3.000000,7.700000);
  \draw (-3.000000,10.266667) node[left] {400};
  \draw (-2.700000,10.266667) -- (-3.000000,10.266667);
  \draw (-3.000000,12.833333) node[left] {500};
  \draw (-2.700000,12.833333) -- (-3.000000,12.833333);
  \draw (-3.000000,15.400000) node[left] {600};
  \draw (-5.200000,15.400000) node[left,rotate=90] {Mass / GeV};

  \draw[-stealth,line width=1.00pt,dashed,color=black!69.99998962000001] (5.800000,9.938502) -- (10.800000,5.278373);
  \draw[-stealth,line width=1.00pt,dashed,color=black!69.9999097] (5.800000,9.734451) -- (10.800000,5.278373);
  \draw[-stealth,line width=1.00pt,dashed,color=black!70.0] (6.200000,5.872964) -- (10.800000,5.278373);
  \draw[-stealth,line width=0.69pt,dashed,color=black!51.43378696] (6.200000,8.370815) -- (6.200000,5.872964);
  \draw[-stealth,line width=0.31pt,dashed,color=black!28.56621304] (6.200000,8.370815) -- (10.800000,5.278373);
  \draw[-stealth,line width=0.05pt,dashed,color=black!12.787397746] (10.800000,12.148047) -- (5.800000,9.938502);
  \draw[-stealth,line width=0.05pt,dashed,color=black!13.26132244] (10.800000,12.148047) -- (5.800000,9.734451);
  \draw[-stealth,line width=0.18pt,dashed,color=black!20.8433365] (10.800000,12.148047) -- (6.200000,5.872964);
  \draw[-stealth,line width=0.12pt,dashed,color=black!17.02303324] (10.800000,12.148047) -- (6.200000,8.370815);
  \draw[-stealth,line width=0.11pt,dashed,color=black!16.53433888] (11.200000,12.152644) -- (5.800000,9.734451);
  \draw[-stealth,line width=0.23pt,dashed,color=black!24.05244682] (11.200000,12.152644) -- (6.200000,8.370815);
  \draw[-stealth,line width=1.00pt,dashed,color=black!69.99998253999999] (5.800000,13.688374) -- (10.800000,5.278373);
  \draw[-stealth,line width=0.09pt,dashed,color=black!15.11089006] (6.200000,13.312326) -- (0.800000,3.174918);
  \draw[-stealth,line width=0.09pt,dashed,color=black!15.11089006] (6.200000,13.312326) -- (6.200000,5.872964);
  \draw[-stealth,line width=0.38pt,dashed,color=black!32.78377588] (6.200000,13.312326) -- (6.200000,8.370815);
  \draw[-stealth,line width=0.15pt,dashed,color=black!19.04938678] (6.200000,13.312326) -- (10.800000,5.278373);
  \draw[-stealth,line width=0.00pt,dashed,color=black!10.092894889] (6.200000,13.312326) -- (10.800000,12.148047);

  \draw[color=blue,thick] (-0.200000,3.174918) -- (1.800000,3.174918);
  \draw[color=blue,thick] (4.800000,9.938502) -- (6.800000,9.938502);
  \draw[color=blue,thick] (4.800000,9.734451) -- (6.800000,9.734451);
  \draw[color=red,thick] (5.200000,5.872964) -- (7.200000,5.872964);
  \draw[color=red,thick] (5.200000,8.370815) -- (7.200000,8.370815);
  \draw[color=blue,thick] (9.800000,5.278373) -- (11.800000,5.278373);
  \draw[color=blue,thick] (9.800000,12.148047) -- (11.800000,12.148047);
  \draw[color=red,thick] (10.200000,12.152644) -- (12.200000,12.152644);
  \draw[color=blue,thick] (4.800000,13.688374) -- (6.800000,13.688374);
  \draw[color=red,thick] (5.200000,13.312326) -- (7.200000,13.312326);

  \draw (-0.400000,3.174918) node[left] {\small $h^0$};
  \draw (4.600000,9.525377) node[left] {\small $\tilde{\nu}_\text{L}$};
  \draw (4.600000,10.147576) node[left] {\small $\tilde{\ell}_\text{L}$};
  \draw (7.400000,5.872964) node[right] {\small $\tilde{\tau}_1$};
  \draw (7.400000,8.370815) node[right] {\small $\tilde{\nu}_\tau$};
  \draw (9.600000,5.278373) node[left] {\small $\tilde{\chi}_1^0$};
  \draw (9.600000,12.148047) node[left] {\small $\tilde{\chi}_2^0$};
  \draw (12.400000,12.152644) node[right] {\small $\tilde{\chi}_1^\pm$};
  \draw (4.600000,13.688374) node[left] {\small $\tilde{\ell}_\text{R}$};
  \draw (7.400000,13.312326) node[right] {\small $\tilde{\tau}_2$};
\end{tikzpicture}

    \end{minipage} 
        \caption{Particle mass spectrum given by \textbf{PySLHA} \cite{Buckley:2013jua} for model (a) (upper panels) and (c) (lower panels). The right panels are zoomed-in to the left panels for low-lying masses. }
    \label{fig:massspectrum}    
\end{figure}
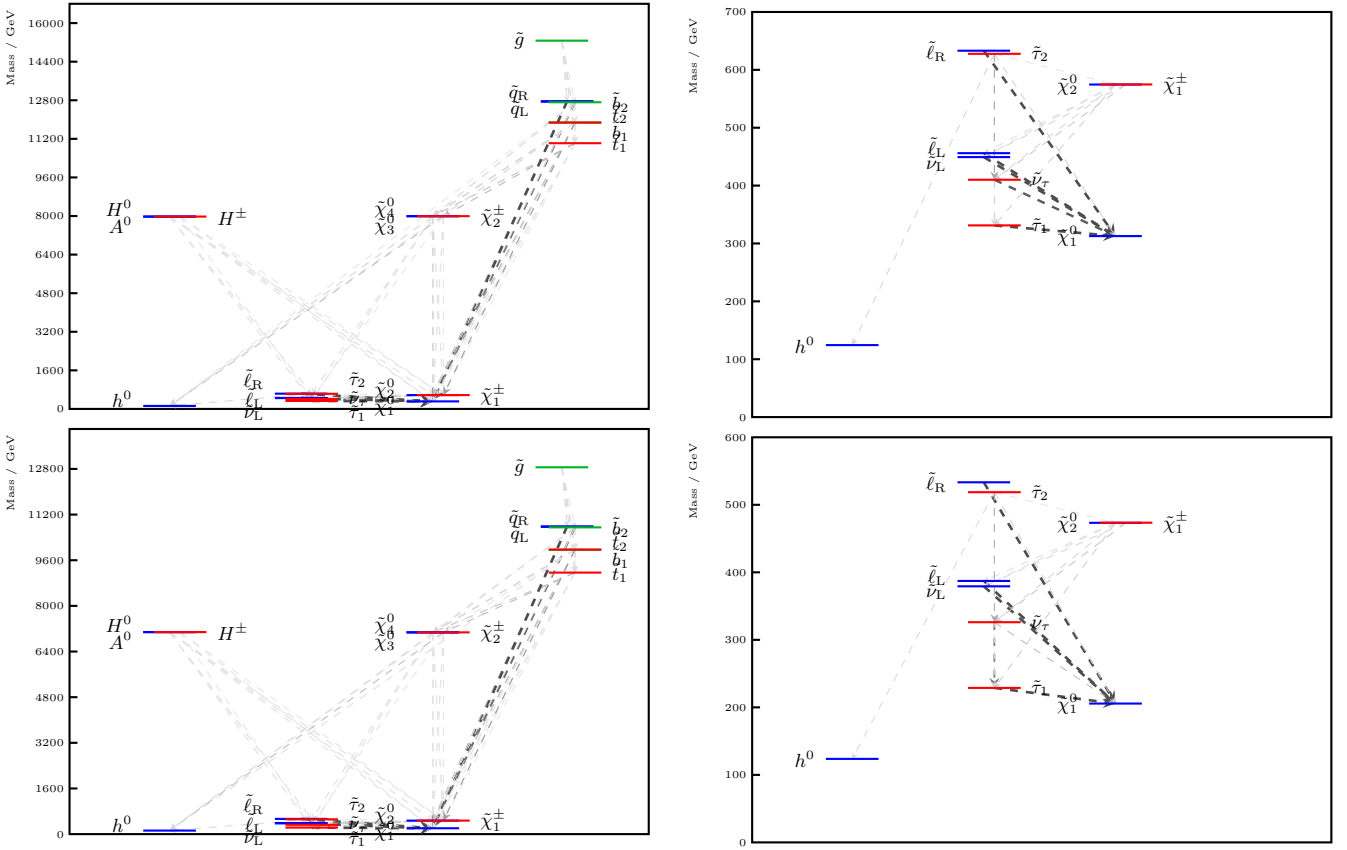

\clearpage
\section{Conclusion\label{sec:7}}
 In this work we have investigated the grand unified group $\mathsf{SO(10)}$
 with the Higgs representations $\mathsf{(560\cdot \overline{560})}$ in the heavy Higgs sector and $2\mathsf{\times 10+320}$ in the light sector sector which
has the property that it simultaneously reduces the rank of
$\mathsf{SO(10)}$ from 5 to 4 as well as reduces the rest of the symmetry to
the Standard Model gauge group. Further, $\mathsf{560+\overline{560}}$
contain five color Higgs triplets/anti-triplets and only four Higgs doublet pairs
while $2\mathsf{\times10+320}$ light Higgs contain 5  Higgs color triplets/anti-triplets
and 5  Higgs doublet pairs. Thus the mixing of the heavy and the light
Higgs sectors leads to all the color triplets/anti-triplets becoming heavy
and all of the Higgs doublets becoming heavy except on Higgs doublet pair
needed for electroweak symmetry breaking. We now list the new elements of the work accomplished 
in this paper. 
\begin{itemize}
\item
While the model was proposed in~\cite{Babu:2011tw}
further development of the model required explicit computations of the couplings which involves construction of the 320-multiplet and its couplings to $560+\overline{560}$  which were not available in previous works.
This has been accomplished in the current paper.  We expand further on the specifics of the new work below.
\item
Construction of the  
 irreducible tensor-spinor $\ket{\Theta_{\mu\nu}^{(\mathsf{560})}}$ starting from the  $\mathsf{720}~(=\mathsf{16}\times\mathsf{45})$ dimensional reducible tensor-spinor $\ket{\Theta_{\mu\nu}^{\mathsf{720}}}$
 is carried out  by removing  $\mathsf{160}~(=\mathsf{16}\times\mathsf{10})$ components
 as shown in Appendix \ref{sec:A}.

\item
Construction of two $\mathsf{320}$ multiplets $(\mathsf{320}_a, \mathsf{320}_s)$ one symmetric and the other anti-symmetric in the
 two indices starting with the 1000 dimensional reducible tensor  $T_{\mu\nu\lambda}$ 
is carried out in Appendix \ref{sec:D} and results exhibited in  Eqs.(\ref{320a}) and (\ref{320s}).

\item
Couplings involving tensor spinors with  tensor Higgs representations are not available in the literature 
and we have developed new techniques to accomplish it. Thus
one of the central accomplishments of this paper is that we have carried out an explicit computation of the couplings of the heavy sector with the light sector, i.e., $\mathsf{(560\cdot 560)\cdot 10}_i$ ($i=1,2$) and 
$\mathsf{(560\cdot 560)\cdot 320}$  using one of the tensor representations of the $\mathsf{320}$, i.e., the one anti-symmetric in two indices as shown in appendices \ref{sec:E}, \ref{sec:F} and \ref{sec:G}. 

\item
Further, we have computed the mass matrix of the Higgs doublets arising from the $\mathsf{560+\overline{560}}$, $\mathsf{320}$, and from $\mathsf{10}_i ~(i=1,2)$ after spontaneous breaking of the GUT symmetry in terms of the parameters 
of the cubic superpotential as shown in Appendix \ref{sec:G} and displayed in 
Eq.(\ref{higgs-matrix-elements}).
\item
After diagonalization of both the kinetic energy and the mass matrix we have identified the pair of light Higgs doublets that enter in the electroweak symmetry breaking. Thus we have deduced the MSSM with a pair of light Higgs doublets in a natural way from a diagonalization involving nine Higgs doublets where all but the electroweak
doublets become superheavy as  shown in Eq.(\ref{doublet mass matrix}) and Eq.(\ref{UV}).

\item
As another major result of the paper we show that the diagonalization of the Higgs mass matrix leads to splitting between the couplings of the up Higgs to the top quark vs the coupling of the down Higgs to the bottom quark due to the light Higgs doublets arising as linear combinations of the Higgses  in 
$\mathsf{320}$ and $\mathsf{10}_i~(i=1,2)$. This allows  
the Yukawa coupling of the top quark becoming larger than the Yukawa coupling to the down quark at the GUT scale which  leads to $b-t-\tau$ unification with low $\tan\beta$ with $\tan\beta$ as low as 5-10
as shown in Table (\ref{tab:yukawa}) and Table (\ref{tab:SUGRA}).

\item 
It is shown that the $\mathsf{SO(10)}$ model with $\tilde g$SUGRA RG evolution leads to split sparticle mass spectrum,
with a set of light sparticles involving the sleptons and the weakinos with masses lying in the few hundred
GeV region consistent with  the \bl{Higgs boson mass}, the relic density and $g_\mu-2$ data constraints. 
The remaining particle spectrum
consisting of the gluino, the squarks and the 
 Higgs $H^0, A^0, H^\pm$ is heavy with masses lying in the TeV region. This is exhibited in Table (\ref{tab:constraints}) and in Fig.(\ref{fig:massspectrum}). 
 
 \item
  The production cross sections for the light 
 sparticles are estimated in  Table(\ref{tab:crosssection}), Table(\ref{tab:crosssection-tau}),
 and in Table(\ref{tab:crosssection-chi2}), and found to be of order few fb which appears encouraging for the discovery of supersymmetry at the LHC. Specifically, the $\tilde \chi^0_2 \tilde \chi_1^\pm$ production cross section can lie in the range 40-80 fb as seen in Table(\ref{tab:crosssection-chi2}) 
and is of considerable interest as it leads to the trileptonic signal which is one of the signature events of
supersymmetry because it typically has low background from the Standard Model processes. 
However, a more dedicated analysis is needed to make definitive predictions after the appropriate detector cuts 
and backgrounds are included in the analysis. 

\end{itemize}
There are significant possibilities for further work in this class of $\mathsf{SO(10)}$ models. Thus, as mentioned above,
a more dedicated analysis of sparticle production signals taking into account detector cuts and backgrounds, would
allow one to make reliable estimates of what one may expect at HL-LHC and HE-LHC. Since all the parameters of the model are now constrained, prediction of proton decay branching ratios into 
$p\to e^+ \pi^0, p \to \bar \nu K^+$ and into other decay modes can be made. Another possible extension of the 
model is a double copy, i.e., MSSM$_2$, which gives four pair of light Higgs. In this case the possibility 
that one pair of light Higgs doublets couple with the third generation of fermions and the other to the first two 
generation of fermions has the potential of explaining the relative mass hierarchy between the third generation and the first generation of fermions. 
Of course such an extension requires a check on gauge coupling unification and
 consistency with the set of experimental constraints discussed in the analysis of this work.\\

\noindent
{\bf Acknowledgements:}
This research is supported in part by the NSF Grant PHY-2209903.\\

\begin{appendices}

{
\section{$\boldsymbol{\mathsf{560}+\mathsf{\overline{560}}}$ tensor-spinor in its $\boldsymbol{\mathsf{SU(5)}}$ oscillator modes \label{sec:A}}
{
{We begin by displaying the the tensorial structure of each $\mathsf{SU(5)}$ multiplets  within the  $\mathsf{560}-$plet.
\begin{eqnarray}\label{tensorstructure1}
   \mathsf{560}\left[\Theta_{\mu\nu}^{\breve{s}~(\mathsf{560})}\right]&=&\mathsf{1}\left[{\mathsf{H}}^{(\mathsf{560})}\right]~+~\mathsf{\overline{5}}
   \left[{\mathsf{H}}^{(\mathsf{560})}_i\right]~+~\mathsf{\overline{10}}\left[{\mathsf{H}}^{(\mathsf{560})}_{ij}\right]~+~\mathsf{10}
   \left[{\mathsf{H}}^{(\mathsf{560})ij}\right]~+~\mathsf{10}\left[\widehat{{\mathsf{H}}}^{(\mathsf{560})ij}\right]\nonumber\\
   &&+~\mathsf{24}\left[{\mathsf{H}}^{(\mathsf{560})i}_j\right]~+~\mathsf{40}\left[{\mathsf{H}}^{(\mathsf{560})ijk}_l\right]~+ ~\mathsf{45}
   \left[{\mathsf{H}}^{(\mathsf{560})ij}_k\right]~+~\mathsf{\overline{45}}\left[{\mathsf{H}}^{(\mathsf{560})k}_{ij}\right]\nonumber\\
   &&+~\mathsf{\overline{50}}
   \left[{\mathsf{H}}^{(\mathsf{560})lm}_{ijk}\right]~+~\mathsf{\overline{70}}\left[{\mathsf{H}}^{(\mathsf{560})k}_{(S)ij}\right]
   ~+~\mathsf{75}\left[{\mathsf{H}}^{(\mathsf{560})ij}_{kl}\right]
   ~+~\mathsf{175}
     \left[{\mathsf{H}}^{(\mathsf{560})m}_{[ijk]l}\right],
\end{eqnarray}
     where the Latin indices $i,j,k,\dots$ taking on the values $1,2,\dots, 5$. From here onward, we suppress the spinor index $\breve{s}$ for brevity.}}

\subsection{
{
{The reducible $\boldsymbol{{720}}$ dimensional tensor-spinor multiplet}}}
In Eqs. (\ref{def560}) it was shown that the irreducible tensor-spinor $\ket{\Theta_{\mu\nu}^{(\mathsf{560})}}$ can be obtained from the $\mathsf{720}~(=\mathsf{16}\times\mathsf{45})$ dimensional reducible tensor-spinor $\ket{\Theta_{\mu\nu}^{\mathsf{720}}}$   by removing  $\mathsf{160}~(=\mathsf{16}\times\mathsf{10})$ components. 
From here onward, we suppress the spinor index for brevity. For the purpose of applications we need to
exhibit the decomposition of reducible 720 and irreducible $\mathsf{560}$ mulitplets in terms of the $\mathsf{SU(5)}$ modes.
To this end
we consider a field theoretic description of the $\mathsf{720}-$plet and define the Clifford elements $\Gamma_{\mu}$ in terms of creation and annihilation operators $b_i$ and $b_i^{\dagger}$ \cite{Mohapatra:1979nn,Wilczek:1981iz} as
 $\Gamma_{2i}=(b_i+ b_i^{\dagger}), \Gamma_{2i-1}= -i(b_i-b_i^{\dagger})$,  where
 $\{b_i,b_j^{\dagger}\}=\delta_{ij}$,
$\{b_i,b_j\}=0$ and $\{b_i^{\dagger}, b_j^{\dagger}\} = 0$ and that the $\mathsf{SU(5)}$ singlet satisfies $b_i\ket{0}=0$.  In terms of $b_i, b_i^{\dagger}$ the $\mathsf{16}$ and $\mathsf{\overline{16}}$ multiplets of $\mathsf{SO(10)}$ have  the following $\mathsf{SU(5)}$ oscillator expansions:
$\ket{\Theta^{(\mathsf{16})}}=\ket{0}M+\frac{1}{2}b_i^{\dagger}b_j^{\dagger}\ket{0}{M}^{ij}+\frac{1}{24}\epsilon^{ijklm}
     b_j^{\dagger}b_k^{\dagger}b_l^{\dagger}b_m^{\dagger}\ket{0}{M}_{i}$ and $\ket{\overline{\Theta}^{(\mathsf{\overline{16}})}}=b_1^{\dagger}b_2^{\dagger}b_3^{\dagger}b_4^{\dagger}b_4^{\dagger}
     \ket{0}N+\frac{1}{12}\epsilon^{ijklm}b_k^{\dagger}b_l^{\dagger}b_m^{\dagger}\ket{0}{N}_{ij}+
     b_i^{\dagger}\ket{0}{N}^{i}$. Thus we have 
     \begin{eqnarray}\label{720plet}
     \ket{\Theta_{\mu\nu}^{({720})}}&=&\ket{0}X_{\mu\nu}+\frac{1}{2}b_i^{\dagger}b_j^{\dagger}\ket{0}X_{\mu\nu}^{ij}+\frac{1}{24}\epsilon^{ijklm}
     b_j^{\dagger}b_k^{\dagger}b_l^{\dagger}b_m^{\dagger}\ket{0}X_{i\mu\nu}.
     \end{eqnarray}
 We now display the $\mathsf{SU(5)}$ decomposition of   $\ket{\Theta_{\mu\nu}^{({720})}}$:
\begin{eqnarray}\label{A1}
\begin{split}
\Theta_{\mu\nu}(720)&={X}_{\mu\nu}~(45)~\oplus~{X}_{i\mu\nu}~(\overline{5}\times45)~\oplus~{X}_{\mu\nu}^{ij}~(10\times 45),\\
X_{\mu\nu}(45)&={X}_{c_ic_j}~(10)~\oplus~{X}_{\overline{c}_i\overline{c}_j}~(\overline{10})~\oplus~{X}_{{c}_i\overline{c}_j}~(25),\\
X_{i\mu\nu}(\overline{5}\times 45)&={X}_{ic_jc_k}~(\overline{5}\times 10)~\oplus~{X}_{i \overline{c}_j\overline{c}_k}~(\overline{5}\times \overline{10})~\oplus~{X}_{i{c}_k\overline{c}_j}~(\overline{5}\times 25),\\
X_{\mu\nu}^{ij}({10}\times 45)&={X}_{c_kc_l}^{ij}~({10}\times 10)~\oplus~{X}_{ \overline{c}_k\overline{c}_l}^{ij}~(10\times \overline{10})~\oplus~{X}_{{c}_k\overline{c}_l}^{ij}~({10}\times 25),
\end{split}
\end{eqnarray}
     where the quantities in the parenthesis represent the dimensionality of the tensor. These $\mathsf{SU(5)}$ reducible tensors are further decomposed into their irreducible parts as follows
     \begin{eqnarray}\label{A2}
     \begin{split}
     \mathsf{10_1}:&~~{X}_{c_ic_j}={}^{(1)}\!{U}^{ij},\\
     \overline{\mathsf{10}}:&~~{X}_{\overline{c}_i\overline{c}_j}={ U}_{ij},\\
     {25}=\mathsf{24_1}+\mathsf{1_1}:&~~ {X}_{{c}_i\overline{c}_j}={}^{(1)}\!{ U}^{i}_j+\frac{1}{5}\delta^i_j{}^{(1)}\!{ U},
     \end{split}
     \end{eqnarray}
      \begin{eqnarray}
       \begin{split}
     \overline{5}\times 10=\mathsf{45}+\mathsf{5}:&~~{X}_{ic_jc_k}={ U}^{jk}_i+\frac{1}{4}\left(\delta^k_i{ U}^{j}-\delta^j_i{ U}^{k}\right),\\
     \overline{{5}}\times \overline{{10}}=\mathsf{40_1}+\mathsf{10_2}:&~~{X}_{i\overline{c}_j\overline{c}_k}=\epsilon_{jklmn}{}^{(1)}\!{ U}^{lmn}_i+\epsilon_{ijklm}{}^{(2)}\!{ U}^{lm},\\
     \overline{5}\times 25=\overline{\mathsf{45_1}}+\overline{\mathsf{5}}_1+\overline{\mathsf{70}}+\overline{\mathsf{5_2}}:&~~
     {X}_{i{c}_k\overline{c}_j}=\frac{1}{2}{}^{(1)}\!{ U}^{k}_{ij}+\frac{1}{8}\left(\delta^k_j{}^{(1)}\!{ U}_i-\delta^k_i{}^{(1)}\!{ U}_j\right)+
     \frac{1}{2}{ U}^{k}_{(S)ij}+\frac{1}{12}\left(\delta^k_j{}^{(2)}\!{ U}_i-\delta^k_i{}^{(2)}\!{ U}_j\right),
     \end{split}
     \end{eqnarray}
         \begin{eqnarray}
         \begin{split}
     {10}\times 10=\overline{\mathsf{50}}+\overline{\mathsf{45_2}}+\overline{\mathsf{5_3}}:&~~{X}_{c_kc_l}^{ij}=
     \epsilon^{ijmno}{ U}^{kl}_{mno}+\frac{3}{2}\left(\epsilon^{ijkmn}{}^{(2)}\!{ U}^{l}_{mn}-\epsilon^{ijlmn}{}^{(2)}\!{ U}^{k}_{mn}\right)
     +\frac{1}{2}\epsilon^{ijklm}{}^{(3)}\!{ U}_{m},\\
    10\times \overline{10}=\mathsf{75}+\mathsf{24}_2+\mathsf{1}_2:&~~ {X}_{ \overline{c}_k\overline{c}_l}^{ij}={ U}^{ij}_{kl}+
    \frac{1}{3}\left(\delta^i_l{}^{(2)}\!{ U}^{j}_{k}-\delta^i_k{}^{(2)}\!{ U}^{j}_{l}+\delta^j_k{}^{(2)}\!{ U}^{i}_{l}-
    \delta^j_l{}^{(2)}\!{ U}^{i}_{k}\right)+\frac{1}{20}\left(\delta^i_l\delta^j_k-\delta^i_k\delta^j_l\right){}\!^{(2)}\!{ U},\\
    {10}\times 25=\mathsf{10_3}+\mathsf{15}+\mathsf{10_4}+\mathsf{40_2}+\mathsf{175}:&~~{X}_{{c}_k\overline{c}_l}^{ij}=
    \frac{1}{6}\left(\delta^i_l{}^{(3)}\!{ U}^{jk}-\delta^j_l{}^{(3)}\!{ U}^{ik}+\delta^k_l{}^{(3)}\!{ U}^{ij}\right)+
    \frac{1}{8}\left(\delta^i_l{ U}^{jk}_{(S)}-\delta^j_l{ U}^{ik}_{(S)}\right)\\
    &~~~~~~~~~~~~~+\frac{1}{3}\left(\delta^i_l{}^{(4)}\!{ U}^{jk}-\delta^j_l{}^{(4)}\!{ U}^{ik}+\delta^k_l{}^{(4)}\!{ U}^{ij}\right)+ {}^{(2)}\!{ U}^{ijk}_{l}+\epsilon^{ijxyz}{ U}^k_{[xyz]l}.
     \end{split}
     \end{eqnarray}

Next, we investigate the implication of the constraint Eq. (\ref{560 Constraint 1}). To that end, contracting $\Gamma_{\mu}$ with the $\mathsf{720}-$plet gives
\begin{eqnarray}
\Gamma_{\mu}\ket{\Theta_{\mu\nu}^{(\mathsf{720})}}&=& \ket{\overline{\Theta}_{\nu}^{(\mathsf{\overline{160}})}},
\end{eqnarray}
where
\begin{subequations}\label{A3}
		\begin{align}
\ket{\overline{\Theta}_{c_n}^{(\mathsf{\overline{160}})}}&=b_1^{\dagger}b_2^{\dagger}b_3^{\dagger}b_4^{\dagger}b_4^{\dagger}
     \ket{0}{ U}^n+b_i^{\dagger}\ket{0}\left({}^{(1)}\!{ U}^{in}-\frac{1}{2}{}^{(3)}\!{ U}^{in}-\frac{1}{2}{ U}_{(S)}^{in}-{}^{(4)}\!{ U}^{in}\right)\nonumber\\
     &~~~+\frac{1}{12}\epsilon^{ijklm}b_k^{\dagger}b_l^{\dagger}b_m^{\dagger}\ket{0}\left[72{}^{(2)}\!{ U}^{n}_{ij}+18\left(\delta^n_i{}^{(3)}\!{ U}_{j}-\delta^n_j{}^{(3)}\!{ U}_{i}\right)+{}^{(1)}\!{ U}^{n}_{ij}+
     \frac{1}{4}\left(\delta^n_i{}^{(1)}\!{ U}_{j}-\delta^n_j{}^{(1)}\!{ U}_{i}\right)\right]\\
    \ket{\overline{\Theta}_{\overline{c}_{n}}^{(\mathsf{\overline{160}})}}&=b_1^{\dagger}b_2^{\dagger}b_3^{\dagger}b_4^{\dagger}b_4^{\dagger}
     \ket{0}\left(-\frac{1}{2}{}^{(1)}\!{ U}_{n}-\frac{1}{2}{}^{(2)}\!{ U}_{n}\right)+ b_i^{\dagger}\ket{0}\left({}^{(1)}\!{ U}_{n}^i+\frac{1}{5}\delta^i_n{}^{(1)}\!{ U}-{}^{(2)}\!{ U}_{n}^i-\frac{1}{5}\delta^i_n{}^{(2)}\!{ U}\right)\nonumber\\
     &~~~+\frac{1}{12}\epsilon^{ijklm}b_k^{\dagger}b_l^{\dagger}b_m^{\dagger}\ket{0}\left[\epsilon_{ijnop}\left({}^{(2)}\!{ U}^{op}+3{}^{(3)}\!{ U}^{op}+6{}^{(4)}\!{ U}^{op}\right)+\epsilon_{ijopq}\left({}^{(1)}\!{ U}^{opq}_n+6{}^{(2)}\!{ U}^{opq}_n\right)\right]
\end{align}
\end{subequations}
To get the $\mathsf{560}$ tensor-spinor, $\ket{\Theta_{\mu\nu}^{(\mathsf{560})}}$, we must impose Eq. (\ref{560 Constraint 1}). This is equivalent to setting $\ket{\Theta_{\nu}^{(\mathsf{\overline{160}})}}=0$. Thus  Eqs. (\ref{A3}) gives the following constraints
\begin{subequations}\label{A4}
		\begin{align}
{ U}^n&=0,\\
{}^{(1)}\!{ U}^{in}-\frac{1}{2}{}^{(3)}\!{ U}^{in}-\frac{1}{2}{ U}_{(S)}^{in}-{}^{(4)}\!{ U}^{in}&=0,\\
72{}^{(2)}\!{ U}^{n}_{ij}+18\left(\delta^n_i{}^{(3)}\!{ U}_{j}-\delta^n_j{}^{(3)}\!{ U}_{i}\right)+{}^{(1)}\!{ U}^{n}_{ij}+\frac{1}{4}\left(\delta^n_i{}^{(1)}\!{ U}_{j}-\delta^n_j{}^{(1)}\!{ U}_{i}\right)&=0,\\
-\frac{1}{2}{}^{(1)}\!{ U}_{n}-\frac{1}{2}{}^{(2)}\!{ U}_{n}&=0,\\
{}^{(1)}\!{ U}_{n}^i+\frac{1}{5}\delta^i_n{}^{(1)}\!{ U}-{}^{(2)}\!{ U}_{n}^i-\frac{1}{5}\delta^i_n{}^{(2)}\!{ U}&=0,\\
\epsilon_{ijnop}\left({}^{(2)}\!{ U}^{op}+3{}^{(3)}\!{ U}^{op}+6{}^{(4)}\!{ U}^{op}\right)+\epsilon_{ijopq}\left({}^{(1)}\!{ U}^{opq}_n+6{}^{(2)}\!{ U}^{opq}_n\right)&=0.
\end{align}
\end{subequations}
The above allows us to write
\begin{eqnarray}
\ket{\Theta_{\mu\nu}^{(\mathsf{560})}}=\ket{\Theta_{\mu\nu}^{(\mathsf{720})}}\left|_{constraint~of ~Eqs.~(\ref{A4})}\right.
\end{eqnarray}
 Next we replace the notation $U$, $U_i, U_{ij}$ etc  by the more tranparent
 particle notation by defining
\[
{}^{(2)}\!{ U}\equiv{\mathsf{H}}^{(\mathsf{560})},~~~{}^{(1)}\!{ U}_{i}\equiv{\mathsf{H}}^{(\mathsf{560})}_{i},~~~
{ U}_{ij}\equiv{\mathsf{H}}^{(\mathsf{560})}_{ij},~~~
{}^{(1)}\!{ U}^{ij}\equiv{\mathsf{H}}^{(\mathsf{560})ij},~~~{}^{(2)}\!{ U}^{ij}\equiv\widehat{\mathsf{H}}^{(\mathsf{560})ij},
~~~{}^{(2)}\!{ U}^{i}_j\equiv{\mathsf{H}}^{(\mathsf{560})i}_j,~~~{}^{(1)}\!{ U}^{ijk}_l\equiv{\mathsf{H}}^{(\mathsf{560})ijk}_l,\]
\[{ U}^{ij}_k\equiv{\mathsf{H}}^{(\mathsf{560})ij}_k,~~~{}^{(1)}\!{ U}^{k}_{ij}\equiv{\mathsf{H}}^{(\mathsf{560})k}_{ij},~~~{ U}^{lm}_{ijk}\equiv
{\mathsf{H}}^{(\mathsf{560})lm}_{ijk},
~~~{ U}^{k}_{(S)ij}={\mathsf{H}}^{(\mathsf{560})k}_{(S)ij},~~~{ U}^{ij}_{kl}={\mathsf{H}}^{(\mathsf{560})ij}_{kl},~~~{ U}^m_{[ijk]l}=
{\mathsf{H}}^{(\mathsf{560})m}_{[ijk]l}.\]
With the above notation and 
 using Eqs. (\ref{A4}), we write the expansion of the $\mathsf{560}-$dimensional tensor-spinor in terms of its $\mathsf{SU(5)}$ oscillator modes:
\begin{eqnarray}
\ket{\Theta_{c_xc_y}^{(\mathsf{560})}}&=&\ket{0}{\mathsf{H}}^{(\mathsf{560})xy}+\frac{1}{2}b_i^{\dagger}b_j^{\dagger}\ket{0}
\left[\epsilon^{ijklm}{\mathsf{H}}^{(\mathsf{560})xy}_{klm}\right.\nonumber\\
&&\left.+\frac{1}{{48}}\left(\epsilon^{ijkly}{\mathsf{H}}^{(\mathsf{560})x}_{kl}
-\epsilon^{ijklx}{\mathsf{H}}^{(\mathsf{560})y}_{kl}\right)-\frac{1}{{144}}\epsilon^{ijkxy}{\mathsf{H}}^{(\mathsf{560})}_k\right]
+\frac{1}{24}\epsilon^{ijklm}
     b_j^{\dagger}b_k^{\dagger}b_l^{\dagger}b_m^{\dagger}\ket{0}{\mathsf{H}}^{(\mathsf{560})xy}_i,\label{tensor1}\\
     \nonumber\\
     \ket{\Theta_{\bar{c}_x\bar{c}_y}^{(\mathsf{560})}}&=&\ket{0}{\mathsf{H}}^{(\mathsf{560})}_{xy}+\frac{1}{2}b_i^{\dagger}b_j^{\dagger}\ket{0}
\left[{\mathsf{H}}^{(\mathsf{560})ij}_{xy}+\frac{1}{3}\left(\delta^i_y{\mathsf{H}}^{(\mathsf{560})j}_x-\delta^i_x{\mathsf{H}}^{(\mathsf{560})j}_y
+\delta^j_x{\mathsf{H}}^{(\mathsf{560})i}_y-\delta^j_y{\mathsf{H}}^{(\mathsf{560})i}_x\right)\right.\nonumber\\
&&+\left.\frac{1}{20}\left(\delta^i_y\delta^j_x-\delta^i_x\delta^j_y\right){\mathsf{H}}^{(\mathsf{560})}\right]+\frac{1}{24}\epsilon^{ijklm}
     b_j^{\dagger}b_k^{\dagger}b_l^{\dagger}b_m^{\dagger}\ket{0}\left[\epsilon_{nopxy}{\mathsf{H}}^{(\mathsf{560})nop}_i\right.\nonumber\\
     &&\left. +\epsilon_{inoxy}{\widehat{\mathsf{H}}}^{(\mathsf{560})no}\right],\label{tensor2}\\
     \nonumber\\
    \ket{\Theta_{{c}_x\bar{c}_y}^{(\mathsf{560})}}&=&\ket{0}\left[{\mathsf{H}}^{(\mathsf{560})x}_y+\frac{1}{5}\delta^{x}_y
    {\mathsf{H}}^{(\mathsf{560})}\right]+\frac{1}{2}b_i^{\dagger}b_j^{\dagger}\ket{0}
\left[\frac{1}{4}\left(\delta^i_y{\mathsf{H}}^{(\mathsf{560})jx}-\delta^j_y{\mathsf{H}}^{(\mathsf{560})ix}\right)\right.\nonumber\\
&&-\left.\frac{1}{72}\left(4\delta^x_y{\widehat{\mathsf{H}}}^{(\mathsf{560})ij}
-\delta^j_y{\widehat{\mathsf{H}}}^{(\mathsf{560})ix}+\delta^i_y
{\widehat{\mathsf{H}}}^{(\mathsf{560})jx}\right)-\frac{1}{6}{\mathsf{H}}^{(\mathsf{560})ijx}_y+\epsilon^{ijklm}
{\mathsf{H}}^{(\mathsf{560})x}_{[klm]y}\right]\nonumber\\
&&+\frac{1}{24}\epsilon^{ijklm}
     b_j^{\dagger}b_k^{\dagger}b_l^{\dagger}b_m^{\dagger}\ket{0}\left[\frac{1}{24}\left(5\delta^x_y{\mathsf{H}}^{(\mathsf{560})}_i-\delta^x_i
     {\mathsf{H}}^{(\mathsf{560})}_y\right)
     -\frac{1}{2}\left({\mathsf{H}}^{(\mathsf{560})x}_{iy}+{\mathsf{H}}^{(\mathsf{560})x}_{(S)iy}\right)
     \right].\label{tensor3}
\end{eqnarray}

Similarly, one can start with  ${\overline{720}}~(=\mathsf{\overline{16}}\times\mathsf{45})$ tensor-spinor as
\[\ket{\overline{\Theta}_{\mu\nu}^{({\overline{720}})}}=b_1^{\dagger}b_2^{\dagger}b_3^{\dagger}b_4^{\dagger}b_4^{\dagger}
     \ket{0}{\overline X}_{\mu\nu}+\frac{1}{12}\epsilon^{ijklm}b_k^{\dagger}b_l^{\dagger}b_m^{\dagger}\ket{0}{\overline X}_{\mu\nu ij}+
     b_i^{\dagger}\ket{0}{\overline X}^{i}_{\mu\nu},\]
  and use the constraint $\Gamma_{\mu}\ket{\overline{\Theta}_{\mu\nu}^{(\mathsf{\overline{720}})}}= \ket{\Theta_{\nu}^{(\mathsf{{160}})}}=0$ (equivalent to  $\Gamma_{\mu}\ket{\overline{\Theta}_{\mu\nu}^{(\mathsf{\overline{560}})}}=0$) to obtain the $\mathsf{\overline{560}}-$plet. Here we find
\begin{eqnarray}
 \ket{\overline{\Theta}_{{c}_x{c}_y}^{(\overline{\mathsf{560}})}}&=&b_1^{\dagger}b_2^{\dagger}b_3^{\dagger}b_4^{\dagger}b_5^{\dagger}\ket{0}
 {\mathsf{H}}^{(\overline{\mathsf{560}})xy}+\frac{1}{12}\epsilon^{ijklm}b_k^{\dagger}b_l^{\dagger}b_m^{\dagger}\ket{0}
\left[{\mathsf{H}}^{(\overline{\mathsf{560}})xy}_{ij}+\frac{1}{3}\left(\delta_i^y{\mathsf{H}}^{(\overline{\mathsf{560}})x}_j
-\delta_i^x{\mathsf{H}}^{(\overline{\mathsf{560}})y}_j+\delta_j^x{\mathsf{H}}^{(\overline{\mathsf{560}})y}_i\right.\right.\nonumber\\
&&\left.\left.-\delta_j^y
{\mathsf{H}}^{(\overline{\mathsf{560}})x}_i\right)+\frac{1}{20}\left(\delta_i^y\delta_j^x-\delta_i^x\delta_j^y\right)
{\mathsf{H}}^{(\overline{\mathsf{560}})}\right]+
    b_i^{\dagger}\ket{0}\left[\epsilon^{nopxy}{\mathsf{H}}^{(\overline{\mathsf{560}})i}_{nop}
    +\epsilon^{inoxy}\widehat{\mathsf{H}}^{(\overline{\mathsf{560}})}_{no}\right],\label{tensor4}\\
     \nonumber\\
\ket{\overline{\Theta}_{\bar{c}_x\bar{c}_y}^{(\overline{\mathsf{560}})}}&=&b_1^{\dagger}b_2^{\dagger}b_3^{\dagger}b_4^{\dagger}b_5^{\dagger}
\ket{0}{\mathsf{H}}^{(\overline{\mathsf{560}})}_{xy}+\frac{1}{12}\epsilon^{ijklm}b_k^{\dagger}b_l^{\dagger}b_m^{\dagger}\ket{0}
\left[\epsilon_{ijpqr}{\mathsf{H}}^{(\overline{\mathsf{560}})pqr}_{xy}\right.\nonumber\\
&&\left.+\frac{1}{{48}}\left(\epsilon_{ijpqy}{\mathsf{H}}^{(\overline{\mathsf{560}})pq}_{x}
-\epsilon_{ijpqx}{\mathsf{H}}^{(\overline{\mathsf{560}})pq}_{y}\right)-\frac{1}{{144}}\epsilon_{ijpxy}
{\mathsf{H}}^{(\overline{\mathsf{560}})p}\right]+b_i^{\dagger}\ket{0}{\mathsf{H}}^{(\overline{\mathsf{560}})i}_{xy},\label{tensor5}\\
     \nonumber\\
     \ket{\overline{\Theta}_{{c}_x\bar{c}_y}^{(\overline{\mathsf{560}})}}&=&b_1^{\dagger}b_2^{\dagger}b_3^{\dagger}b_4^{\dagger}b_5^{\dagger}
     \ket{0}\left[{\mathsf{H}}^{(\overline{\mathsf{560}})x}_y+\frac{1}{5}\delta^{x}_y
     {\mathsf{H}}^{(\overline{\mathsf{560}})}\right]+\frac{1}{12}\epsilon^{ijklm}b_k^{\dagger}b_l^{\dagger}b_m^{\dagger}\ket{0}
\left[\frac{1}{4}\left(\delta^x_i{\mathsf{H}}^{(\overline{\mathsf{560}})}_{jy}-\delta_j^x{\mathsf{H}}^{(\overline{\mathsf{560}})}_{iy}
\right)\right.\nonumber\\
&&-\left.\frac{1}{72}\left(4\delta^x_y
\widehat{\mathsf{H}}^{(\overline{\mathsf{560}})}_{ij}
-\delta_j^x\widehat{\mathsf{H}}^{(\overline{\mathsf{560}})}_{iy}+\delta_i^x\widehat{\mathsf{H}}^{(\overline{\mathsf{560}})}_{jy}\right)-\frac{1}{6}
{\mathsf{H}}^{(\overline{\mathsf{560}})x}_{ijy}+\epsilon_{ijpqr}{\mathsf{H}}^{(\overline{\mathsf{560}})[pqr]x}_{y}\right]\nonumber\\
&&+b_i^{\dagger}\ket{0}\left[\frac{1}{24}\left(5\delta^x_y{\mathsf{H}}^{(\overline{\mathsf{560}})i}-\delta_y^i
{\mathsf{H}}^{(\overline{\mathsf{560}})x}\right)
     -\frac{1}{2}\left({\mathsf{H}}^{(\overline{\mathsf{560}})ix}_{y}+{\mathsf{H}}^{(\overline{\mathsf{560}})ix}_{(S)y}\right)
     \right],\label{tensor6}
\end{eqnarray}
where
\begin{eqnarray}\label{tensorstructure2}
   \mathsf{\overline{560}}\left[\overline{\Theta}_{\mu\nu}^{(\mathsf{\overline{560}})}\right]&=&\mathsf{\overline{1}}(5)
   \left[{\mathsf{H}}^{(\overline{\mathsf{560}})}\right]~+~\mathsf{{5}}(-3)\left[{\mathsf{H}}^{(\overline{\mathsf{560}})i}\right]~+~\mathsf{{10}}(9)
   \left[{\mathsf{H}}^{(\overline{\mathsf{560}})ij}\right]~+~\overline{\mathsf{10}}(1)\left[{\mathsf{H}}^{(\overline{\mathsf{560}})}_{ij}\right]
   ~+~\overline{\mathsf{10}}(1)\left[\widehat{\mathsf{H}}^{(\overline{\mathsf{560}})}_{ij}\right]\nonumber\\
   &&+~\overline{\mathsf{24}}(5)\left[{\mathsf{H}}^{(\overline{\mathsf{560}})i}_j\right]~+~\overline{\mathsf{40}}(1)
   \left[{\mathsf{H}}^{(\overline{\mathsf{560}})l}_{ijk}\right]~+ ~\overline{\mathsf{45}}(-7)
   \left[{\mathsf{H}}^{(\overline{\mathsf{560}})k}_{ij}\right]~+~\mathsf{{45}}(-3)\left[{\mathsf{H}}^{(\overline{\mathsf{560}})ij}_k\right]
   \nonumber\\
   &&+~\mathsf{{50}}(-3)\left[{\mathsf{H}}^{(\overline{\mathsf{560}})ijk}_{lm}\right]~+~\mathsf{{70}}(-3)
   \left[{\mathsf{H}}^{(\overline{\mathsf{560}})ij}_{(S)k}\right]~+~\overline{\mathsf{75}}(5)
   \left[{\mathsf{H}}^{(\overline{\mathsf{560}})ij}_{kl}\right]~+~\overline{\mathsf{175}}(1)
     \left[{\mathsf{H}}^{(\overline{\mathsf{560}})[ijk]l}_m\right].
\end{eqnarray}

Eqs. (\ref{tensor1} - \ref{tensor6}) show how the various  $\mathsf{SU(5)}$  components given in Eqs. (\ref{tensorstructure1}) and (\ref{tensorstructure2}) enter in the oscillator decomposition of the constrained $\mathsf{560}+\mathsf{\overline{560}}$ multiplet.
}

{
\subsection{Normalization of kinetic energies of irreducible  $\boldsymbol{\mathsf{SU(5)}}$
multiplets arising in the decomposition of 
 $\boldsymbol{\mathsf{560}+\mathsf{\overline{560}}}$ multiplets}
 We exhibit below the decomposition of the
kinetic energy of the $\mathsf{560}-$plet in terms of the kinetic energies
of $\mathsf{SU(5)}$ multiplets residing in $\mathsf{560}$.
\begin{eqnarray}
\mathcal{L}_{{\textsc{ke}}}^{(\mathsf{560})}&=&- \braket{\partial_A\Theta_{\mu\nu}^{(\mathsf{560})}|\partial^A\Theta_{\mu\nu}^{(\mathsf{560})}}\nonumber\\
&=&-\partial_AX_{\mu\nu}^{\dagger}\partial^AX_{\mu\nu}-\frac{1}{2}\partial_AX_{ij\mu\nu}^{\dagger}\partial^AX_{\mu\nu}^{ij}
-\partial_AX_{\mu\nu}^{i\dagger}\partial^AX_{i\mu\nu}  \nonumber\\
&=&-\partial_AX_{c_ic_j}^{\dagger}\partial^AX_{c_ic_j} -\partial_AX_{\bar{c}_i\bar{c}_j}^{\dagger}\partial^AX_{\bar{c}_i\bar{c}_j}
-2\partial_AX_{{c}_i\bar{c}_j}^{\dagger}\partial^AX_{{c}_i\bar{c}_j} \nonumber\\
&&-\frac{1}{2}\partial_AX_{ijc_kc_l}^{\dagger}\partial^AX_{c_kc_l}^{ij} -\frac{1}{2}\partial_AX_{ij\bar{c}_k\bar{c}_l}^{\dagger}\partial^AX^{ij}_{\bar{c}_k\bar{c}_l}
-\partial_AX_{ij{c}_k\bar{c}_l}^{\dagger}\partial^AX_{{c}_k\bar{c}_l}^{ij} \nonumber\\
&&-\partial_AX_{c_jc_k}^{i\dagger}\partial^AX_{ic_jc_k} -\partial_AX_{\bar{c}_j\bar{c}_k}^{i\dagger}\partial^AX_{i\bar{c}_j\bar{c}_k}
-2\partial_AX_{{c}_j\bar{c}_k}^{i\dagger}\partial^AX_{i{c}_j\bar{c}_k} \nonumber\\
&=&-\frac{9}{20}{\partial_A}{\mathsf{H}}^{(\mathsf{560})\dagger}{\partial^A}{\mathsf{H}}^{(\mathsf{560})}-\frac{721}{1728}
{\partial_A}{\mathsf{H}}^{(\mathsf{560})i\dagger}{\partial^A}{\mathsf{H}}^{(\mathsf{560})}_i
-{\partial_A}{\mathsf{H}}^{(\mathsf{560})ij\dagger}{\partial^A}{\mathsf{H}}^{(\mathsf{560})}_{ij}-\frac{3}{2}
{\partial_A}{\mathsf{H}}^{(\mathsf{560})\dagger}_{ij}{\partial^A}{\mathsf{H}}^{(\mathsf{560})ij}
\nonumber\\
&&-\frac{865}{72}{\partial_A}\widehat{\mathsf{H}}^{(\mathsf{560})\dagger}_{ij}{\partial^A}\widehat{\mathsf{H}}^{(\mathsf{560})ij}
-{\frac{8}{3}}{\partial_A}{\mathsf{H}}^{(\mathsf{560})i\dagger}_{j}{\partial^A}{\mathsf{H}}^{(\mathsf{560})j}_i
-\frac{433}{36}{\partial_A}{\mathsf{H}}^{(\mathsf{560})l\dagger}_{ijk}{\partial^A}{\mathsf{H}}^{(\mathsf{560})ijk}_l
-{\partial_A}{\mathsf{H}}^{(\mathsf{560})i\dagger}_{jk}{\partial^A}{\mathsf{H}}^{(\mathsf{560})jk}_i\nonumber\\
&&
-\frac{145}{288}{\partial_A}{\mathsf{H}}^{(\mathsf{560})ij\dagger}_{k}{\partial^A}{\mathsf{H}}^{(\mathsf{560})k}_{ij}-6
{\partial_A}{\mathsf{H}}^{(\mathsf{560})ijk\dagger}_{lm}{\partial^A}{\mathsf{H}}^{(\mathsf{560})lm}_{ijk}
-\frac{1}{2}{\partial_A}{\mathsf{H}}^{(\mathsf{560})ij\dagger}_{(S)k}{\partial^A}{\mathsf{H}}^{(\mathsf{560})k}_{(S)ij}
-\frac{1}{2}{\partial_A}{\mathsf{H}}^{(\mathsf{560})kl\dagger}_{ij}{\partial^A}{\mathsf{H}}^{(\mathsf{560})ij}_{kl}
~~~~~~~~~~~\nonumber\\
&&
-12{\partial_A}{\mathsf{H}}^{(\mathsf{560})[ijk]l\dagger}_{m}{\partial^A}{\mathsf{H}}^{(\mathsf{560})m}_{[ijk]l},\nonumber\\
\nonumber          
\end{eqnarray}
where $A=0,\cdots 3$ is the Dirac index.
In terms of  normalized $\mathsf{SU(5)}$ fields , the kinetic energy of $\mathsf{560}$ tensor-spinor takes the form
\begin{eqnarray}
\mathcal{L}_{{\textsc{ke}}}^{(\mathsf{560})}&=&-{\partial_A}{\mathbf{H}}^{(\mathsf{1_{560}})\dagger}{\partial^A}{\mathbf{H}}^{(\mathsf{1_{560}})}
-{\partial_A}{\mathbf{H}}^{(\mathsf{\overline{5}_{560}})i\dagger}{\partial^A}{\mathbf{H}}^{(\mathsf{\overline{5}_{560}})}_i
-\frac{1}{2!}{\partial_A}{\mathbf{H}}^{(\mathsf{\overline{10}_{560}})ij\dagger}{\partial^A}{\mathbf{H}}^{(\mathsf{\overline{10}_{560}})}_{ij}
-\frac{1}{2!}{\partial_A}{\mathbf{H}}^{(\mathsf{{10}_{560}})\dagger}_{ij}{\partial^A}{\mathbf{H}}^{(\mathsf{{10}_{560}})ij}
\nonumber\\
&&-\frac{1}{2!}{\partial_A}\widehat{\mathbf{H}}^{(\mathsf{{10}_{560}})\dagger}_{ij}{\partial^A}\widehat{\mathbf{H}}^{(\mathsf{{10}_{560}})ij}-
{\partial_A}{\mathbf{H}}^{(\mathsf{{24}_{560}})i\dagger}_{j}{\partial^A}{\mathbf{H}}^{(\mathsf{{24}_{560}})j}_i
-\frac{1}{3!}{\partial_A}{\mathbf{H}}^{(\mathsf{{40}_{560}})l\dagger}_{ijk}{\partial^A}{\mathbf{H}}^{(\mathsf{{40}_{560}})ijk}_l
-\frac{1}{2!}{\partial_A}{\mathbf{H}}^{(\mathsf{{45}_{560}})i\dagger}_{jk}{\partial^A}{\mathbf{H}}^{(\mathsf{{45}_{560}})jk}_i
~~~~~~~~~~~\nonumber\\
&&-\frac{1}{2!}{\partial_A}{\mathbf{H}}^{(\mathsf{\overline{45}_{560}})ij\dagger}_{k}{\partial^A}{\mathbf{H}}^{(\mathsf{\overline{45}_{560}})k}_{ij}
-\frac{1}{2!}\frac{1}{3!}{\partial_A}{\mathbf{H}}^{(\mathsf{\overline{50}_{560}})ijk\dagger}_{lm}{\partial^A}{\mathbf{H}}^{(\mathsf{\overline{50}_{560}})lm}_{ijk}
-\frac{1}{2!}{\partial_A}{\mathbf{H}}^{(\mathsf{{70}_{560}})ij\dagger}_{k}{\partial^A}{\mathbf{H}}^{(\mathsf{{}_{560}})k}_{ij}
~~~~~~~~~~~\nonumber\\
&&-\frac{1}{2!}\frac{1}{2!}{\partial_A}{\mathbf{H}}^{(\mathsf{{75}_{560}})kl\dagger}_{ij}{\partial^A}{\mathbf{H}}^{(\mathsf{{75}_{560}})ij}_{kl}
-\frac{1}{3!}{\partial_A}{\mathbf{H}}^{(\mathsf{{175}_{560}})[ijk]l\dagger}_{m}{\partial^A}{\mathbf{H}}^{(\mathsf{{175}_{560}})m}_{[ijk]l}.
\end{eqnarray}
In a similar fashion, one can now normalize the $\mathsf{SU(5)}$ tensors in  the $\mathsf{\overline{560}}-$plet. Thus together for the $\mathsf{560}$ and $\overline{\mathsf{560}}$
multiplets, we have the following relationship among the un-normalized fields
$\mathsf{H}$ and normalized fields $\mathbf{H}$ as follows            
\begin{eqnarray}\label{Normalized SU(5) fields in 560}
{{\mathsf{H}}^{(\mathsf{560})}\choose {\mathsf{H}}^{(\overline{\mathsf{560}})}}=
\frac{2}{3}\sqrt{5}{{\mathbf{H}}^{(\mathsf{{1}_{560}})}\choose {\mathbf{H}}^{(\mathsf{\overline{1}_{\overline{560}}})}},~~~~&&
{{\mathsf{H}}^{(\mathsf{560})}_{i}\choose {\mathsf{H}}^{(\overline{\mathsf{560}})i}}=24\sqrt{\frac{3}{721}}{{\mathbf{H}}^{(\mathsf{\overline{5}_{560}})}_{i}\choose {\mathbf{H}}^{(\mathsf{{5}_{\overline{560}}})i}},\nonumber\\
{{\mathsf{H}}^{(\mathsf{560})}_{ij}\choose {\mathsf{H}}^{(\overline{\mathsf{560}})ij}}=\frac{1}{\sqrt{2}}{{\mathbf{H}}^{(\mathsf{\overline{10}_{560}})}_{ij}\choose {\mathbf{H}}^{(\mathsf{{10}_{\overline{560}}})ij}},~~~~&&
{{\mathsf{H}}^{(\mathsf{560})ij}\choose {\mathsf{H}}^{(\overline{\mathsf{560}})}_{ij}}=\frac{1}{\sqrt{3}}{{\mathbf{H}}^{(\mathsf{{10}_{560}})ij}\choose {\mathbf{H}}^{(\mathsf{\overline{10}_{\overline{560}}})}_{ij}},\nonumber\\
{\widehat{\mathsf{H}}^{(\mathsf{560})ij}\choose \widehat{\mathsf{H}}^{(\overline{\mathsf{560}})}_{ij}}=\frac{6}{\sqrt{865}}{\widehat{\mathbf{H}}^{(\mathsf{{10}_{560}})ij}\choose \widehat{\mathbf{H}}^{(\mathsf{\overline{10}_{\overline{560}}})}_{ij}},~~~~&&
{{\mathsf{H}}^{(\mathsf{560})i}_j\choose {\mathsf{H}}^{(\overline{\mathsf{560}})i}_{j}}={\frac{1}{2}\sqrt{\frac{3}{2}}}{{\mathbf{H}}^{(\mathsf{{24}_{560}})i}_j\choose {\mathbf{H}}^{(\mathsf{\overline{24}_{\overline{560}}})i}_{j}},\nonumber\\
{{\mathsf{H}}^{(\mathsf{560})ijk}_l\choose {\mathsf{H}}^{(\overline{\mathsf{560}})l}_{ijk}}=\sqrt{\frac{6}{433}}{{\mathbf{H}}^{(\mathsf{{40}_{560}})ijk}_l\choose {\mathbf{H}}^{(\mathsf{\overline{40}_{\overline{560}}})l}_{ijk}},~~~~&&
{{\mathsf{H}}^{(\mathsf{560})ij}_k\choose {\mathsf{H}}^{(\overline{\mathsf{560}})k}_{ij}}=\frac{1}{\sqrt{2}}{{\mathbf{H}}^{(\mathsf{{45}_{560}})ij}_k
\choose  {\mathbf{H}}^{(\mathsf{\overline{45}_{\overline{560}}})k}_{ij}},\nonumber\\
{{\mathsf{H}}^{(\mathsf{560})k}_{ij}\choose {\mathsf{H}}^{(\overline{\mathsf{560}})ij}_{k}}=\frac{12}{\sqrt{145}}{{\mathbf{H}}^{(\mathsf{\overline{45}_{560}})k}_{ij}\choose {\mathbf{H}}^{(\mathsf{{45}_{\overline{560}}})ij}_{k}},~~~~&&{{\mathsf{H}}^{(\mathsf{560})lm}_{ijk}\choose {\mathsf{H}}^{(\overline{\mathsf{560}})ijk}_{lm}}=\frac{1}{6\sqrt{2}}{{\mathbf{H}}^{(\mathsf{\overline{50}_{560}})lm}_{ijk}\choose {\mathbf{H}}^{(\mathsf{{50}_{\overline{560}}})ijk}_{lm}},\nonumber\\
{{\mathsf{H}}^{(\mathsf{560})ij}_{kl}\choose {\mathsf{H}}^{(\overline{\mathsf{560}})ij}_{kl}}=\frac{1}{\sqrt{2}}{{\mathbf{H}}^{(\mathsf{{75}_{560}})ij}_{kl}\choose {\mathbf{H}}^{(\mathsf{\overline{75}_{\overline{560}}})ij}_{kl}},~~~~&&{{\mathsf{H}}^{(\mathsf{560})m}_{[ijk]l}\choose
{{\mathsf{H}}^{(\overline{\mathsf{560}})[ijk]l}_{m}}}=\frac{1}{6\sqrt{2}}{{\mathbf{H}}^{(\mathsf{{175}_{560}})m}_{[ijk]l}\choose {\mathbf{H}}^{(\mathsf{\overline{175}_{\overline{560}}})[ijk]l}_{m}},\nonumber\\
{{\mathsf{H}}^{(\mathsf{560})k}_{(S)ij}\choose {\mathsf{H}}^{(\overline{\mathsf{560}})ij}_{(S)k}}={{\mathbf{H}}^{(\mathsf{{70}_{560}})k}_{ij}\choose {\mathbf{H}}^{(\mathsf{\overline{70}_{\overline{560}}})jk}_{k}}.~~~~&&
\end{eqnarray}
}

\section {Couplings of the $\boldsymbol{\mathsf{SU(5)}}$ components $\boldsymbol{\mathsf{1}}$, $\boldsymbol{\mathsf{24}}$, $\boldsymbol{\mathsf{75}}$ 
that enter in spontaneous breaking 
 of $\boldsymbol{\mathsf{SO(10)}}$ GUT symmetry\label{sec:B}}
GUT symmetry breaking is triggered by the superpotential (see Eq. (\ref{GUTsymmetry1})):
\begin{eqnarray}\label{GUTsymmetryB1}
W_{\textsc{gut}}&=&\frac{1}{2!}M_{45}\Phi^{(45)}_{\mu\nu}\Phi^{(45)}_{\mu\nu}+\lambda_{45}\braket{\Theta_{\mu\nu}^{(\mathsf{560})*}|B|
\overline{\Theta}_{\nu\sigma}^{(\overline{\mathsf{560}})}}\Phi^{(45)}_{\mu\sigma}+ M_{560}\braket{\Theta_{\mu\nu}^{(\mathsf{560})*}|B|
\overline{\Theta}_{\mu\nu}^{(\overline{\mathsf{560}})}}
\end{eqnarray}
Here few remarks about the second term appearing in Eq. (\ref{GUTsymmetryB1}) are in order. First note that there are  seemingly other viable interactions of $\mathsf{560}+\mathsf{\overline{560}}$ with the $\mathsf{45}$ multiplet that can appear in Eq. (\ref{GUTsymmetryB1}) such as
\begin{itemize}
\item $\lambda^{\prime}_{45}\braket{\Theta_{\mu\nu}^{(\mathsf{560})*}|B\Gamma_{\rho}\Gamma_{\sigma}|\overline{\Theta}_{\rho\sigma}^{(\overline{\mathsf{560}})}}
    \Phi^{(45)}_{\mu\nu}$,
\item $\lambda^{\prime\prime}_{45}\braket{\Theta_{\mu\nu}^{(\mathsf{560})*}|B\Gamma_{\nu}\Gamma_{\rho}|\overline{\Theta}_{\rho\sigma}^{(\overline{\mathsf{560}})}}
    \Phi^{(45)}_{\mu\sigma}$, etc.
    \item $\lambda^{\prime\prime\prime}_{45}~\epsilon_{\mu_1\mu_1\dots \mu_9\mu_{10}}\braket{\Theta_{\mu_1\mu_2}^{(\mathsf{560})*}|B\Gamma_{\mu_3}\Gamma_{\mu_4}\Gamma_{\mu_5}\Gamma_{\mu_6}
|\overline{\Theta}_{\mu_7\mu_8}^{(\overline{\mathsf{560}})}}\Phi^{(45)}_{\mu_9\mu_{10}}$
\end{itemize}
The first two couplings are  zero. This is due to the constraint $\Gamma_{\alpha}\ket{\overline{\Theta}_{\alpha\beta}^{(\overline{\mathsf{560}})}}=0$. 
The last term above is non-vanishing but we would not take it into account and in
this analysis we consider only the
 simplest coupling of  $\mathsf{560}+\mathsf{\overline{560}}$ with the $\mathsf{45}$ 
 \bl{as given} in Eq. (\ref{GUTsymmetryB1}).

Further, the second term appearing in Eq. (\ref{GUTsymmetryB1}) should actually read $\displaystyle\frac{\lambda_{45}}{2!}\left(\braket{\Theta_{\mu\nu}^{(\mathsf{560})*}|B|
\overline{\Theta}_{\nu\sigma}^{(\overline{\mathsf{560}})}}-\braket{\Theta_{\sigma\nu}^{(\mathsf{560})*}|B|
\overline{\Theta}_{\nu\mu}^{(\overline{\mathsf{560}})}}\right)\Phi^{(45)}_{\mu\sigma}$. This is due to the fact that only the piece antisymmetric in indices $(\mu,\sigma)$ of the product $\braket{\Theta_{\mu\nu}^{(\mathsf{560})*}|B|
\overline{\Theta}_{\nu\sigma}^{(\overline{\mathsf{560}})}}$ will couple to the antisymmetric tensor $\Phi^{(45)}_{\mu\sigma}$.

Eliminating the $\mathsf{45}$ multiplet through $\displaystyle\frac{\partial W_{\textsc{gut}} }{\partial\Phi^{45}_{\alpha\beta}}=0$, we get
${\Phi^{(45)}_{\alpha\beta}}=\displaystyle -\frac{\lambda_{45}}{2M_{45}}\left(\braket{\Theta_{\alpha\nu}^{(\mathsf{560})*}|B|
\overline{\Theta}_{\nu\beta}^{(\overline{\mathsf{560}})}}-\braket{\Theta_{\beta\nu}^{(\mathsf{560})*}|B|
\overline{\Theta}_{\nu\alpha}^{(\overline{\mathsf{560}})}}\right)$, which correctly exhibits antisymmetry in indices $(\alpha,\beta)$. Finally,
{\allowdisplaybreaks
\begin{eqnarray}\label{GUTsymmetryB2}
W_{\textsc{gut}}&=&-\frac{\lambda_{45}^2}{4M_{45}}\left[\braket{\Theta_{\alpha\nu}^{(\mathsf{560})*}|B|
\overline{\Theta}_{\nu\beta}^{(\overline{\mathsf{560}})}}\braket{\Theta_{\alpha\mu}^{(\mathsf{560})*}|B|
\overline{\Theta}_{\mu\beta}^{(\overline{\mathsf{560}})}}-\braket{\Theta_{\alpha\nu}^{(\mathsf{560})*}|B|
\overline{\Theta}_{\nu\beta}^{(\overline{\mathsf{560}})}}\braket{\Theta_{\beta\mu}^{(\mathsf{560})*}|B|
\overline{\Theta}_{\mu\alpha}^{(\overline{\mathsf{560}})}}\right]
+ M_{560}\braket{\Theta_{\mu\nu}^{(\mathsf{560})*}|B|
\overline{\Theta}_{\mu\nu}^{(\overline{\mathsf{560}})}}~~~~~~~~~~~~~\\
&=&\frac{\lambda_{45}^2}{4M_{45}}\left[-2\braket{\Theta_{c_i\bar c_k}^{(\mathsf{560})*}|B|
\overline{\Theta}_{c_kc_j}^{(\overline{\mathsf{560}})}}\braket{\Theta_{\bar c_i\bar c_l}^{(\mathsf{560})*}|B|
\overline{\Theta}_{c_l\bar c_j}^{(\overline{\mathsf{560}})}}-2\braket{\Theta_{c_i\bar c_k}^{(\mathsf{560})*}|B|
\overline{\Theta}_{c_k\bar c_j}^{(\overline{\mathsf{560}})}}\braket{\Theta_{c_l\bar c_i}^{(\mathsf{560})*}|B|
\overline{\Theta}_{c_j\bar c_l}^{(\overline{\mathsf{560}})}}\right.\nonumber\\
&&~~~~~~~~~-2\left.\braket{\Theta_{c_i\bar c_k}^{(\mathsf{560})*}|B|
\overline{\Theta}_{c_k\bar c_j}^{(\overline{\mathsf{560}})}}\braket{\Theta_{\bar c_i\bar c_l}^{(\mathsf{560})*}|B|
\overline{\Theta}_{c_lc_j}^{(\overline{\mathsf{560}})}}+2\braket{\Theta_{c_i\bar c_k}^{(\mathsf{560})*}|B|
\overline{\Theta}_{c_k c_j}^{(\overline{\mathsf{560}})}}\braket{\Theta_{\bar c_j\bar c_l}^{(\mathsf{560})*}|B|
\overline{\Theta}_{c_l\bar c_i}^{(\overline{\mathsf{560}})}}\right.\nonumber\\
&&~~~~~~~~~+2\left.\braket{\Theta_{c_k\bar c_i}^{(\mathsf{560})*}|B|
\overline{\Theta}_{c_j\bar c_k}^{(\overline{\mathsf{560}})}}\braket{\Theta_{\bar c_j\bar c_l}^{(\mathsf{560})*}|B|
\overline{\Theta}_{c_lc_i}^{(\overline{\mathsf{560}})}}+\braket{\Theta_{c_i\bar c_k}^{(\mathsf{560})*}|B|
\overline{\Theta}_{c_k \bar c_j}^{(\overline{\mathsf{560}})}}\braket{\Theta_{c_j\bar c_l}^{(\mathsf{560})*}|B|
\overline{\Theta}_{c_l\bar c_i}^{(\overline{\mathsf{560}})}}\right.\nonumber\\
&&~~~~~~~~~+\left.\braket{\Theta_{c_k\bar c_i}^{(\mathsf{560})*}|B|
\overline{\Theta}_{c_j\bar c_k}^{(\overline{\mathsf{560}})}}\braket{\Theta_{c_l\bar c_j}^{(\mathsf{560})*}|B|
\overline{\Theta}_{c_i\bar c_l}^{(\overline{\mathsf{560}})}}+\braket{\Theta_{\bar c_i\bar c_k}^{(\mathsf{560})*}|B|
\overline{\Theta}_{c_k c_j}^{(\overline{\mathsf{560}})}}\braket{\Theta_{\bar c_j\bar c_l}^{(\mathsf{560})*}|B|
\overline{\Theta}_{c_l c_i}^{(\overline{\mathsf{560}})}}+\dots\right]\nonumber\\
&&-M_{560}\left[\braket{\Theta_{\bar c_i\bar c_j}^{(\mathsf{560})*}|B|
\overline{\Theta}_{c_j c_i}^{(\overline{\mathsf{560}})}}+2\braket{\Theta_{c_i\bar c_j}^{(\mathsf{560})*}|B|
\overline{\Theta}_{c_j\bar c_i}^{(\overline{\mathsf{560}})}}+\dots\right]\label{GUTsymmetryB3}
\end{eqnarray}
Using Eqs. (\ref{tensor1} - \ref{tensor6}) and Eq. (\ref{Normalized SU(5) fields in 560}) in Eq. (\ref{GUTsymmetryB3}) we get
\begin{eqnarray}\label{GUTsymmetryB4}
W_{\textsc{gut}}&=&\frac{\lambda_{45}^2}{4M_{45}}\left[-\frac{1}{16}{\mathbf{H}}^{(\mathsf{{75}_{560}})xy}_{iz}
{\mathbf{H}}^{(\mathsf{\overline{75}_{\overline{560}}})zj}_{xy}{\mathbf{H}}^{(\mathsf{{75}_{560}})pq}_{jr}
{\mathbf{H}}^{(\mathsf{\overline{75}_{\overline{560}}})ri}_{pq}\right.\nonumber\\
&&~~~~~~~~~\left.
-{\frac{1}{8\sqrt{3}}}{\mathbf{H}}^{(\mathsf{{75}_{560}})xy}_{iz}{\mathbf{H}}^{(\mathsf{\overline{75}_{\overline{560}}})zj}_{xy}
{\mathbf{H}}^{(\mathsf{{75}_{560}})iq}_{jp}{\mathbf{H}}^{(\mathsf{\overline{24}_{\overline{560}}})p}_{q}-
{\frac{1}{8\sqrt{3}}}{\mathbf{H}}^{(\mathsf{{75}_{560}})xy}_{iz}{\mathbf{H}}^{(\mathsf{\overline{75}_{\overline{560}}})zj}_{xy}
{\mathbf{H}}^{(\mathsf{{24}_{560}})p}_{q}{\mathbf{H}}^{(\mathsf{\overline{75}_{\overline{560}}})qi}_{pj}\right.\nonumber\\
&&~~~~~~~~~\left.
-{\frac{5}{24}}{\mathbf{H}}^{(\mathsf{{75}_{560}})xy}_{iz}{\mathbf{H}}^{(\mathsf{\overline{75}_{\overline{560}}})zj}_{xy}{\mathbf{H}}^{(\mathsf{{24}_{560}})i}_{p}
{\mathbf{H}}^{(\mathsf{\overline{24}_{\overline{560}}})p}_{j}
+{\frac{11}{48}}{\mathbf{H}}^{(\mathsf{{75}_{560}})xy}_{iz}{\mathbf{H}}^{(\mathsf{\overline{75}_{\overline{560}}})zj}_{xy}
{\mathbf{H}}^{(\mathsf{{24}_{560}})p}_{j}{\mathbf{H}}^{(\mathsf{\overline{24}_{\overline{560}}})i}_{p}\right.\nonumber\\
&&~~~~~~~~~\left.
+{\frac{1}{48}}{\mathbf{H}}^{(\mathsf{{75}_{560}})xy}_{iz}{\mathbf{H}}^{(\mathsf{\overline{75}_{\overline{560}}})zi}_{xy}
{\mathbf{H}}^{(\mathsf{{24}_{560}})p}_{q}
{\mathbf{H}}^{(\mathsf{\overline{24}_{\overline{560}}})q}_{p}
-{\frac{1}{48}}{\mathbf{H}}^{(\mathsf{{75}_{560}})jx}_{iy}{\mathbf{H}}^{(\mathsf{\overline{24}_{\overline{560}}})y}_{x}
{\mathbf{H}}^{(\mathsf{{75}_{560}})ip}_{jq}
{\mathbf{H}}^{(\mathsf{\overline{24}_{\overline{560}}})q}_{p}\right.\nonumber\\
&&~~~~~~~~~\left.
-{\frac{1}{48}}{\mathbf{H}}^{(\mathsf{{24}_{560}})x}_{y}
{\mathbf{H}}^{(\mathsf{\overline{75}_{\overline{560}}})yj}_{xi}{\mathbf{H}}^{(\mathsf{{24}_{560}})p}_{q}
{\mathbf{H}}^{(\mathsf{\overline{75}_{\overline{560}}})qi}_{pj}
-{\frac{1}{24}}{\mathbf{H}}^{(\mathsf{{75}_{560}})jx}_{iy}
{\mathbf{H}}^{(\mathsf{\overline{24}_{\overline{560}}})y}_{x}{\mathbf{H}}^{(\mathsf{{24}_{560}})p}_{q}
{\mathbf{H}}^{(\mathsf{\overline{75}_{\overline{560}}})qi}_{pj}\right.\nonumber\\
&&~~~~~~~~~\left.
+{\frac{1}{8\sqrt{30}}}{\mathbf{H}}^{(\mathsf{{75}_{560}})xy}_{iz}{\mathbf{H}}^{(\mathsf{\overline{75}_{\overline{560}}})zj}_{xy}
{\mathbf{H}}^{(\mathsf{{24}_{560}})i}_{j}{\mathbf{H}}^{(\mathsf{\overline{1}_{\overline{560}}})}+
{\frac{1}{8\sqrt{30}}}{\mathbf{H}}^{(\mathsf{{75}_{560}})xy}_{iz}
{\mathbf{H}}^{(\mathsf{\overline{75}_{\overline{560}}})zj}_{xy}{\mathbf{H}}^{(\mathsf{{1}_{560}})}
{\mathbf{H}}^{(\mathsf{\overline{24}_{\overline{560}}})i}_{j}\right.\nonumber\\
&&~~~~~~~~~\left.
+\frac{1}{90}{\mathbf{H}}^{(\mathsf{{75}_{560}})xy}_{iz}{\mathbf{H}}^{(\mathsf{\overline{75}_{\overline{560}}})zi}_{xy}
{\mathbf{H}}^{(\mathsf{{1}_{560}})}{\mathbf{H}}^{(\mathsf{\overline{1}_{\overline{560}}})}\right.\nonumber\\
&&~~~~~~~~~\left.+{\frac{1}{48\sqrt{3}}}{\mathbf{H}}^{(\mathsf{{75}_{560}})jx}_{iy}{\mathbf{H}}^{(\mathsf{\overline{24}_{\overline{560}}})y}_{x}
{\mathbf{H}}^{(\mathsf{{24}_{560}})i}_{p}{\mathbf{H}}^{(\mathsf{\overline{24}_{\overline{560}}})p}_{j}+
{\frac{1}{48\sqrt{3}}}{\mathbf{H}}^{(\mathsf{{24}_{560}})x}_{y}
{\mathbf{H}}^{(\mathsf{\overline{75}_{\overline{560}}})yj}_{xi}{\mathbf{H}}^{(\mathsf{{24}_{560}})i}_{p}
{\mathbf{H}}^{(\mathsf{\overline{24}_{\overline{560}}})p}_{j}\right.\nonumber\\
&&~~~~~~~~~\left.
+{\frac{1}{24\sqrt{10}}}{\mathbf{H}}^{(\mathsf{{75}_{560}})jx}_{iy}{\mathbf{H}}^{(\mathsf{\overline{24}_{\overline{560}}})y}_{x}
{\mathbf{H}}^{(\mathsf{{24}_{560}})i}_{j}{\mathbf{H}}^{(\mathsf{\overline{1}_{\overline{560}}})}
+{\frac{1}{24\sqrt{10}}}{\mathbf{H}}^{(\mathsf{{75}_{560}})jx}_{iy}
{\mathbf{H}}^{(\mathsf{\overline{24}_{\overline{560}}})y}_{x}{\mathbf{H}}^{(\mathsf{{1}_{560}})}
{\mathbf{H}}^{(\mathsf{\overline{24}_{\overline{560}}})i}_{j}\right.\nonumber\\
&&~~~~~~~~~\left.
+{\frac{1}{24\sqrt{10}}}{\mathbf{H}}^{(\mathsf{{24}_{560}})x}_{y}
{\mathbf{H}}^{(\mathsf{\overline{75}_{\overline{560}}})yj}_{xi}{\mathbf{H}}^{(\mathsf{{24}_{560}})i}_{j}
{\mathbf{H}}^{(\mathsf{\overline{1}_{\overline{560}}})}
+{\frac{1}{24\sqrt{10}}}{\mathbf{H}}^{(\mathsf{{24}_{560}})x}_{y}{\mathbf{H}}^{(\mathsf{\overline{75}_{\overline{560}}})yj}_{xi}
{\mathbf{H}}^{(\mathsf{{1}_{560}})}{\mathbf{H}}^{(\mathsf{\overline{24}_{\overline{560}}})i}_{j}\right.\nonumber\\
&&~~~~~~~~~\left.
-{\frac{221}{576}}{\mathbf{H}}^{(\mathsf{{24}_{560}})j}_{x}{\mathbf{H}}^{(\mathsf{\overline{24}_{\overline{560}}})x}_{i}
{\mathbf{H}}^{(\mathsf{{24}_{560}})i}_{y}{\mathbf{H}}^{(\mathsf{\overline{24}_{\overline{560}}})y}_{j}
+{\frac{55}{144}}{\mathbf{H}}^{(\mathsf{{24}_{560}})j}_{x}{\mathbf{H}}^{(\mathsf{\overline{24}_{\overline{560}}})x}_{i}
{\mathbf{H}}^{(\mathsf{{24}_{560}})y}_{j}{\mathbf{H}}^{(\mathsf{\overline{24}_{\overline{560}}})i}_{y}\right.\nonumber\\
&&~~~~~~~~~\left.
-{\frac{7}{576}}{\mathbf{H}}^{(\mathsf{{24}_{560}})x}_{y}{\mathbf{H}}^{(\mathsf{\overline{24}_{\overline{560}}})y}_{x}
{\mathbf{H}}^{(\mathsf{{24}_{560}})i}_{j}{\mathbf{H}}^{(\mathsf{\overline{24}_{\overline{560}}})j}_{i}\right.\nonumber\\
&&~~~~~~~~~\left.
-{\frac{1}{48\sqrt{30}}}{\mathbf{H}}^{(\mathsf{{24}_{560}})i}_{x}{\mathbf{H}}^{(\mathsf{\overline{24}_{\overline{560}}})x}_{j}
{\mathbf{H}}^{(\mathsf{{24}_{560}})j}_{i}{\mathbf{H}}^{(\mathsf{\overline{1}_{\overline{560}}})}
-{\frac{1}{48\sqrt{30}}}{\mathbf{H}}^{(\mathsf{{24}_{560}})i}_{x}
{\mathbf{H}}^{(\mathsf{\overline{24}_{\overline{560}}})x}_{j}{\mathbf{H}}^{(\mathsf{{1}_{560}})}
{\mathbf{H}}^{(\mathsf{\overline{24}_{\overline{560}}})j}_{i}\right.\nonumber\\
&&~~~~~~~~~\left.
-{\frac{1}{480}}{\mathbf{H}}^{(\mathsf{{24}_{560}})j}_{i}{\mathbf{H}}^{(\mathsf{\overline{1}_{\overline{560}}})}
{\mathbf{H}}^{(\mathsf{{24}_{560}})i}_{j}{\mathbf{H}}^{(\mathsf{\overline{1}_{\overline{560}}})}-{\frac{1}{480}}
{\mathbf{H}}^{(\mathsf{{1}_{560}})}{\mathbf{H}}^{(\mathsf{\overline{24}_{\overline{560}}})j}_{i}
{\mathbf{H}}^{(\mathsf{{1}_{560}})}{\mathbf{H}}^{(\mathsf{\overline{24}_{\overline{560}}})i}_{j}\right.\nonumber\\
&&~~~~~~~~~\left.
+{\frac{1}{144}}{\mathbf{H}}^{(\mathsf{{24}_{560}})j}_{i}
{\mathbf{H}}^{(\mathsf{\overline{1}_{\overline{560}}})}{\mathbf{H}}^{(\mathsf{{1}_{560}})}
{\mathbf{H}}^{(\mathsf{\overline{24}_{\overline{560}}})i}_{j}\right.\nonumber\\
&&~~~~~~~~~\left.
-\frac{1}{405}{\mathbf{H}}^{(\mathsf{{1}_{560}})}{\mathbf{H}}^{(\mathsf{\overline{1}_{\overline{560}}})}
{\mathbf{H}}^{(\mathsf{{1}_{560}})}{\mathbf{H}}^{(\mathsf{\overline{1}_{\overline{560}}})}+\dots\right]\nonumber\\
&&+~iM_{560}\left[\frac{1}{4}{\mathbf{H}}^{(\mathsf{{75}_{560}})ij}_{kl}{\mathbf{H}}^{(\mathsf{\overline{75}_{\overline{560}}})kl}_{ij}
+{\mathbf{H}}^{(\mathsf{{24}_{560}})i}_{j}{\mathbf{H}}^{(\mathsf{\overline{24}_{\overline{560}}})j}_{i}+
{\mathbf{H}}^{(\mathsf{{1}_{560}})}{\mathbf{H}}^{(\mathsf{\overline{1}_{\overline{560}}})}+\dots\right]
\end{eqnarray}
}
The ellipsis above denote terms containing  $\mathsf{SU(5)}$ fields other than $\mathsf{75}(-5)$, $\mathsf{24}(-5)$ and $\mathsf{1}(-5)$. Those unrepresented fields do not enter into GUT Symmetry breaking.

We identify the $\mathsf{SU(3)_C\times SU(2)_L\times U(1)_Y}$ singlets through
\begin{eqnarray}\label{GUTsymmetryB5}
{{\mathbf{H}}^{(\mathsf{{1}_{560}})}\choose {\mathbf{H}}^{(\mathsf{\overline{1}_{\overline{560}}})}}\equiv{{{\mathcal S}}_{1}\choose {\overline{\mathcal S}}_{1}},~~~~&&
{{\mathbf{H}}^{(\mathsf{{24}_{560}})\alpha}_{\beta}\choose {\mathbf{H}}^{(\mathsf{\overline{24}_{\overline{560}}})\alpha}_{\beta}}=\frac{1}{3}\delta^{\alpha}_{\beta}{{{\mathcal S}}_{24}\choose {  \overline{\mathcal S}}_{24}}+\dots,\nonumber\\
{{\mathbf{H}}^{(\mathsf{{24}_{560}})a}_{b}\choose {\mathbf{H}}^{(\mathsf{\overline{24}_{\overline{560}}})a}_{b}}=-\frac{1}{2}\delta^{a}_{b}{{{\mathcal S}}_{24}\choose {  \overline{\mathcal S}}_{24}}+\dots,~~~~&&
{{\mathbf{H}}^{(\mathsf{{75}_{560}})\alpha\beta}_{\gamma\sigma}\choose{\mathbf{H}}^{(\mathsf{\overline{75}_{\overline{560}}})\alpha\beta}
_{\gamma\sigma}}=\frac{1}{6}\left(\delta^{\alpha}_{\gamma}\delta^{\beta}_{\sigma}-
\delta^{\alpha}_{\sigma}\delta^{\beta}_{\gamma}\right){{{\mathcal S}}_{75}\choose { \overline{\mathcal S}}_{75}}+\dots\nonumber\\
{{\mathbf{H}}^{(\mathsf{{75}_{560}})ab}_{cd}\choose {\mathbf{H}}^{(\mathsf{\overline{75}_{\overline{560}}})ab}_{cd}}=\frac{1}{2}\left(\delta^{a}_{c}\delta^{b}_{d}-
\delta^{a}_{d}\delta^{b}_{c}\right){{{\mathcal S}}_{75}\choose { \overline{\mathcal S}}_{75}},~~~~&&
{{\mathbf{H}}^{(\mathsf{{75}_{560}})\alpha a}_{\beta b}\choose {\mathbf{H}}^{(\mathsf{\overline{75}_{\overline{560}}})\alpha a}_{\beta b}}=-\frac{1}{6}\delta^{a}_{b}\delta^{\alpha}_{\beta}{{{\mathcal S}}_{75}\choose {\overline{\mathcal S}}_{75}}+\dots,
\end{eqnarray}
where $\alpha,~\beta,~\gamma,~\sigma= 1,~2,~3$ are $\mathsf{SU(3)}$ color indices and $a,~b,~c,~d=4,~5$ are $\mathsf{SU(2)}$ weak indices. The ellipsis above represent terms containing fields which are not $\mathsf{SU(3)_C\times\ SU(2)_L\times U(1)_Y}$ singlets $(\mathsf{1,1})(0)$.
The above singlets are normalized according to \cite{Nath:2015kaa}
\begin{eqnarray}\label{GUTsymmetryB6}
{{{\mathcal S}}_{1}\choose {  \overline{\mathcal S}}_{1}}={{{\mathbf S}}_{1}\choose {  \overline{\mathbf S}}_{1}},~~~~~{{{\mathcal S}}_{24}\choose {  \overline{\mathcal S}}_{24}}=\sqrt{\frac{6}{5}}
{{{\mathbf S}}_{24}\choose {  \overline{\mathbf S}}_{24}},~~~~~{{{\mathcal S}}_{75}\choose {  \overline{\mathcal S}}_{75}}=\frac{1}{\sqrt{2}}
{{{\mathbf S}}_{75}\choose {  \overline{\mathbf S}}_{75}}.
\end{eqnarray}
Finally, we denote the the VEVs of the normalized fields as follows
\begin{eqnarray}\label{GUTsymmetryB7}
{{{\mathcal V}}_{1}\choose {\overline{\mathcal V}_{1}}}\equiv{\braket{{{\mathbf S}}_{1}}\choose \braket{{\overline{\mathbf S}}_{1}}},~~~~~{{{\mathcal V}}_{24}\choose {\overline{\mathcal V}_{24}}}\equiv
{\braket{{{\mathbf S}}_{24}}\choose \braket{{\overline{\mathbf S}}_{24}}},~~~~~{{{\mathcal V}}_{75}\choose {\overline{\mathcal V}_{75}}}\equiv
{\braket{{{{\mathbf S}}_{75}}}\choose \braket{{\overline{\mathbf S}}_{75}}}.
\end{eqnarray}
Inserting Eqs. (\ref{GUTsymmetryB5} - \ref{GUTsymmetryB7}) into Eq. (\ref{GUTsymmetryB4}) gives Eq. (\ref{GUTsymmetryvevs}).\\
{

{
 \section{Extraction and normalization of $\boldsymbol{\mathsf{SU(2)_L}}$ doublet fields in the model\label{sec:C}}
 \subsection{$\boldsymbol{\mathsf{5}+\mathsf{\overline{5}}}$ and $\boldsymbol{\mathsf{45}+\mathsf{\overline{45}}}$ of  $\boldsymbol{\mathsf{SU(5)}}$}
 The $\mathsf{5}+\mathsf{\overline{5}}$ plets of $\mathsf{SU(5)}$ are normalized according to (see \cite{Nath:2001uw})
\begin{eqnarray}\label{Normalized 5+bar5 fields in 10plet}
\mathsf{H}^{(\mathsf{10}_r)i}=\sqrt{2}\mathbf{H}^{(\mathsf{5_{10_r}})i},&~~~~~ \mathsf{H}^{(\mathsf{10}_r)}_i=\sqrt{2}\mathbf{H}^{(\mathsf{\overline{5}_{10_r}})}_i
\end{eqnarray}
The identification of  $\mathsf{SU(2)}$  doublets, denoted by $\mathsf D$'s, contained in $\mathsf{5}+\mathsf{\overline{5}}$ and $\mathsf{45}+\mathsf{\overline{45}}$ , are done through (see \cite{Nath:2015kaa})
\begin{eqnarray}
\mathbf{H}^{({\mathsf{5}}_{\#})a}\equiv {}^{({\mathsf{5}}_{\#})}\!{{\mathsf D}}^{a};&& 
\mathbf{H}^{(\overline{\mathsf{5}}_{\#})}_a\equiv {}^{(\overline{\mathsf{5}}_{\#})}\!{{\mathsf D}}_{a},\label{Extraction of 5+bar5 doublets}\\
\nonumber\\
\mathbf{H}^{({\mathsf{45}}_{\#})\alpha a}_\beta=\frac{1}{3}\delta^{\alpha}_{\beta}~{}^{({\mathsf{45}}_{\#})}\!{{\mathsf D}}^{a}+\dots;
&&\mathbf{H}^{({\mathsf{45}}_{\#})ab }_c=\delta^{b}_{c}~{}^{({\mathsf{45}}_{\#})}\!{{\mathsf D}}^{a}-\delta^{a}_{c}~{}^{({\mathsf{45}}_{\#})}\!{{\mathsf D}}^{b},\nonumber\\
\mathbf{H}^{({\overline{\mathsf{45}}}_{\#})\beta}_{\alpha a}=\frac{1}{3}\delta^{\beta}_{\alpha}~{}^{({\overline{\mathsf{45}}}_{\#})}\!{{\mathsf D}}_a\dots;&&\mathbf{H}_{ab }^{({\overline{\mathsf{45}}}_{\#})c}=\delta^{c}_{b}~{}^{({\overline{\mathsf{45}}}_{\#})}\!{{\mathsf D}}_{a}-\delta^{c}_{a}~{}^{({\overline{\mathsf{45}}}_{\#})}\!{{\mathsf D}}_{b},\label{Extraction of 45+bar45 doublets}
\end{eqnarray}
where $a,~b,~c=4,~5$ are reserved for $\mathsf{SU(2)}$ weak indices and $\alpha,~\beta,~\gamma,=1,~2, ~3$ are $\mathsf{SU(3)}$ color indices. 
The ellipsis above represent terms containing fields which are not $\mathsf{SU(3)_C\times\ SU(2)_L\times U(1)_Y}$ doublets $(\mathsf{1,2})(3)$.
The symbol $\#$ refers to the $\mathsf{10}_r, ~\mathsf{320},~ \mathsf{560+\overline{560}}$ fields of $\mathsf{SO(10)}$.  Note, however, that $\mathsf D$'s are un-normalized. To normalize the $\mathsf{SU(2)}$ doublets contained in the $\mathsf{SO(10)}$ tensors, we carry out the following field redefinition (\cite{Nath:2015kaa}):
\begin{eqnarray}\label{Normalized doublets in 5+bar5 and 45+bar45}
{}^{({5}_{{\#}})}\!{\mathsf D}^{a}= {}^{({5}_{{\#}})}\!{\mathbf D}^{a};&&{}^{(\overline{5}_{{\#}})}\!{\mathsf D}_{a}={}^{(\overline{5}_{{\#}})}\!{\mathbf D}_{a},\nonumber\\
{}^{({\mathsf{45}}_{{\#}})}\!{\mathsf D}^{a}=\frac{1}{2}\sqrt{\frac{3}{2}}~{}^{({\mathsf{45}}_{{\#}})}\!{\mathbf D}^{a};&&{}^{(\overline{\mathsf{45}}_{\#})}\!{\mathsf D}_{a}=\frac{1}{2}\sqrt{\frac{3}{2}}~{}^{(\overline{\mathsf{45}}_{\#})}\!{\mathbf D}_{a},
\end{eqnarray}
where $\mathbf D$'s represent the doublet fields with canonically normalized kinetic energy terms.
\subsection{$\boldsymbol{\mathsf{70}+\mathsf{\overline{70}}}$ of  $\boldsymbol{\mathsf{SU(5)}}$}
The $\mathsf{70}$-plet of $\mathsf{SU(5)}$ has the following  $\mathsf{SU(3)_C\times SU(2)_L \times U(1)_Y}$ decomposition \cite{Slansky:1981yr}
 \begin{eqnarray}\label{70 plet under SM}
\mathsf{70} \left[\mathbf{H}^{(\mathsf{70}_{\#})ij}_{k}\right]&=&(\mathsf{1,2})(3)\left[{}^{(\mathsf{70}_{\#})}\!{\mathsf D}^{a}\right]+(\mathsf{3,1})(-2)\left[{}^{(\mathsf{70}_{\#})}\!{\mathsf T}^{\alpha}\right]+
(\mathsf{1,4})(3)\left[{}^{(\mathsf{70}_{\#})}\!\psi^{\{ab\}}_c\right]+(\mathsf{3,3})(-2)\left[{}^{(\mathsf{70}_{\#})}\!\psi^{a\alpha}_b\right]\nonumber\\
&&+(\mathsf{\bar{3},3})(8)\left[{}^{(\mathsf{70}_{\#})}\!\psi^{\{ab\}}_{\alpha}\right]+(\mathsf{8,2})(3)\left[{}^{(\mathsf{70}_{\#})}\!\psi^{\alpha a}_{\beta}\right]+
(\mathsf{6,2})(-7)\left[{}^{(\mathsf{70}_{\#})}\!\psi^{\{\alpha\beta\}}_a\right]\nonumber\\
&&+(\mathsf{15,1})(-2)\left[{}^{(\mathsf{70}_{\#})}\!\psi^{\{\alpha\beta\}}_{\gamma}\right],
 \end{eqnarray}
where we have defined
  \begin{eqnarray}\label{tracelessness condition on 70plet}
       \begin{split}
 \mathbf{H}^{({\mathsf{70}}_{\#})\alpha a}_{\alpha}= -\mathbf{H}^{({\mathsf{70}}_{\#})b a}_{b}
\equiv{}^{({\mathsf{70}}_{\#})}\!{\mathsf D}^{a};&&~~~~\mathbf{H}^{({\mathsf{70}}_{\#})\beta \alpha}_{\beta}= -\mathbf{H}^{({\mathsf{70}}_{\#})a \alpha}_{a}
\equiv{}^{({\mathsf{70}}_{\#})}\!{\mathsf T}^{\alpha};~~~~\\
\mathbf{H}^{({\mathsf{70}}_{\#})ab}_{\alpha}\equiv {}^{(\mathsf{70}_{\#})}\!\psi^{\{ab\}}_{\alpha};&&~~~~\mathbf{H}^{({\mathsf{70}}_{\#})\alpha\beta}_{a}\equiv {}^{(\mathsf{70}_{\#})}\!\psi^{\{\alpha\beta\}}_{a}.
\end{split}
\end{eqnarray}
Here $\#$ refers to $\mathsf{320}$ and $\mathsf{\overline{560}}$.
The first two relationships in Eq. (\ref{tracelessness condition on 70plet}) follow from the tracelessness condition on the $\mathsf{SU(5)}$ irreducible tensor $\mathbf{H}^{({\mathsf{70}}_{\#})ij}_{k}$.
The reducible tensors of the $\mathsf{70}$-plet can be expressed in terms of the irreducible ones as follows:
\begin{eqnarray}\label{Extraction of 70+bar70 doublets}
\mathbf{H}^{({\mathsf{70}}_{\#})a\alpha}_{b}={}^{(\mathsf{70}_{\#})}\!\psi^{a\alpha}_{b}-\frac{1}{2}\delta^a_b~{}^{({\mathsf{70}}_{\#})}\!{\mathsf T}^{\alpha};&&\mathbf{H}^{({\mathsf{70}}_{\#})\alpha a}_{\beta}={}^{(\mathsf{70}_{\#})}\!\psi^{\alpha a}_{\beta}+\frac{1}{3}\delta^{\alpha}_{\beta}~{}^{({\mathsf{70}}_{\#})}\!{\mathsf D}^{a};\nonumber\\
\mathbf{H}^{({\mathsf{70}}_{\#})ab}_{c}={}^{(\mathsf{70}_{\#})}\!\psi^{\{ab\}}_{c}-\frac{1}{3}\left(\delta^b_c~{}^{({\mathsf{70}}_{\#})}\!{\mathsf D}^{a}+\delta^a_c{}~^{({\mathsf{70}}_{\#})}\!{\mathsf D}^{b}\right);&&\mathbf{H}^{({\mathsf{70}}_{\#})\alpha\beta}_{\gamma}={}^{(\mathsf{70}_{\#})}\!\psi^{\{\alpha\beta\}}_{\gamma}+\frac{1}{4}\left(\delta^{\alpha}_{\gamma}~
{}^{({\mathsf{70}}_{\#})}\!{\mathsf T}^{\beta}+\delta^{\beta}_{\gamma}~{}^{({\mathsf{70}}_{\#})}\!{\mathsf T}^{\alpha}\right).\nonumber\\
\end{eqnarray}
The kinetic energy of the $\mathsf{70}$-plet is given by
\begin{eqnarray}\label{C4}
\mathcal{L}_{\textsc{ke}}^{(\mathsf{\mathsf{70}}_{\#})}&=&-\partial_A\mathbf{H}^{({\mathsf{70}}_{\#})ij}_{k}
\partial^A\mathbf{H}^{({\mathsf{70}}_{\#})ij\dagger}_{k}\nonumber\\
&=&-\left[\partial_A{}^{(\mathsf{70}_{\#})}\!\psi^{\{\alpha\beta\}}_{\gamma}~\partial^A{}^{(\mathsf{70}_{\#})}\!\psi^{\{\alpha\beta\}\dagger}_{\gamma}
+\partial_A{}^{(\mathsf{70}_{\#})}\!\psi^{\{ab\}}_{c}~\partial^A{}^{(\mathsf{70}_{\#})}\!\psi^{\{ab\}\dagger}_{c}+
\partial_A{}^{(\mathsf{70}_{\#})}\!\psi^{\{ab\}}_{\alpha}~\partial^A{}^{(\mathsf{70}_{\#})}\!\psi^{\{ab\}\dagger}_{\alpha}\right.\nonumber\\
&&\left.~~~~+\partial_A{}^{(\mathsf{70}_{\#})}\!\psi^{\{\alpha\beta\}}_{a}~\partial^A{}^{(\mathsf{70}_{\#})}\!\psi^{\{\alpha\beta\}\dagger}_{a}
+2~\partial_A{}^{(\mathsf{70}_{\#})}\!\psi^{\{\alpha a\}}_{\beta}~\partial^A{}^{(\mathsf{70}_{\#})}\!\psi^{\{\alpha a\}\dagger}_{\beta}+2~\partial_A{}^{(\mathsf{70}_{\#})}\!\psi^{\{a\alpha\}}_{b}~\partial^A{}^{(\mathsf{70}_{\#})}\!\psi^{\{a\alpha\}\dagger}_{b}\right.\nonumber\\
&&\left.~~~~+\frac{4}{3}~
\partial^A{}^{(\mathsf{70}_{\#})}\!{\mathsf D}^{a}~\partial_A{}^{(\mathsf{70}_{\#})}\!{\mathsf D}^{a\dagger}+\frac{3}{2}~
\partial^A{}^{(\mathsf{70}_{\#})}\!{\mathsf T}^{\alpha}~\partial_A{}^{(\mathsf{70}_{\#})}\!{\mathsf T}^{\alpha\dagger}\right],
\end{eqnarray}
In terms of  normalized $\mathsf{SU(3)_C}\times\mathsf{SU(2)_L}\times\mathsf{U(1)_Y}$ fields, the kinetic energy of $\mathsf{70}$ tensor takes the form
\begin{eqnarray}
\mathcal{L}_{\textsc{ke}}^{(\mathsf{70}_{\#})}
&=&-\left[\frac{1}{2!}\partial_A {}^{(\mathsf{70}_{\#})}\!{\boldsymbol\Psi}^{\{\alpha\beta\}}_{\gamma}~\partial^A{}^{(\mathsf{70}_{\#})}\!{\boldsymbol\Psi}^{\{\alpha\beta\}\dagger}_{\gamma}
+\frac{1}{2!}\partial_A{}^{(\mathsf{70}_{\#})}\!{\boldsymbol\Psi}^{\{ab\}}_{c}~\partial^A{}^{(\mathsf{70}_{\#})}\!{\boldsymbol\Psi}^{\{ab\}\dagger}_{c}+
\frac{1}{2!}\partial_A{}^{(\mathsf{70}_{\#})}\!{\boldsymbol\Psi}^{\{ab\}}_{\alpha}~\partial^A{}^{(\mathsf{70}_{\#})}\!{\boldsymbol\Psi}^{\{ab\}\dagger}_{\alpha}\right.\nonumber\\
&&\left.~~~~+\frac{1}{2!}\partial_A{}^{(\mathsf{70}_{\#})}\!{\boldsymbol\Psi}^{\{\alpha\beta\}}_{a}~
\partial^A{}^{(\mathsf{70}_{\#})}\!{\boldsymbol\Psi}^{\{\alpha\beta\}\dagger}_{a}+\partial_A{}^{(\mathsf{70}_{\#})}\!{\boldsymbol\Psi}^{\{\alpha a\}}_{\beta}~\partial^A{}^{(\mathsf{70}_{\#})}\!{\boldsymbol\Psi}^{\{\alpha a\}\dagger}_{\beta}+\partial_A{}^{(\mathsf{70}_{\#})}\!{\boldsymbol\Psi}^{\{a\alpha\}}_{b}~\partial^A{}^{(\mathsf{70}_{\#})}\!{\boldsymbol\Psi}^{\{a\alpha\}\dagger}_{b}
\right.\nonumber\\
&&\left.~~~~
+\partial^A{}^{(\mathsf{70}_{\#})}\!{\mathbf D}^{a}~\partial_A{}^{(\mathsf{70}_{\#})}\!{\mathbf D}^{a\dagger}+
\partial^A{}^{(\mathsf{70}_{\#})}\!{\mathbf T}^{\alpha}~\partial_A{}^{(\mathsf{70}_{\#})}\!{\mathbf T}^{\alpha\dagger}\right],
\end{eqnarray}
 One can now extend the above results to $\mathsf{\overline{70}}$ of $\mathsf{SU(5)}$ contained in ${\mathsf{320}}-$plet and $\mathsf{560}-$plet of $\mathsf{SO(10)}$. Therefore,
\begin{eqnarray}\label{Normalized doublets in 70+bar70}
{{}^{(\mathsf{70}_{\#})}\!\psi^{\{\alpha\beta\}}_{\gamma}\choose {}^{(\mathsf{\overline{70}}_{\#})}\!\psi_{\{\alpha\beta\}}^{\gamma}}=
\frac{1}{\sqrt{2}}{{}^{(\mathsf{70}_{\#})}\!{\boldsymbol\Psi}^{\{\alpha\beta\}}_{\gamma}\choose {}^{(\mathsf{\overline{70}}_{\#})}\!{\boldsymbol\Psi}_{\{\alpha\beta\}}^{\gamma}},~~~~&&
{{}^{(\mathsf{70}_{\#})}\!\psi^{\{ab\}}_{c}\choose {}^{(\mathsf{\overline{70}}_{\#})}\!\psi_{\{ab\}}^{c}}=
\frac{1}{\sqrt{2}}{{}^{(\mathsf{70}_{\#})}\!{\boldsymbol\Psi}^{\{ab\}}_{c}\choose {}^{(\mathsf{\overline{70}}_{\#})}\!{\boldsymbol\Psi}_{\{ab\}}^{c}},\nonumber\\
{{}^{(\mathsf{70}_{\#})}\!\psi^{\{ab\}}_{\alpha}\choose {}^{(\mathsf{\overline{70}}_{\#})}\!\psi_{\{ab\}}^{\alpha}}=
\frac{1}{\sqrt{2}}{{}^{(\mathsf{70}_{\#})}\!{\boldsymbol\Psi}^{\{ab\}}_{\alpha}\choose {}^{(\mathsf{\overline{70}}_{\#})}\!{\boldsymbol\Psi}_{\{ab\}}^{\alpha}},~~~~&&
{{}^{(\mathsf{70}_{\#})}\!\psi^{\{\alpha\beta\}}_{a}\choose {}^{(\mathsf{\overline{70}}_{\#})}\!\psi_{\{\alpha\beta\}}^{a}}=
\frac{1}{\sqrt{2}}{{}^{(\mathsf{70}_{\#})}\!{\boldsymbol\Psi}^{\{\alpha\beta\}}_{a}\choose {}^{(\mathsf{\overline{70}}_{\#})}\!{\boldsymbol\Psi}_{\{\alpha\beta\}}^{a}},\nonumber\\
{{}^{(\mathsf{70}_{\#})}\!\psi^{\alpha a}_{\beta}\choose {}^{(\mathsf{\overline{70}}_{\#})}\!\psi_{\alpha a}^{\beta}}=
\frac{1}{\sqrt{2}}{{}^{(\mathsf{70}_{\#})}\!{\boldsymbol\Psi}^{\alpha a}_{\beta}\choose {}^{(\mathsf{\overline{70}}_{\#})}\!{\boldsymbol\Psi}_{\alpha a}^{\beta}},~~~~&&
{{}^{(\mathsf{70}_{\#})}\!\psi^{a\alpha}_{b}\choose {}^{(\mathsf{\overline{70}}_{\#})}\!\psi_{a\alpha}^{b}}=
\frac{1}{\sqrt{2}}{{}^{(\mathsf{70}_{\#})}\!{\boldsymbol\Psi}^{a\alpha}_{b}\choose {}^{(\mathsf{\overline{70}}_{\#})}\!{\boldsymbol\Psi}_{a\alpha}^{b}},\nonumber\\
{{}^{(\mathsf{70}_{\#})}\!{\mathsf D}^{a}\choose {}^{(\mathsf{\overline{70}}_{\#})}\!{\mathsf D}_{a}}=
\frac{\sqrt{3}}{2}{{}^{(\mathsf{70}_{\#})}\!{\mathbf D}^{a}\choose {}^{(\mathsf{\overline{70}}_{\#})}\!{\mathbf D}_{a}},~~~~&&
{{}^{(\mathsf{70}_{\#})}\!{\mathsf T}^{\alpha}\choose {}^{(\mathsf{\overline{70}}_{\#})}\!{\mathsf T}_{\alpha}}=
\sqrt{\frac{2}{3}}{{}^{(\mathsf{70}_{\#})}\!{\mathbf T}^{\alpha}\choose {}^{(\mathsf{\overline{70}}_{\#})}\!{\mathbf T}_{\alpha}}
\end{eqnarray}
}}
{
\section {The $\boldsymbol{\mathsf{320}}$ dimensional tensors of 
$\boldsymbol{\mathsf{SO(10)}}$\label{sec:D}}
There are two $\mathsf{320}$ multiplets in $\mathsf{SO(10)}$ and both are three index ones
and to construct these we start with the 1000 dimensional reducible multiplet 
$T_{\mu\nu\lambda}$. This can be decomposed as follows 
\[1000 ~\big[T_{\mu\nu\lambda}\big]= \mathsf{120}~ \big[A_{[\mu\nu\lambda]}\big] + \mathsf{210^{\prime}}~\big[S_{(\mu\nu\lambda)}\big]+ \mathsf{320}~\big[\Lambda^{(\mathsf{320}_a)}_{[\mu\nu]\lambda}\big] + \mathsf{320}~\big[\Lambda^{(\mathsf{320}_s)}_{(\mu\nu)\lambda}\big] + \mathsf{10}+
\mathsf{10} + \mathsf{10}\]
Here $\Lambda^{(\mathsf{320}_a)}_{[\mu\nu]\lambda}$
 is the $\mathsf{320}$ multiplet which is anti-symmetric in the first two indices while 
$\Lambda^{(\mathsf{320}_s)}_{(\mu\nu)\lambda}$
is the tensor multiplet which is symmetric in the first two indices. We now show
how the various irreducible components of $T_{\mu\nu\lambda}$ arise. 
{


{\subsection{Construction of $\boldsymbol{\Lambda^{(\mathsf{320}_a)}_{[\mu\nu]\lambda}}$ tensor}}}
The three index tensor $A_{\mu\nu\lambda}$ is totally anti-symmetric 
in the three indices and can be constructed from $T_{\mu\nu\lambda}$ in a
direct way so that 
\begin{align}
 A_{[\mu\nu\lambda]}&\equiv \frac{1}{3!} \left(T_{\mu\nu\lambda} +
 T_{\nu\lambda\mu} + T_{\lambda\mu\nu}
- T_{\nu\mu\lambda} - T_{\lambda\nu\mu} -T_{\mu\lambda\nu}\right)\nonumber
\end{align}
It has $^{10}C_3=120$ number of independent components. Next let us consider the tensor $\Lambda''_{[\mu\nu]\lambda}
=\frac{1}{2}[T_{\mu\nu\lambda}-T_{\nu\mu\lambda}]$ which has $450$ components
and from it construct a $330$ multiplet $\Lambda'_{\mu\nu\lambda}$ using $450-120$ 
which gives  
 \begin{align}
 {\Lambda'}_{[\mu\nu]\lambda} 
 &=\frac{1}{3} \left( T_{\mu\nu\lambda}- T_{\nu\mu\lambda}\right) +\frac{1}{6}\left(-T_{\nu\lambda\mu}
+ T_{\lambda\nu\mu} - T_{\lambda\mu\nu} +T_{\mu\lambda\nu}\right)\nonumber
 \end{align}
Finally, we construct the irreducible $(\mathsf{330}-\mathsf{10}=)~ \mathsf{320}-$dimensional tensor, $\Lambda^{(\mathsf{320}_a)}_{[\mu\nu]\lambda}$, from  ${\Lambda'}_{[\mu\nu]\lambda}$ by subtracting of its trace. Thus we write: $\Lambda^{(\mathsf{320}_a)}_{[\mu\nu]\lambda}={\Lambda'}_{[\mu\nu]\lambda}+\kappa\left(\delta_{\nu\lambda}~{\Lambda'}_{[\mu\alpha]\alpha}-\delta_{\mu\lambda}
~{\Lambda'}_{[\nu\alpha]\alpha}\right)$, where $\displaystyle {\Lambda'}_{[\mu\alpha]\alpha}=\frac{1}{2}\left(T_{\mu\alpha\alpha}-T_{\alpha\mu\alpha}\right)$.
Contracting this last equation with $\delta_{\nu\lambda}$ leads to $\displaystyle \kappa=-\frac{1}{9}$ which gives the first of the two $\mathsf{320}$ multiplets in $1000$,
\begin{align}
\Lambda^{(\mathsf{320}_a)}_{[\mu\nu]\lambda}&=\frac{1}{6} \Big[2 \big(T_{\mu\nu\lambda}- T_{\nu\mu\lambda}\big)+ T_{\lambda\nu\mu}+T_{\mu\lambda\nu}- T_{\nu\lambda\mu}-T_{\lambda\mu\nu}\Big]-\frac{1}{18}\Big[\delta_{\nu\lambda}\big(T_{\mu\alpha\alpha}-T_{\alpha\mu\alpha}\big)
-\delta_{\mu\lambda}\big(T_{\nu\alpha\alpha}-T_{\alpha\nu\alpha}\big)\Big]
\end{align}
{

{\subsection{Construction of $\mathbf{\Lambda^{(\mathsf{320}_s)}_{(\mu\nu)\lambda}}$ tensor}}}
Next we construct the $\mathsf{320}$ multiplet which is symmetric in the first two 
indices. Here the $\mathsf{320}$ arises from the 1000-plet as follows: $550-220-10$
which we now exhibit in detail. 
Thus from the 1000-multiplet we define a tensor symmetric in the first two indices
so that
\begin{align}
\displaystyle \widetilde{\Lambda''}_{(\mu\nu)\lambda}&\equiv\frac{1}{2}\left[T_{\mu\nu\lambda}+T_{\nu\mu\lambda}\right]\nonumber
\end{align}
which has a dimensionality of $^{11}C_2 \times 10=550$. We remove from 
it the  tensor $S_{(\mu\nu\lambda)}$ 
which is totally symmetric in all three indices, has a dimensionality of $^{12}C_3
=220$ and the tensor form 
  \begin{align}
 S_{(\mu\nu\lambda)}&\equiv \frac{1}{3!} \left(T_{\mu\nu\lambda} +
 T_{\nu\lambda\mu} + T_{\lambda\mu\nu}
+T_{\nu\mu\lambda} +T_{\lambda\nu\mu} +T_{\mu\lambda\nu}\right)\nonumber
\end{align}
Removing the totally symmetric part from  $\widetilde{\Lambda''}_{(\mu\nu)\lambda}$ gives us the tensor $\widetilde{\Lambda'}_{(\mu\nu)\lambda}$ of dimensionality ${330 ~(=550-220)}$ where
\begin{align}
\widetilde{\Lambda'}_{(\mu\nu)\lambda}&=\widetilde{\Lambda''}_{(\mu\nu)\lambda}-S_{(\mu\nu\lambda)}
= \frac{1}{3}\left(T_{\mu\nu\lambda} +T_{\nu\mu\lambda}\right)-\frac{1}{6}\left(T_{\nu\lambda\mu}+ T_{\lambda\mu\nu}
 +T_{\lambda\nu\mu} +T_{\mu\lambda\nu}\right)\nonumber
\end{align}
As noted the tensor  $\widetilde{\Lambda'}_{(\mu\nu)\lambda}$ has 330 components and to get the second irreducible tensor $\mathsf{320}$ we need to make
it traceless. 
\begin{align}
\Lambda^{(\mathsf{320}_s)}_{(\mu\nu)\lambda}&=\widetilde{\Lambda'}_{(\mu\nu)\lambda}+\delta_{\nu\lambda}U_{\mu}
+\delta_{\mu\lambda}V_{\nu}+\delta_{\mu\nu}W_{\lambda}.\nonumber
\end{align}
We determine $U_{\mu}$, $V_{\nu}$ and $W_{\lambda}$ such that $\Lambda^{(\mathsf{320}_s)}_{(\mu\nu)\lambda}$ is traceless. Contracting the last equation with
$\delta_{\nu\lambda}$, $\delta_{\mu\lambda}$, and $\delta_{\mu\nu}$ gives
\[0=\widetilde{\Lambda'}_{(\mu\alpha)\alpha}+10U_{\mu}+V_{\mu}+W_{\mu},~~~0=\widetilde{\Lambda'}_{(\alpha\nu)\alpha}+U_{\nu}+10V_{\nu}+W_{\nu}, ~~~0=\widetilde{\Lambda'}_{(\alpha\alpha)\lambda}+U_{\lambda}+V_{\lambda}+10W_{\lambda}.\]
The solution to these equations are $\displaystyle U_{\mu}=\frac{1}{54}\left(-T_{\mu\alpha\alpha}+2T_{\alpha\alpha\mu}-T_{\alpha\mu\alpha}\right)$,~  $\displaystyle V_{\nu}=\frac{1}{54}\left(-T_{\nu\alpha\alpha}+2T_{\alpha\alpha\nu}-T_{\alpha\nu\alpha}\right)$ and  $\displaystyle W_{\lambda}=\frac{1}{27}\left(T_{\lambda\alpha\alpha}-2T_{\alpha\alpha\lambda}+T_{\alpha\lambda\alpha}\right)$. Thus the second $\mathsf{320}$ plet is given by
\begin{align}
\Lambda^{(\mathsf{320}_s)}_{(\mu\nu)\lambda}&= \frac{1}{6} \Big[2 \big(T_{\mu\nu\lambda}+ T_{\nu\mu\lambda}\big)- \big(T_{\lambda\nu\mu}+T_{\mu\lambda\nu}+ T_{\nu\lambda\mu}+T_{\lambda\mu\nu}\big)\Big]\nonumber\\
&~~~+\frac{1}{54}\Big[\delta_{\nu\lambda}\big(-T_{\mu\alpha\alpha}+2T_{\alpha\alpha\mu}-T_{\alpha\mu\alpha}\big)
+\delta_{\mu\lambda}\big(-T_{\nu\alpha\alpha}+2T_{\alpha\alpha\nu}-T_{\alpha\nu\alpha}\big)-2
\delta_{\mu\nu}\big(-T_{\lambda\alpha\alpha}+2T_{\alpha\alpha\lambda}-T_{\alpha\lambda\alpha}\big)\Big]
\end{align}
 
The decomposition of the $\mathsf{320}-$dimensional tensor under $\mathsf{SU(5)\times U(1)}$ is given by
\begin{eqnarray}\label{tensorstructure3}
\mathsf{320}\left[{\Lambda^{(\mathsf{320}_a)}_{[\mu\nu]\lambda}\choose \Lambda^{(\mathsf{320}_s)}_{(\mu\nu)\lambda}}\right]&=&\mathsf{5}(2)\left[\mathsf{H}^{(\mathsf{320})i}\right] + \overline{\mathsf{5}}(-2)\left[\mathsf{H}^{(\mathsf{320})}_i\right]+\mathsf{40}(-6)\left[\mathsf{H}^{(\mathsf{320})ijk}_l\right] + \overline{\mathsf{40}}(6) \left[\mathsf{H}^{(\mathsf{320})l}_{ijk}\right] +\mathsf{45}(2)\left[\mathsf{H}^{(\mathsf{320})ij}_k\right]\nonumber\\
&&+\overline{\mathsf{45}}(-2)\left[\mathsf{H}^{(\mathsf{320})k}_{ij}\right]
+\mathsf{70}(2)\left[\mathsf{H}^{(\mathsf{320})ij}_{(S)k}\right]+\overline{\mathsf{70}}(-2)\left[\mathsf{H}^{(\mathsf{320})k}_{(S)ij}\right]
\end{eqnarray}
 }

 {
 {
\section{Possible trilinear couplings of the $\boldsymbol{\mathsf{320}-}$plet with the $\boldsymbol{\mathsf{560}-}$plet and their generational symmetries.}\label{sec:E}
{\allowdisplaybreaks
\subsection{Couplings that involve {three} gammas}
Consider the coupling defined through
\begin{eqnarray}
J_{\grave{a}\grave{b}}&=&\Theta_{\alpha\beta}^{(\mathsf{560})\grave{a}\mathsf{T}}B\Gamma_{\mu}\Gamma_{\nu}\Gamma_{\lambda}
{\Theta}_{\alpha\beta}^{({\mathsf{560}})\grave{b}}~X_{\mu\nu\lambda},\label{start3gamma}
\end{eqnarray}
where for clarity, we write the expression $\braket{\Theta_{\alpha\beta}^{(\mathsf{560})\grave{a}*}|B\Gamma_{\mu}\Gamma_{\nu}\Gamma_{\lambda}|{\Theta}_{\alpha\beta}^{({\mathsf{560}})\grave{b}}}$ as
$\Theta_{\alpha\beta}^{(\mathsf{560})\grave{a}\mathsf{T}}B\Gamma_{\mu}\Gamma_{\nu}\Gamma_{\lambda}{\Theta}_{\alpha\beta}^{({\mathsf{560}})\grave{b}}$.
Here $^\mathsf{T}$ stands for the transpose, $X_{\mu\nu\lambda}$ is a generic three-index tensor and  $\grave{a},~\grave{b}$ are generation indices. Therefore
\begin{eqnarray}
J_{\grave{a}\grave{b}}&=& \Big(\Theta_{\alpha\beta}^{(\mathsf{560})\grave{a}\mathsf{T}}B\Gamma_{\mu}\Gamma_{\nu}\Gamma_{\lambda}~
\Theta_{\alpha\beta}^{(\mathsf{560})\grave{b}}\Big)^{\mathsf{T}}X_{\mu\nu\lambda}\nonumber\\
&=&\Theta_{\alpha\beta}^{(\mathsf{560})\grave{b}\mathsf{T}}\Big(\Gamma_{\mu}\Gamma_{\nu}\Gamma_{\lambda}\Big)^{\mathsf{T}}
\underbrace{B^{\mathsf{T}}}_{=-B}\Big(\Theta_{\alpha\beta}^{(\mathsf{560})\grave{a}\mathsf{T}}\Big)^{\mathsf{T}}X_{\mu\nu\lambda}
\nonumber\\
&=&-\Theta_{\alpha\beta}^{(\mathsf{560})\grave{b}\mathsf{T}}
\Gamma_{\lambda}^{\mathsf{T}}\Gamma_{\nu}^{\mathsf{T}}\Gamma_{\mu}^{\mathsf{T}}B\Theta_{\alpha\beta}^{(\mathsf{560})\grave{a}}X_{\mu\nu\lambda}\nonumber\\
&=&-\Theta_{\alpha\beta}^{(\mathsf{560})\grave{b}\mathsf{T}}B\underbrace{\Big(B^{-1}\Gamma_{\lambda}^{\mathsf{T}}B\Big)}_{=-\Gamma_{\lambda}}
\underbrace{\Big(B^{-1}~\Gamma_{\nu}^{\mathsf{T}}B\Big)}_{=-\Gamma_{\nu}}
\underbrace{\Big(B^{-1}~\Gamma_{\mu}^{\mathsf{T}}B\Big)}_{=-\Gamma_{\mu}}
\Theta_{\alpha\beta}^{(\mathsf{560})\grave{a}}X_{\mu\nu\lambda}\nonumber\\
&=&-(-1)^3\Theta_{\alpha\beta}^{(\mathsf{560})\grave{b}\mathsf{T}}B\Gamma_{\lambda}\Gamma_{\nu}\Gamma_{\mu}
\Theta_{\alpha\beta}^{(\mathsf{560})\grave{a}}X_{\mu\nu\lambda}\nonumber\\
&=&\Theta_{\alpha\beta}^{(\mathsf{560})\grave{b}\mathsf{T}}B
\underbrace{\Gamma_{\lambda}\Gamma_{\nu}\Gamma_{\mu}}_{2\delta_{\nu\lambda}\Gamma_{\mu}-2\delta_{\mu\lambda}
\Gamma_{\nu}+\Gamma_{\nu}\Gamma_{\mu}\Gamma_{\lambda}~~[\text{see}~ \bm{*}]}\Theta_{\alpha\beta}^{(\mathsf{560})\grave{a}}X_{\mu\nu\lambda}\nonumber\\
 &=&2\Theta_{\alpha\beta}^{(\mathsf{560})\grave{b}\mathsf{T}}B\Gamma_{\mu}\Theta_{\alpha\beta}^{(\mathsf{560})\grave{a}}X_{\mu\sigma\sigma}
 -2\Theta_{\alpha\beta}^{(\mathsf{560})\grave{b}\mathsf{T}}B\Gamma_{\nu}\Theta_{\alpha\beta}^{(\mathsf{560})\grave{a}}X_{\sigma\nu\sigma}
  +\Theta_{\alpha\beta}^{(\mathsf{560})\grave{b}\mathsf{T}}B\Gamma_{\nu}\Gamma_{\mu}\Gamma_{\lambda}
  \Theta_{\alpha\beta}^{(\mathsf{560})\grave{a}}X_{\mu\nu\lambda}~~~~~~~~\label{end3gamma}\\
\nonumber
   \end{eqnarray}
 {{CASE 1}}: ~~~$X_{\mu\nu\lambda}=\Lambda^{(\mathsf{320}_a)}_{[\mu\nu]\lambda}$ 
\begin{align}
J_{\grave{a}\grave{b}}&= \Theta_{\alpha\beta}^{(\mathsf{560})\grave{a}\mathsf{T}}
B\Gamma_{\mu}\Gamma_{\nu}\Gamma_{\lambda}\Theta_{\alpha\beta}^{(\mathsf{560})\grave{b}}\Lambda^{(\mathsf{320}_a)}_{[\mu\nu]\lambda}~~\text{[From Eq. (\ref{start3gamma})]}\nonumber\\
&=2\Theta_{\alpha\beta}^{(\mathsf{560})\grave{b}\mathsf{T}}B\Gamma_{\mu}\Theta_{\alpha\beta}^{(\mathsf{560})\grave{a}}
\underbrace{\Lambda^{(\mathsf{320}_a)}_{[\mu\sigma]\sigma}}_{=0~(\text{traceless})}
 -2\Theta_{\alpha\beta}^{(\mathsf{560})\grave{b}\mathsf{T}}B\Gamma_{\nu}\Theta_{\alpha\beta}^{(\mathsf{560})\grave{a}}
 \underbrace{\Lambda^{(\mathsf{320}_a)}_{[\sigma\nu]\sigma}}_{=0~(\text{traceless})}
 +\Theta_{\alpha\beta}^{(\mathsf{560})\grave{b}\mathsf{T}}B\Gamma_{\nu}\Gamma_{\mu}\Gamma_{\lambda}\Theta_{\alpha\beta}^{(\mathsf{560})\grave{a}}
 \underbrace{\Lambda^{(\mathsf{320}_a)}_{[\mu\nu]\lambda}}_{
  -\Lambda^{(\mathsf{320}_a)}_{[\nu\mu]\lambda}}~~\text{[From Eq. (\ref{end3gamma})]}
 \nonumber\\
&=-\Theta_{\alpha\beta}^{(\mathsf{560})\grave{b}\mathsf{T}}~B~\Gamma_{\nu}\Gamma_{\mu}\Gamma_{\lambda}\Theta_{\alpha\beta}^{(\mathsf{560})\grave{a}}
\Lambda^{(\mathsf{320}_a)}_{[\nu\mu]\lambda}\nonumber\\
&=-J_{\grave{b}\grave{a}} ~\Longrightarrow ~\text{Conclusion:}~ J_{aa}= 0.~\emph{This coupling vanishes  {only} for single generation.}
\nonumber\\
\nonumber
\end{align}
{{CASE 2}}: ~~~$X_{\mu\nu\lambda}=\Lambda^{(\mathsf{320}_s)}_{(\mu\nu)\lambda}$
\begin{flalign}
J_{\grave{a}\grave{b}}&= \Theta_{\alpha\beta}^{(\mathsf{560})\grave{a}\mathsf{T}}B\Gamma_{\mu}\Gamma_{\nu}\Gamma_{\lambda}
\Theta_{\alpha\beta}^{(\mathsf{560})\grave{b}}\Lambda^{(\mathsf{320}_s)}_{(\mu\nu)\lambda}~~\text{[From Eq. (\ref{start3gamma})]} \nonumber\\
&=2\Theta_{\alpha\beta}^{(\mathsf{560})\grave{b}\mathsf{T}}B\Gamma_{\mu}\Theta_{\alpha\beta}^{(\mathsf{560})\grave{a}}
\underbrace{\Lambda^{(\mathsf{320}_s)}_{(\mu\sigma)\sigma}}_{=0~(\text{traceless})}
 -2\Theta_{\alpha\beta}^{(\mathsf{560})\grave{b}\mathsf{T}}B\Gamma_{\nu}\Theta_{\alpha\beta}^{(\mathsf{560})\grave{a}}
 \underbrace{\Lambda^{(\mathsf{320}_s)}_{(\sigma\nu)\sigma}}_{=0~(\text{traceless})}
 +\Theta_{\alpha\beta}^{(\mathsf{560})\grave{b}\mathsf{T}}B\Gamma_{\nu}\Gamma_{\mu}\Gamma_{\lambda}
 \Theta_{\alpha\beta}^{(\mathsf{560})\grave{a}}\Lambda^{(\mathsf{320}_s)}_{(\mu\nu)\lambda}~~\text{[From Eq. (\ref{end3gamma})]}
\nonumber\\
&=\Theta_{\alpha\beta}^{(\mathsf{560})\grave{b}\mathsf{T}}B\underbrace{\Gamma_{\nu}\Gamma_{\mu}}_{=\delta_{\nu\mu}+\Sigma_{\nu\mu}~~[\text{see}~ \bm{**}]}\Gamma_{\lambda}\Theta_{\alpha\beta}^{(\mathsf{560})\grave{a}}\Lambda^{(\mathsf{320}_s)}_{(\mu\nu)\lambda}
\nonumber\\
&=\Theta_{\alpha\beta}^{(\mathsf{560})\grave{b}\mathsf{T}}B\Gamma_{\lambda}\Theta_{\alpha\beta}^{(\mathsf{560})\grave{a}}
\underbrace{\Lambda^{(\mathsf{320}_s)}_{(\sigma\sigma)\lambda}}_{=0~(\text{traceless})}+
\Theta_{\alpha\beta}^{(\mathsf{560})\grave{b}\mathsf{T}}B\Sigma_{\nu\mu}\Gamma_{\lambda}\Theta_{\alpha\beta}^{(\mathsf{560})\grave{a}}
\Lambda^{(\mathsf{320}_s)}_{(\mu\nu)\lambda}
\nonumber\\
&=\Theta_{\alpha\beta}^{(\mathsf{560})\grave{b}\mathsf{T}}B\Sigma_{\nu\mu}\Gamma_{\lambda}\Theta_{\alpha\beta}^{(\mathsf{560})\grave{a}}
\Lambda^{(\mathsf{320}_s)}_{(\mu\nu)\lambda}
\nonumber\\
&=0 ~~[\text{since}~\Sigma~\text{matrix is antisymmetric in its indices and}~\mathsf{320}~\text{tensor is symmetric in its first two indices}]
\nonumber\\
&\text{Conclusion:}~ {J_{\grave{a}\grave{b}}= 0, \forall \grave{a},\grave{b}. ~\emph{This coupling vanishes for {both} single \& multiple generations}}\nonumber
\end{flalign}
{\scriptsize{\[\bm{*}~~~
 \Gamma_{\lambda}\Gamma_{\nu}\Gamma_{\mu} = \big(-\Gamma_{\nu}  \Gamma_{\lambda}+2\delta_{\nu\lambda}\big)\Gamma_{\mu}
=2\delta_{\nu\lambda}\Gamma_{\mu}-\Gamma_{\nu}\big(2\delta_{\mu\lambda}-\Gamma_{\mu}\Gamma_{\lambda}\big)=2\delta_{\nu\lambda}\Gamma_{\mu}
-2\delta_{\mu\lambda}
\Gamma_{\nu}+\Gamma_{\nu}\Gamma_{\mu}\Gamma_{\lambda}
\]
\[\bm{**}~~~\Gamma_{\nu}\Gamma_{\mu}=\underbrace{\frac{1}{2}\big(\Gamma_{\nu}\Gamma_{\mu}+\Gamma_{\mu}\Gamma_{\nu}\big)}_{=\delta_{\nu\mu}}+
\underbrace{\frac{1}{2}\big(\Gamma_{\nu}\Gamma_{\mu}-\Gamma_{\mu}\Gamma_{\nu}\big)}_{=\Sigma_{\nu\mu}}=\delta_{\nu\mu}+\underbrace{\Sigma_{\nu\mu}}
_{=-\Sigma_{\mu\nu}}\]}}
\subsection{Couplings that involve a {single} gamma}
Consider the coupling defined through
\begin{eqnarray}
I_{\grave{a}\grave{b}}&=& \Theta_{\mu\sigma}^{(\mathsf{560})\grave{a}\mathsf{T}}B\Gamma_{\lambda}\Theta_{\nu\sigma}^{(\mathsf{560})\grave{b}}X_{\mu\nu\lambda},\label{start1gamma}
\end{eqnarray}
where $X_{\mu\nu\lambda}$ is a generic three-index tensor. Therefore
\begin{eqnarray}
I_{\grave{a}\grave{b}}&=& \Big(\Theta_{\mu\sigma}^{(\mathsf{560})\grave{a}\mathsf{T}}B\Gamma_{\lambda}~\Theta_{\nu\sigma}^{(\mathsf{560})\grave{b}}\Big)^{\mathsf{T}}
X_{\mu\nu\lambda}\nonumber\\
&=&\Theta_{\nu\sigma}^{(\mathsf{560})\grave{b}\mathsf{T}}\Gamma_{\lambda}^{\mathsf{T}}\underbrace{B^{\mathsf{T}}}_{=-B}
\Big(\Theta_{\mu\sigma}^{(\mathsf{560})\grave{a}{\mathsf{T}}}\Big)^{\mathsf{T}}~X_{\mu\nu\lambda}\nonumber\\
&=&-\Theta_{\nu\sigma}^{(\mathsf{560})\grave{b}\mathsf{T}}\Gamma_{\lambda}^{\mathsf{T}}  B
\Theta_{\mu\sigma}^{(\mathsf{560})\grave{a}}X_{\mu\nu\lambda}\nonumber\\
&=&-\Theta_{\nu\sigma}^{(\mathsf{560})\grave{b}\mathsf{T}}B\underbrace{\Big(B^{-1}\Gamma_{\lambda}^{\mathsf{T}}B\Big)}_{=-\Gamma_{\lambda}}
\Theta_{\mu\sigma}^{(\mathsf{560})\grave{a}}X_{\mu\nu\lambda}\nonumber\\
&=&+\Theta_{\nu\sigma}^{(\mathsf{560})\grave{b}\mathsf{T}}B\Gamma_{\lambda}
\Theta_{\mu\sigma}^{(\mathsf{560})\grave{a}}X_{\mu\nu\lambda}\nonumber\\
&=&\Theta_{\mu\sigma}^{(\mathsf{560})\grave{b}\mathsf{T}}B\Gamma_{\lambda}~
\Theta_{\nu\sigma}^{(\mathsf{560})\grave{a}}X_{\nu\mu\lambda}~~~\text{$\mu,\nu$ (dummy indices) interchange}\label{end1gamma}\\
\nonumber
\end{eqnarray}
{{CASE 1(a)}}: ~~~$X_{\mu\nu\lambda}=\Lambda^{(\mathsf{320}_s)}_{(\mu\nu)\lambda}$ 
\begin{align}
I_{\grave{a}\grave{b}}&= \Theta_{\mu\sigma}^{(\mathsf{560})\grave{a}\mathsf{T}}B\Gamma_{\lambda}\Theta_{\nu\sigma}^{(\mathsf{560})\grave{b}}\Lambda^{(\mathsf{320}_s)}_{(\mu\nu)\lambda}~~~~~~~~~~ \text{[From Eq. (\ref{start1gamma})]}\nonumber\\
&=\Theta_{\mu\sigma}^{(\mathsf{560})\grave{b}\mathsf{T}}B\Gamma_{\lambda}\Theta_{\nu\sigma}^{(\mathsf{560})\grave{a}}
\Lambda^{(\mathsf{320}_s)}_{(\nu\mu)\lambda}~~~~~~~~~~ \text{[From Eq. (\ref{end1gamma})]}\nonumber\\
&=\Theta_{\mu\sigma}^{(\mathsf{560})\grave{b}\mathsf{T}}B\Gamma_{\lambda}\Theta_{\nu\sigma}^{(\mathsf{560})\grave{a}}
\Lambda^{(\mathsf{320}_s)}_{(\mu\nu)\lambda}\nonumber\\
&=+~I_{\grave{b}\grave{a}} ~\Longrightarrow ~\text{Conclusion:}~ {I_{\grave{a}\grave{a}}\neq 0.~~\emph{This coupling is non-vanishing for {both} single \& multiple generations}}
\nonumber\\
\nonumber
\end{align}
{{CASE 1(b)}}: ~~~$X_{\mu\nu\lambda}=\Lambda^{(\mathsf{320}_s)}_{(\mu\lambda)\nu}$ 
\begin{align}
I_{\grave{a}\grave{b}}& =\Theta_{\mu\sigma}^{(\mathsf{560})\grave{a}\mathsf{T}}B\Gamma_{\lambda}\Theta_{\nu\sigma}^{(\mathsf{560})\grave{b}}\Lambda^{(\mathsf{320}_s)}_{(\mu\lambda)\nu}~~~~~~~~~~ \text{[From Eq. (\ref{start1gamma})]}\nonumber\\
&=\Theta_{\mu\sigma}^{(\mathsf{560})\grave{b}\mathsf{T}}B\Gamma_{\lambda}\Theta_{\nu\sigma}^{(\mathsf{560})\grave{a}}
\Lambda^{(\mathsf{320}_s)}_{(\nu\lambda)\mu}~~~~~~~~~~ \text{[From Eq. (\ref{end1gamma})]}\nonumber\\
&\text{{{No symmetry}} in the exchange of 560's}\nonumber\\
&\text{Conclusion: ~{\emph{This coupling is non-vanishing for {both} single \& multiple generations}}}
\nonumber\\
\nonumber
\end{align}
{{CASE 2(a)}}: ~~~$X_{\mu\nu\lambda}=\Lambda^{(\mathsf{320}_a)}_{[\mu\nu]\lambda}$ 
\begin{align}
I_{\grave{a}\grave{b}}&= \Theta_{\mu\sigma}^{(\mathsf{560})\grave{a}\mathsf{T}}B\Gamma_{\lambda}\Theta_{\nu\sigma}^{(\mathsf{560})\grave{b}}
\Lambda^{(\mathsf{320}_a)}_{[\mu\nu]\lambda}~~~~~~~~~~ \text{[From Eq. (\ref{start1gamma})]}\nonumber\\
&=\Theta_{\mu\sigma}^{(\mathsf{560})\grave{b}\mathsf{T}}B\Gamma_{\lambda}\Theta_{\nu\sigma}^{(\mathsf{560})\grave{a}}
\Lambda^{(\mathsf{320}_a)}_{[\nu\mu]\lambda}~~~~~~~~~~ \text{[From Eq. (\ref{end1gamma})]}\nonumber\\
&=-\Theta_{\mu\sigma}^{(\mathsf{560})\grave{b}\mathsf{T}}B\Gamma_{\lambda}\Theta_{\nu\sigma}^{(\mathsf{560})\grave{a}}
\Lambda^{(\mathsf{320}_a)}_{[\mu\nu]\lambda}\nonumber\\
&=-I_{\grave{b}\grave{a}} ~\Longrightarrow ~\text{Conclusion:}~{I_{aa}= 0.~~\emph{This coupling vanishes {only} for single generation}}\nonumber\\
\nonumber
\end{align}
{{CASE 2(b)}}: ~~~$X_{\mu\nu\lambda}=\Lambda^{(\mathsf{320}_a)}_{[\mu\lambda]\nu}$ 
\begin{align}
I_{\grave{a}\grave{b}}&= \Theta_{\mu\sigma}^{(\mathsf{560})\grave{a}\mathsf{T}}B\Gamma_{\lambda}\Theta_{\nu\sigma}^{(\mathsf{560})\grave{b}}
\Lambda^{(\mathsf{320}_a)}_{[\mu\lambda]\nu}~~~~~~~~~~ \text{[From Eq. (\ref{start1gamma})]}\nonumber\\
&=\Theta_{\mu\sigma}^{(\mathsf{560})\grave{b}\mathsf{T}}B\Gamma_{\lambda}\Theta_{\nu\sigma}^{(\mathsf{560})\grave{a}}
\Lambda^{(\mathsf{320}_a)}_{[\nu\lambda]\mu}~~~~~~~~~~ \text{[From Eq. (\ref{end1gamma})]}\nonumber\\
&\text{{{No symmetry}} in the exchange of 560's}\nonumber\\
&\text{Conclusion: {\emph{This coupling is non-vanishing for {both} single \& multiple generations}}}\nonumber\\
\nonumber
\end{align}
}}
}

{

{
\section{Construction of normalized $\boldsymbol{\mathsf{70}-}$, $\boldsymbol{\mathsf{45}-}$ and $\boldsymbol{\mathsf{5}-}$plets with diagonal kinetic energy terms contained in $\boldsymbol{\Lambda^{\mathsf{320}_s}_{(\mu\nu)\lambda}}$ of $\boldsymbol{\mathsf{SO(10)}}$}\label{sec:F}
\allowdisplaybreaks{
\subsection{STEP 1: Identifying {three} $\boldsymbol{\mathsf{70}}$'s, {two} $\boldsymbol{\mathsf{45}}$'s and {eight} $\boldsymbol{\mathsf{5}}$'s in $\boldsymbol{\Lambda^{\mathsf{320}_s}_{(\mu\nu)\lambda}}$}
\begin{align}\label{compdef320}
\Lambda^{\mathsf{320}_s}_{(\mu\nu)\lambda} &= t_{(\mu\nu)\lambda}
+\frac{1}{54}
(\delta_{\nu\lambda} \delta_{\mu\beta} + \delta_{\mu\lambda} \delta_{\nu\beta}
- 2 \delta_{\mu\nu} \delta_{\lambda\beta})\tau_{\beta}
\end{align}
where
\begin{subequations}
{\begin{align}
t_{(\mu\nu)\lambda}&\equiv \frac{1}{6}\left[2\left(T_{\mu\nu\lambda} + T_{\nu\mu\lambda}\right)
-\left( T_{\lambda\mu\nu}+T_{\lambda\nu\mu}
 +T_{\mu\lambda\nu} + T_{\nu\lambda\mu}\right)\right]\label{traceless part}\\
 \tau_{\beta}&\equiv -T_{\beta\alpha\alpha}+2T_{\alpha\alpha\beta}-T_{\alpha\beta\alpha}\label{trace part}
\end{align}}
\end{subequations}
We define the eight distinct $\mathsf{5}-$plets of $\mathsf{SU(5)}$ as follows:
\begin{eqnarray}\label{all5pletsin320}
\begin{split}
 T_{c_ic_n\bar c_n} + T_{c_nc_i\bar c_n}&\equiv H^{(5_1)i},\\
 T_{\bar c_nc_ic_n} + T_{\bar c_nc_nc_i}&\equiv H^{(5_2)i},\\
T_{c_i\bar c_nc_n} + T_{c_n\bar c_nc_i}&\equiv H^{(5_3)i} ,\\
\\
 T_{\bar c_nc_nc_i} - T_{\bar c_nc_ic_n}&\equiv H^{(5_2^{\prime})i},\\
 T_{c_n\bar c_nc_i} - T_{c_i\bar c_nc_n}&\equiv H^{(5_3^{\prime})i},\\
 \\
 T_{c_i c_n\bar c_n}+T_{c_i\bar c_n c_n}&\equiv H^{({5}_{1}^{\prime\prime})i},\\
  T_{c_nc_i \bar c_n}+T_{\bar c_nc_i c_n}&\equiv H^{({5}_{2}^{\prime\prime})i},\\
  T_{c_n \bar c_n c_i}+T_{\bar c_n c_n c_i}&\equiv H^{({5}_{3}^{\prime\prime})i}.
\end{split}
\end{eqnarray}
while the three $\mathsf{{70}}-$plets of $\mathsf{SU(5)}$ are defined through
\begin{eqnarray}\label{all70pletsin320}
\begin{split}
T_{c_ic_j\bar c_k} + T_{c_jc_i\bar c_k}&=H^{(70_1)ij}_k+\frac{1}{6}\left[\delta^i_kH^{(5_1)j}+\delta^j_kH^{(5_1)i}\right],\\
T_{\bar c_kc_ic_j} + T_{\bar c_kc_jc_i}&=H^{(70_2)ij}_k+\frac{1}{6}\left[\delta^i_kH^{(5_2)j}+\delta^j_kH^{(5_2)i}\right],\\
T_{c_i\bar c_kc_j} + T_{c_j\bar c_kc_i}&=H^{(70_3)ij}_k+\frac{1}{6}\left[\delta^i_kH^{(5_3)j}+\delta^j_kH^{(5_3)i}\right].
\end{split}
\end{eqnarray}
and finally the two distinct $\mathsf{{45}}-$plets of $\mathsf{SU(5)}$ are defined by 
\begin{eqnarray}\label{all45pletsin320}
\begin{split}
 T_{\bar c_ic_jc_k} - T_{\bar c_ic_kc_j}&=H^{(45_2)jk}_i+\frac{1}{4}\left[\delta^j_iH^{(5_2^{\prime})k}+\delta^k_iH^{(5_2^{\prime})j}\right],\\
 T_{c_j\bar c_ic_k} - T_{ c_k\bar c_ic_j}&=H^{(45_3)jk}_i+\frac{1}{4}\left[\delta^j_iH^{(5_3^{\prime})k}+\delta^k_iH^{(5_3^{\prime})j}\right].
\end{split}
\end{eqnarray}
One may now very easily define $\mathsf{\overline{5}}-$plets, $\mathsf{\overline{45}}-$plets, $\mathsf{\overline{70}}-$plets of $\mathsf{SU(5)}$. 
The reducible $\mathsf{SU(5)}$ tensors given by Eq.(\ref{traceless part})
\begin{eqnarray}\label{SU(5)tracelessparts320}
\begin{split}
t_{(c_ic_j)\bar c_k}&=\frac{1}{6}\left\{2\left[H^{(70_1)ij}_k+\frac{1}{6}\left(\delta^i_kH^{(5_1)j}+\delta^j_kH^{(5_1)i}\right)\right]
-\left[H^{(70_2)ij}_k+\frac{1}{6}\left(\delta^i_kH^{(5_2)j}+\delta^j_kH^{(5_2)i}\right)\right]\right.\\
&\left.~~~~~~~-\left[H^{(70_3)ij}_k+\frac{1}{6}\left(\delta^i_kH^{(5_3)j}+\delta^j_kH^{(5_3)i}\right)\right]\right\},\\
\\
t_{(\bar c_ic_j)c_k}&=\frac{1}{6}\left\{\frac{3}{2}\left[H^{(45_2)jk}_i+\frac{1}{4}\left(\delta^j_iH^{(5_2^{\prime})k}-\delta^k_iH^{(5_2^{\prime})j}\right)\right]
+\frac{3}{2}\left[H^{(45_3)jk}_i+\frac{1}{4}\left(\delta^j_iH^{(5_3^{\prime})k}-\delta^k_iH^{(5_3^{\prime})j}\right)\right]\right.\\
&\left.~~~~~~~-\left[H^{(70_1)jk}_i+\frac{1}{6}\left(\delta^j_iH^{(5_1)k}+\delta^k_iH^{(5_1)j}\right)\right]+\frac{1}{2}
\left[H^{(70_2)jk}_i+\frac{1}{6}\left(\delta^j_iH^{(5_2)k}+\delta^k_iH^{(5_2)j}\right)\right]\right.\\
&\left.~~~~~~~+\frac{1}{2}\left[H^{(70_3)jk}_i+\frac{1}{6}\left(\delta^j_iH^{(5_3)k}+\delta^k_iH^{(5_3)j}\right)\right]\right\}.
\end{split}
\end{eqnarray}
Note that the tensors $t_{(\bar c_i \bar c_j)c_k}$ and $t_{(c_i\bar c_j)\bar c_k}$ would have same expressions except the unbarred $\mathsf{SU(5)}$ fields would be replaced by barred ones. Finally, we do not compute $t_{(c_i c_j)c_k}$ and $t_{(\bar c_i \bar c_j)\bar c_k}$ as they contain $\mathsf{40}-$plet and
$\mathsf{\overline{40}}-$plet of $\mathsf{SU(5)}$ which does not participate in the doublet-triplet splitting. This is because $\mathsf{40}+\overline{\mathsf{40}}$ multiplets do not share the same $\mathsf{U(1)_Y}$ hypercharge of $\pm 3$ as the $\mathsf{5}+\overline{\mathsf{5}}$ multiplets of $\mathsf{SU(5)}$.
\subsection{STEP 2: Construction of the kinetic energy of $\boldsymbol{\Lambda^{\mathsf{320}_s}_{(\mu\nu)\lambda}}$ multiplet}
\begin{align}\label{constr ke 320}
\mathcal{L}_{\textsc{ke}}^{(\mathsf{\mathsf{320}_s})}&=-\partial_A\Lambda^{\mathsf{320}_s}_{(\mu\nu)\lambda}
\partial^A\Lambda^{\mathsf{320}_s\dagger}_{(\mu\nu)\lambda}\nonumber\\
&=-\bigg[\partial_At_{(\mu\nu)\lambda}\partial^At^{\dagger}_{(\mu\nu)\lambda}
+\frac{1}{54^2}\bigg(\delta_{\nu\lambda}\partial_A\tau_{\mu} + \delta_{\mu\lambda} \partial_A\tau_{\nu}- 2 \delta_{\mu\nu} \partial_A\tau_{\lambda}\bigg)
\bigg(\delta_{\nu\lambda}\partial^A\tau^{\dagger}_{\mu} + \delta_{\mu\lambda} \partial^A\tau^{\dagger}_{\nu}
- 2 \delta_{\mu\nu} \partial^A\tau^{\dagger}_{\lambda}\bigg)\nonumber\\
&~~~+\frac{1}{54}\partial_At_{(\mu\nu)\lambda}\bigg(\delta_{\nu\lambda}\partial^A\tau^{\dagger}_{\mu} + \delta_{\mu\lambda} \partial^A\tau^{\dagger}_{\nu}
- 2 \delta_{\mu\nu} \partial^A\tau^{\dagger}_{\lambda}\bigg)
+\frac{1}{54}\bigg(\delta_{\nu\lambda}\partial_A\tau_{\mu} + \delta_{\mu\lambda} \partial_A\tau_{\nu}
- 2 \delta_{\mu\nu}\partial_A \tau_{\lambda}\bigg)\partial^At^{\dagger}_{(\mu\nu)\lambda}\bigg]\nonumber\\
&= -\frac{1}{8}\bigg[\partial_At_{(c_ic_j)c_k}\partial^At^{\dagger}_{(c_ic_j)c_k} + \partial_At_{(\bar c_i\bar c_j)\bar c_k}\partial^At^{\dagger}_{(\bar c_i\bar c_j)\bar c_k}
+\partial_At_{(c_ic_j)\bar c_k}\partial^At^{\dagger}_{(c_ic_j)\bar c_k}+\partial_At_{(\bar c_i\bar c_j)c_k}\partial^At^{\dagger}_{(\bar c_i\bar c_j)c_k}\nonumber\\
&~~~+2\partial_At_{(\bar c_ic_j)c_k}\partial^At^{\dagger}_{(\bar c_ic_j)c_k}+2\partial_At_{(\bar c_ic_j)\bar c_k}\partial^At^{\dagger}_{(\bar c_ic_j)\bar c_k}
\bigg]+\frac{1}{54} \partial_A\tau_{\mu}\partial^A\tau^{\dagger}_{\mu}\nonumber\\
&=\mathcal{L}_{\textsc{ke}}^{(\mathsf{5_{320}})}+\mathcal{L}_{\textsc{ke}}^{(\mathsf{45_{320}})}+\mathcal{L}_{\textsc{ke}}^{(\mathsf{70_{320}})}+\cdots,
\end{align}
where $\cdots$ represent $\mathcal{L}_{\textsc{ke}}^{(\mathsf{\overline{5}_{320}})}$, $\mathcal{L}_{\textsc{ke}}^{(\mathsf{\overline{45}_{320}})}$, 
 $\mathcal{L}_{\textsc{ke}}^{(\mathsf{\overline{70}_{320}})}$,  $\mathcal{L}_{\textsc{ke}}^{(\mathsf{{40}_{320}})}$ and  $\mathcal{L}_{\textsc{ke}}^{(\mathsf{\overline{40}_{320}})}$. As mentioned earlier, the doublets contained in $\mathsf{40+\overline{40}}$ do not participate in doublet triplet splitting as their hypercharge are different from $\mathsf{5+\overline{5}}$. Thus,
 \begin{subequations}
\begin{align}
\mathcal{L}_{\textsc{ke}}^{(\mathsf{70_{320}})}&=-\frac{1}{192}\bigg[4\partial_AH^{(70_{1})ij}_k\partial^AH^{(70_{1})ij{\dagger}}_k
+\partial_AH^{(70_{2})ij}_k\partial^AH^{(70_{2})ij{\dagger}}_k+\partial_AH^{(70_{3})ij}_k\partial^AH^{(70_{3})ij{\dagger}}_k\nonumber\\
&~~~~~~~~~~~-2\partial_AH^{(70_{1})ij}_k\partial^AH^{(70_{2})ij{\dagger}}_k
-2\partial_AH^{(70_{2})ij}_k\partial^AH^{(70_{1})ij{\dagger}}_k
-2\partial_AH^{(70_{1})ij}_k\partial^AH^{(70_{3})ij{\dagger}}_k\nonumber\\
&~~~~~~~~~~~-2\partial_AH^{(70_{3})ij}_k\partial^AH^{(70_{1})ij{\dagger}}
+\partial_AH^{(70_{2})ij}_k\partial^AH^{(70_{3})ij{\dagger}}_k
+\partial_AH^{(70_{3})ij}_kH^{(70_{2})ij{\dagger}}_k\bigg]\label{ke70}\\
\nonumber\\
\mathcal{L}_{\textsc{ke}}^{(\mathsf{45_{320}})}&=-\frac{1}{64}\bigg[\partial_AH^{(45_{2})ij}_k\partial^AH^{(45_{2})ij\dagger}_k
+\partial_AH^{(45_{3})ij}_k\partial^AH^{(45_{3})ij{\dagger}}_k\nonumber\\
&~~~~~~~~~~+\partial_AH^{(45_{2})ij}_k\partial^AH^{(45_{3})ij{\dagger}}_k
+\partial_AH^{(45_{3})ij}_k\partial^AH^{(45_{2})ij{\dagger}}_k
\bigg]\label{ke45}\\
\nonumber\\
\mathcal{L}_{\textsc{ke}}^{(\mathsf{5_{320}})}&=-\frac{1}{3456}\bigg[-8\partial_AH^{({5}_{1}^{\prime\prime})i}\partial^AH^{({5}_{1}^{\prime\prime})i\dagger}-8
\partial_AH^{({5}_{2}^{\prime\prime})i}\partial^AH^{({5}_{2}^{\prime\prime})i\dagger}
-32\partial_AH^{({5}_{3}^{\prime\prime})i}\partial^AH^{({5}_{3}^{\prime\prime})i\dagger} \nonumber\\
&~~~~~~~~~~~~~
+24\partial_AH^{(5_{1})i}\partial^AH^{(5_{1})i{\dagger}}+ 6\partial_AH^{(5_{2})i}\partial^AH^{(5_{2})i{\dagger}}
+ 6\partial_AH^{(5_{3})i}\partial^AH^{(5_{3})i{\dagger}}\nonumber\\
&~~~~~~~~~~~~~+27\partial_AH^{(5_{2}^{\prime})i}\partial^AH^{(5_{2}^{\prime})i{\dagger}}
+27\partial_AH^{(5_{3}^{\prime})i}\partial^AH^{(5_{3}^{\prime})i{\dagger}} \nonumber\\
&~~~~~~~~~~~~~-8\partial_AH^{({5}_{1}^{\prime\prime})i}\partial^AH^{({5}_{2}^{\prime\prime})i\dagger}
-8\partial_AH^{({5}_{2}^{\prime\prime})i}\partial^AH^{({5}_{1}^{\prime\prime})i\dagger}
+16\partial_AH^{({5}_{1}^{\prime\prime})i}\partial^AH^{({5}_{3}^{\prime\prime})i\dagger}\nonumber\\
&~~~~~~~~~~~~~+16\partial_AH^{({5}_{3}^{\prime\prime})i}\partial^AH^{({5}_{1}^{\prime\prime})i\dagger}
+16\partial_AH^{({5}_{2}^{\prime\prime})i}\partial^AH^{({5}_{3}^{\prime\prime})i\dagger}
+16\partial_AH^{({5}_{3}^{\prime\prime})i}\partial^AH^{({5}_{2}^{\prime\prime})i\dagger} \nonumber\\
&~~~~~~~~~~~~~-12\partial_AH^{(5_{1})i}\partial^AH^{(5_{2})i{\dagger}}
-12\partial_AH^{(5_{2})i}\partial^AH^{(5_{1})i{\dagger}}-12\partial_AH^{(5_{1})i}\partial^AH^{(5_{3})i{\dagger}}\nonumber\\
&~~~~~~~~~~~~~-12\partial_AH^{(5_{3})i}\partial^AH^{(5_{1})i{\dagger}}
+6\partial_AH^{(5_{2})i}\partial^AH^{(5_{3})i{\dagger}}+6\partial_AH^{(5_{3})i}\partial^AH^{(5_{2})i{\dagger}}\nonumber\\
&~~~~~~~~~~~~~+27\partial_AH^{(5_{2}^{\prime})i}\partial^AH^{(5_{3}^{\prime})i{\dagger}}
+27\partial_AH^{(5_{3}^{\prime})i}\partial^AH^{(5_{2}^{\prime})i{\dagger}}
\bigg]\label{ke5}
\end{align}
\end{subequations}
\subsection{General case of normalization of the kinetic energy
 after diagonalization.} 
Consider a set of fields  $\phi_i$ ($i=1,...,n$) which have mixed kinetic terms.
\begin{align}
\mathcal{L}_{\textsc{ke}}= -\partial_A\phi_i^* K_{ij} \partial^A\phi_j= - \partial_A\Phi^\dagger K \partial^A\Phi,\nonumber
\end{align}
where,
\begin{align}
\Phi^\dagger= \bigg(\phi_1^*~~~~~ \phi_2^*~~~~~\phi_3^*~~~~~\cdots~~~~~\phi_n^*\bigg)\nonumber
\end{align}
We make the transformation
\begin{align}
\Phi= V\xi.\nonumber
\end{align}
In this case we have
\begin{align}
\mathcal{L}_{\textsc{ke}}&=- \partial_A\xi^* V^\dagger K V \partial^A\xi \nonumber\\
&\equiv - \partial_A\xi^\dagger K_D \partial^A\xi,\nonumber
\end{align}
where,
\begin{align}
K_D&= V^\dagger K V =\text{diag}(k_1, k_2, ...,k_n).\nonumber
\end{align}
Next we make the scale transformation where
\begin{displaymath}
    \xi_i=\left\{
     \begin{array}{ll}
      \displaystyle \frac{ \displaystyle 1}{ \displaystyle \sqrt k_i}\chi_i, &~\textnormal{for non-vanishing}~\displaystyle k_i\\
        \displaystyle\chi_i, &~\textnormal{for vanishing}~\displaystyle k_i
     \end{array}
   \right.
\end{displaymath}
\begin{align}
\text{and}~~~k_i\xi_i^*\xi_i&= \chi_i^*\chi_i.\nonumber
\end{align}
Thus,
\begin{align}
\mathcal{L}_{\textsc{ke}}&= -\partial_A \chi_i^* \partial^A \chi_i,\nonumber
\end{align}
where,
\begin{align}
\phi_i&= V_{ij} \xi_j= \frac{1}{\sqrt k_j}V_{ij} \chi_j.\nonumber
\end{align}
Here $\chi_i$ are the fields which have diagonal and normalized
kinetic energy terms.
\subsection{STEP 3: Diagonalization and normalization of kinetic energy of $\boldsymbol{\mathsf{70}}$, $\boldsymbol{\mathsf{45}}$ and $\boldsymbol{\mathsf{5}}$ multiplets.}
\subsubsection{$\boldsymbol{\chi^{(\mathsf{70_{320}})ij}_k}$: Normalized $\boldsymbol{\mathsf{70}-}$plet with diagonal kinetic energy terms}
Write Eq.(\ref{ke70}) as
\begin{align}
{\mathcal L}_{kin}^{\mathsf{(70_{320})}}&=-\partial_A\Phi^{\mathsf{(70)}\dagger}K^{\mathsf{(70)}}\partial^A\Phi^{\mathsf{(70)}},\nonumber\\
\text{where}~~~~~~~~~~~~~~~~~~~~~~~~~~~\Phi^{\mathsf{(70)}\dagger}&=\bigg(H^{(70_{1})ij\dagger}_k~~~H^{(70_{2})ij\dagger}_k~~~H^{(70_{3})ij\dagger}_k\bigg)\nonumber\\
\text{and}~~~~K^{\mathsf{(70)}}&=\frac{1}{192}\begin{pmatrix}
4 & -2 & -2 \\
-2 & 1 & 1  \\
-2 & 1 & 1
\end{pmatrix}.\nonumber\\
\text{The diagonal matrix is}~~~~K_{D}^{\mathsf{(70)}}&=\displaystyle\begin{pmatrix}
\displaystyle\frac{1}{32} & \displaystyle 0 & \displaystyle 0\\
\displaystyle 0 & \displaystyle 0 & \displaystyle 0\\
\displaystyle 0 & \displaystyle 0 & \displaystyle 0\\
\end{pmatrix}\nonumber\\
\text{while the unitary {diagonalizing matrix is}}~~~~ V^{\mathsf{(70)}}&= \displaystyle\begin{pmatrix}
\displaystyle-\sqrt{\frac{2}{3}} & \displaystyle\frac{1}{\sqrt{5}} & \displaystyle\sqrt{\frac{2}{15}}\\
\displaystyle\frac{1}{\sqrt{6}} & \displaystyle 0 & \displaystyle\sqrt{\frac{5}{6}}\\
\displaystyle\frac{1}{\sqrt{6}} & \displaystyle\frac{2}{\sqrt{5}} &
-\displaystyle\frac{1}{\sqrt{30}} &
\end{pmatrix}.\nonumber
\end{align}
The scaled unitary transformed fields are
\begin{eqnarray}\label{Normdiag70in320}
\begin{split}
H^{(70_{1})ij}_k&=4\sqrt{2}~V^{\mathsf{(70)}}_{11}~\chi^{(\mathsf{70_{320}})ij}_k,~~~~V^{\mathsf{(\mathsf{70})}}_{11}=-\sqrt{\frac{2}{3}}\\
H^{(70_{2})ij}_k&=4\sqrt{2}~V^{\mathsf{(70)}}_{21}~\chi^{(\mathsf{70_{320}})ij}_k,~~~~V^{\mathsf{(70)}}_{21}=\frac{1}{\sqrt{6}}\\
H^{(70_{3})ij}_k&=4\sqrt{2}~V^{\mathsf{(70)}}_{31}~\chi^{(\mathsf{70_{320}})ij}_k,~~~~V^{\mathsf{(70)}}_{31}=\frac{1}{\sqrt{6}}
\end{split}
\end{eqnarray}
\subsubsection{$\boldsymbol{\chi^{(\mathsf{45_{320}})ij}_k}$: Normalized $\boldsymbol{\mathsf{45}-}$plet with diagonal kinetic energy terms}
Similarly, Eq.(\ref{ke45}) can be written as
\begin{align}
{\mathcal L}_{kin}^{\mathsf{(45_{320})}}&=-\partial_A\Phi^{\mathsf{(45)}\dagger}K^{\mathsf{(45)}}\partial^A\Phi^{\mathsf{(45)}},\nonumber\\
\text{where,}~~~~\Phi^{\mathsf{(45)}\dagger}&=\bigg(H^{(45_{3})ij\dagger}_k~~~H^{(45_{3})ij\dagger}_k\bigg)\nonumber\\
\text{and}~~~~K^{\mathsf{(45)}}&=\frac{1}{64}\begin{pmatrix}
1 & 1 \\
1 & 1
\end{pmatrix}.\nonumber\\
\text{The diagonal matrix is}~~~~K_{D}^{\mathsf{(45)}}&=\displaystyle\begin{pmatrix}
\displaystyle\frac{1}{32} & \displaystyle 0 \\
\displaystyle 0 & \displaystyle 0
\end{pmatrix}\nonumber\\
\text{while the unitary {diagonalizing matrix is}}~~~~ V^{\mathsf{(45)}}&= \displaystyle\begin{pmatrix}
\displaystyle\frac{1}{\sqrt{2}} & \displaystyle-\frac{1}{\sqrt{2}} \\
\displaystyle\frac{1}{\sqrt{2}} & \displaystyle\frac{1}{\sqrt{2}}
\end{pmatrix}\nonumber
\end{align}
The scaled unitary transformed fields are
\begin{eqnarray}\label{Normdiag45in320}
\begin{split}
H^{(45_{2})ij}_k&=4\sqrt{2}~V^{\mathsf{(45)}}_{11}~\chi^{(\mathsf{45_{320}})ij}_k,~~~~V^{\mathsf{(45)}}_{11}=\frac{1}{\sqrt{2}}\\
H^{(45_{3})ij}_k&=4\sqrt{2}~V^{\mathsf{(45)}}_{21}~\chi^{(\mathsf{45_{320}})ij}_k,~~~~V^{\mathsf{(45)}}_{21}=\frac{1}{\sqrt{2}}
\end{split}
\end{eqnarray}
\subsubsection{$\boldsymbol{\chi^{(\mathsf{5_{320}})i}}$: Normalized $\boldsymbol{\mathsf{5}-}$plet with diagonal kinetic energy terms}
Before we diagonalize the kinetic energy matrix for the $\mathsf{5}-$plet, let us define 
\begin{align}
H^{({5}_{1}^{\prime\prime})}_i =T_{c_i c_n\bar c_n}+T_{c_i\bar c_n c_n}\equiv h_1^i+h_2^i,~~~~~~H^{({5}_{2}^{\prime\prime})}_i =T_{c_nc_i \bar c_n}+T_{\bar c_nc_i c_n}\equiv h_3^i+h_4^i,~~~~~~H^{({5}_{3}^{\prime\prime})}_i =T_{c_n \bar c_nc_i}+T_{\bar c_n c_nc_i}\equiv h_5^i+h_6^i
\nonumber,
\end{align}
then
\begin{align}
H^{(5_1)i}= T_{c_ic_n\bar c_n} + T_{c_nc_i\bar c_n}= h_1^i+h_3^i,~~~~~~
H^{(5_2)i}= T_{\bar c_nc_ic_n} + T_{\bar c_nc_nc_i} &=h_4^i+h_6^i,~~~~~~
H^{(5_3)i}= T_{c_i\bar c_nc_n} + T_{c_n\bar c_nc_i}= h_2^i+h_5^i,\nonumber\\
H^{(5_2^{\prime})k}= T_{\bar c_nc_nc_k} - T_{\bar c_nc_kc_n}= h_6^i-h_4^i,~~~&~~~
H^{(5_3^{\prime})i}= T_{c_n\bar c_nc_k} - T_{c_k\bar c_nc_n}= h_5^i-h_2^i\nonumber
\end{align}
In terms of $h^i$ fields, the kinetic energy of the $\mathsf{5}-$plets from Eq.(\ref{ke5}) can be rewritten as
\begin{align}
{\mathcal L}_{kin}^{\mathsf{(5_{320})}}&=-\frac{1}{3456}\Big[
16\partial_Ah_1^i\partial^Ah_1^{i\dagger}-20\partial_Ah_2^i\partial^Ah_1^{i\dagger}+16\partial_Ah_3^i\partial^Ah_1^{i\dagger}
-20\partial_Ah_4^i\partial^Ah_1^{i\dagger}+4\partial_Ah_5^i\partial^Ah_1^{i\dagger}+4\partial_Ah_6^i\partial^Ah_1^{i\dagger}
\nonumber\\
&~~~~~~~~~~~~-20\partial_Ah_1^i\partial^Ah_2^{i\dagger}+25\partial_Ah_2^i\partial^Ah_2^{i\dagger}-20\partial_Ah_3^i\partial^Ah_2^{i\dagger}
+25\partial_Ah_4^i\partial^Ah_2^{i\dagger}-5\partial_Ah_5^i\partial^Ah_2^{i\dagger}-5\partial_Ah_6^i\partial^Ah_2^{i\dagger}
\nonumber\\
&~~~~~~~~~~~~+16\partial_Ah_1^i\partial^Ah_3^{i\dagger}-20\partial_Ah_2^i\partial^Ah_3^{i\dagger}+16\partial_Ah_3^i\partial^Ah_3^{i\dagger}
-20\partial_Ah_4^i\partial^Ah_3^{i\dagger}+4\partial_Ah_5^i\partial^Ah_3^{i\dagger}+4\partial_Ah_6^i\partial^Ah_3^{i\dagger}
\nonumber\\
&~~~~~~~~~~~~-20\partial_Ah_1^i\partial^Ah_4^{i\dagger}+25\partial_Ah_2^i\partial^Ah_4^{i\dagger}-20\partial_Ah_3^i\partial^Ah_4^{i\dagger}
+25\partial_Ah_4^i\partial^Ah_4^{i\dagger}-5\partial_Ah_5^i\partial^Ah_4^{i\dagger}-5\partial_Ah_6^i\partial^Ah_4^{i\dagger}
\nonumber\\
&~~~~~~~~~~~~+4\partial_Ah_1^i\partial^Ah_5^{i\dagger}-5\partial_Ah_2^i\partial^Ah_5^{i\dagger}+4\partial_Ah_3^i\partial^Ah_5^{i\dagger}
-5\partial_Ah_4^i\partial^Ah_5^{i\dagger}+\partial_Ah_5^i\partial^Ah_5^{i\dagger}+\partial_Ah_6^i\partial^Ah_5^{i\dagger}
\nonumber\\
&~~~~~~~~~~~~+4\partial_Ah_1^i\partial^Ah_6^{i\dagger}-5\partial_Ah_2^i\partial^Ah_6^{i\dagger}+4\partial_Ah_3^i\partial^Ah_6^{i\dagger}
-5\partial_Ah_4^i\partial^Ah_6^{i\dagger}+\partial_Ah_5^i\partial^Ah_6^{i\dagger}+\partial_Ah_6^i\partial^Ah_6^{i\dagger}
\Big]\nonumber
\end{align}
\begin{align}
\text{Now write}~~~~{\mathcal L}_{kin}^{\mathsf{(5_{320})}}&=-\partial_A\Phi^{\mathsf{(5)}\dagger}K^{\mathsf{(5)}}\partial^A\Phi^{\mathsf{(5)}},\nonumber\\
\text{where,}~~~~\Phi^{\mathsf{(5)}\dagger}&=\Big(h_1^{i\dagger}~~~h_2^{i\dagger}~~~h_3^{i\dagger}~~~h_4^{i*}~~~h_5^{i\dagger}~~~h_6^{i\dagger}\Big)\nonumber\\
\text{and}~~~~K^{\mathsf{(5)}}&=\frac{1}{3456}\begin{pmatrix}
16 & -20 & 16 & -20 & 4 & 4 \\
-20 & 25 & -20 & 25 & -5 & -5 \\
16 & -20 & 16 & -20 & 4 & 4 \\
-20 & 25 & -20 & 25 & -5 & -5 \\
4 & -5 & 4 & -5 & 1 & 1 \\
4 & -5 & 4 & -5 & 1 & 1
\end{pmatrix}.\nonumber\\
\text{The diagonal matrix is}~~~~K_{D}^{\mathsf{(5)}}&=\displaystyle\begin{pmatrix}
\displaystyle\frac{7}{288} & \displaystyle 0 & \displaystyle 0 & \displaystyle 0 & \displaystyle 0 & \displaystyle 0\\
\displaystyle 0 & \displaystyle 0 & \displaystyle 0 & \displaystyle 0 & \displaystyle 0 & \displaystyle 0\\
\displaystyle 0 & \displaystyle 0 & \displaystyle 0 & \displaystyle 0 & \displaystyle 0 & \displaystyle 0\\
\displaystyle 0 & \displaystyle 0 & \displaystyle 0 & \displaystyle 0 & \displaystyle 0 & \displaystyle 0\\
\displaystyle 0 & \displaystyle 0 & \displaystyle 0 & \displaystyle 0 & \displaystyle 0 & \displaystyle 0\\
\displaystyle 0 & \displaystyle 0 & \displaystyle 0 & \displaystyle 0 & \displaystyle 0 & \displaystyle 0\\
\end{pmatrix}\nonumber\\
\text{while the unitary {diagonalizing matrix} is}~~~~ V^{\mathsf{(5)}}&= \begin{pmatrix}
 \frac{2}{\sqrt{21}} & -\frac{1}{\sqrt{17}} & -\frac{2}{3} \sqrt{\frac{2}{17}} & \frac{10}{3} \sqrt{\frac{2}{43}} & -\frac{16}{\sqrt{2537}} & \frac{10}{\sqrt{1239}} \\
 -\frac{5}{2 \sqrt{21}} & 0 & 0 & 0 & 0 & \frac{1}{2}\sqrt{\frac{59}{21}} \\
 \frac{2}{\sqrt{21}} & 0 & 0 & 0 & \sqrt{\frac{43}{59}} & \frac{10}{\sqrt{1239}} \\
 -\frac{5}{2 \sqrt{21}} & 0 & 0 & 3 \sqrt{\frac{2}{43}} & \frac{20}{\sqrt{2537}} & -\frac{25}{2 \sqrt{1239}} \\
 \frac{1}{2 \sqrt{21}} & 0 & \frac{1}{3}\sqrt{\frac{17}{2}} & \frac{5}{3 \sqrt{86}} & -\frac{4}{\sqrt{2537}} & \frac{5}{2 \sqrt{1239}} \\
 \frac{1}{2 \sqrt{21}} & \frac{4}{\sqrt{17}} & -\frac{1}{3 \sqrt{34}} & \frac{5}{3 \sqrt{86}} & -\frac{4}{\sqrt{2537}} & \frac{5}{2 \sqrt{1239}} \\
\end{pmatrix}
\nonumber\\
\text{Make the scaled unitary transformation}:~h_1^i&=12\sqrt{\frac{2}{7}}~V^{\mathsf{(5)}}_{11}~\chi^{(\mathsf{5_{320}})i},~~~~V^{\mathsf{(5)}}_{11}=\frac{2}{\sqrt{21}}\nonumber\\
h_2^i&=12\sqrt{\frac{2}{7}}~V^{\mathsf{(5)}}_{21}~\chi^{(\mathsf{5_{320}})i},~~~~V^{\mathsf{(5)}}_{21}= -\frac{5}{2 \sqrt{21}}\nonumber\\
h_3^i&=12\sqrt{\frac{2}{7}}~V^{\mathsf{(5)}}_{31}~\chi^{(\mathsf{5_{320}})i},~~~~V^{\mathsf{(5)}}_{31}= \frac{2}{\sqrt{21}}\nonumber\\
h_4^i&=12\sqrt{\frac{2}{7}}~V^{\mathsf{(5)}}_{41}~\chi^{(\mathsf{5_{320}})i},~~~~V^{\mathsf{(5)}}_{41}= -\frac{5}{2 \sqrt{21}}\nonumber\\
h_5^i&=12\sqrt{\frac{2}{7}}~V^{\mathsf{(5)}}_{51}~\chi^{(\mathsf{5_{320}})i},~~~~V^{\mathsf{(5)}}_{51}=  \frac{1}{2 \sqrt{21}}\nonumber\\
h_6^i&=12\sqrt{\frac{2}{7}}~V^{\mathsf{(5)}}_{61}~\chi^{(\mathsf{5_{320}})i},~~~~V^{\mathsf{(5)}}_{61}=  \frac{1}{2 \sqrt{21}}\nonumber\
\end{align}
\begin{eqnarray}\label{Normdiag5in320}
\begin{split}
H^{({5}_{1}^{\prime\prime})}_i =12\sqrt{\frac{2}{7}}\bigg(V^{\mathsf{(5)}}_{11}+V^{\mathsf{(5)}}_{21}\bigg)\chi^{(\mathsf{5_{320}})i},
~~~H^{({5}_{2}^{\prime\prime})}_i =12\sqrt{\frac{2}{7}}\bigg(V^{\mathsf{(5)}}_{31}+V^{\mathsf{(5)}}_{41}\bigg)\chi^{(\mathsf{5_{320}})i},
~~~H^{({5}_{3}^{\prime\prime})}_i =12\sqrt{\frac{2}{7}}\bigg(V^{\mathsf{(5)}}_{51}+V^{\mathsf{(5)}}_{61}\bigg)\chi^{(\mathsf{5_{320}})i},
\\
H^{({5}_{1})}_i =12\sqrt{\frac{2}{7}}\bigg(V^{\mathsf{(5)}}_{11}+V^{\mathsf{(5)}}_{31}\bigg)\chi^{(\mathsf{5_{320}})i},
~~~H^{({5}_{2})}_i =12\sqrt{\frac{2}{7}}\bigg(V^{\mathsf{(5)}}_{41}+V^{\mathsf{(5)}}_{61}\bigg)\chi^{(\mathsf{5_{320}})i},
~~~H^{({5}_{3})}_i =12\sqrt{\frac{2}{7}}\bigg(V^{\mathsf{(5)}}_{21}+V^{\mathsf{(5)}}_{51}\bigg)\chi^{(\mathsf{5_{320}})i},
\\
H^{({5}_{2}^{\prime})}_i =12\sqrt{\frac{2}{7}}\bigg(V^{\mathsf{(5)}}_{61}-V^{\mathsf{(5)}}_{41}\bigg)\chi^{(\mathsf{5_{320}})i},
~~~H^{({5}_{3}^{\prime})}_i =12\sqrt{\frac{2}{7}}\bigg(V^{\mathsf{(5)}}_{51}-V^{\mathsf{(5)}}_{21}\bigg)\chi^{(\mathsf{5_{320}})i}.~~~~~~~~~~~~~~~~~~~~~~~~~~~~
\end{split}
\end{eqnarray}
}}
}
{
{\allowdisplaybreaks{
\section{Details of doublet Couplings}\label{sec:G}
\subsection{Computation of $\boldsymbol{\mathsf{560\cdot560\cdot320_s}}$ and $\boldsymbol{\mathsf{\overline{560}\cdot\overline{560}\cdot320_s}}$ interactions}
We will first compute in $\mathsf{560\cdot560\cdot320_s}$: $\beta M_{\mu\nu\lambda}\Lambda^{(\mathsf{320}_s)}_{(\mu\nu)\lambda}$, where  
\[M_{\mu\nu\lambda}\equiv\beta\braket{\Theta_{\mu\sigma}^{(\mathsf{560})*}|B\Gamma_{\lambda}|\Theta_{\nu\sigma}^{(\mathsf{560})}}\]
Using Eq. (\ref{compdef320}), we get
\begin{align}\label{intermevaldoubtrp}
\beta\braket{\Theta_{\mu\sigma}^{(\mathsf{560})*}|B\Gamma_{\lambda}|\Theta_{\nu\sigma}^{(\mathsf{560})}}
\Lambda^{(\mathsf{320}_s)}_{(\mu\nu)\lambda}&=\beta\Biggl\{M_{\mu\nu\lambda}t_{(\mu\nu)\lambda}+\frac{1}{54}\bigg(M_{\beta\rho\rho}
+M_{\rho\beta\rho}-2M_{\rho\rho\beta}\bigg) \bigg(-T_{\beta \alpha\alpha} + 2 T_{\alpha\alpha\beta}
-T_{\alpha\beta\alpha}\bigg)\Biggr\}\nonumber\\
&=\beta\Biggl\{M_{\mu\nu\lambda}t_{(\mu\nu)\lambda}+\frac{1}{27}M_{\rho\rho c_i} \bigg(T_{\bar c_i \alpha\alpha}+T_{\alpha\bar c_i \alpha} - 2 T_{\alpha\alpha\bar c_i}\bigg)
+\frac{1}{27}M_{\rho\rho \bar c_i} \bigg(T_{c_i \alpha\alpha}+T_{\alpha c_i \alpha} - 2 T_{\alpha\alpha c_i}\bigg)\Biggr\}\nonumber\\
&=\beta\Biggl\{\frac{1}{108}\biggl[M_{\rho\rho c_i} \Bigl(H^{(\overline{5}_{1}^{\prime\prime})}_i+H^{(\overline{5}_{2}^{\prime\prime})}_i - 2 H^{(\overline{5}_{3}^{\prime\prime})}_i\Bigr)+M_{\rho\rho \bar c_i}  \Bigl(H^{({5}_{1}^{\prime\prime})i}+H^{({5}_{2}^{\prime\prime})i} - 2 H^{({5}_{3}^{\prime\prime})i}\Bigr)\biggr]\nonumber\\
&~~~~~~~~~+M_{\bar c_i \bar c_j c_k}~t_{(c_ic_j)\bar c_k}+M_{c_i c_j \bar c_k}~t_{(\bar c_i\bar c_j)c_k}+2M_{c_i \bar c_j \bar c_k}~t_{(\bar c_ic_j)c_k}+2M_{\bar c_i c_j c_k}~t_{(c_i\bar c_j)\bar c_k}+\dots\Biggr\},
\end{align}
where we have used Eq. (\ref{all5pletsin320}). The ellipsis above represent terms containing fields which do not contain $\mathsf{SU(2)_L}$ doublets. Evaluating each of the terms in Eq.(\ref{intermevaldoubtrp}) using Eqs. (\ref{tensor1} - \ref{tensor3}, \ref{Normalized SU(5) fields in 560}, \ref{SU(5)tracelessparts320}, \ref{Normdiag70in320} - \ref{Normdiag5in320}) we get
\begin{align*}
&\frac{1}{108}\biggl[M_{\rho\rho c_i} \Bigl(H^{(\overline{5}_{1}^{\prime\prime})}_i+H^{(\overline{5}_{2}^{\prime\prime})}_i - 2 H^{(\overline{5}_{3}^{\prime\prime})}_i\Bigr)+M_{\rho\rho \bar c_i}  \Bigl(H^{({5}_{1}^{\prime\prime})i}+H^{({5}_{2}^{\prime\prime})i} - 2 H^{({5}_{3}^{\prime\prime})i}\Bigr)\biggr]\\
&=\frac{1}{108}\biggl\{\braket{\Theta_{\rho\sigma}^{(\mathsf{560})*}|B~b_i|\Theta_{\rho\sigma}^{(\mathsf{560})}} \Bigl[H^{(\overline{5}_{1}^{\prime\prime})}_i+H^{(\overline{5}_{2}^{\prime\prime})}_i - 2 H^{(\overline{5}_{3}^{\prime\prime})}_i\Bigr]+\braket{\Theta_{\rho\sigma}^{(\mathsf{560})*}|B~b_i^{\dagger}|\Theta_{\rho\sigma}^{(\mathsf{560})}}  \Bigl[H^{({5}_{1}^{\prime\prime})i}+H^{({5}_{2}^{\prime\prime})i} - 2 H^{({5}_{3}^{\prime\prime})i}\Bigr]\biggr\}\\
&={\frac{i}{126\sqrt{2}}}~ \mathbf{H}^{(\mathsf{24_{560}})k}_i~ \mathbf{H}^{(\mathsf{45_{560}})ij}_k ~\chi^{(\mathsf{\overline{5}_{320}})}_j
~-~\frac{i}{42\sqrt{6}}~ \mathbf{H}^{(\mathsf{75_{560}})kl}_{ij}~ \mathbf{H}^{(\mathsf{45_{560}})ij}_k ~\chi^{(\mathsf{\overline{5}_{320}})}_l
~-~\frac{97i}{63\sqrt{7210}}~ \mathbf{H}^{(\mathsf{1_{560}})}~ \mathbf{H}^{(\mathsf{\overline{5}_{560}})}_i ~\chi^{(\mathsf{{5}_{320}})i}\\
&~~~+~{\frac{i}{12\sqrt{2163}}}~\mathbf{H}^{(\mathsf{24_{560}})i}_j ~ \mathbf{H}^{(\mathsf{\overline{5}_{560}})}_i~ \chi^{(\mathsf{{5}_{320}})j}
~-~{\frac{19i}{126\sqrt{145}}}~ \mathbf{H}^{(\mathsf{24_{560}})i}_k ~ \mathbf{H}^{(\mathsf{\overline{45}_{560}})k}_{ij}~\chi^{(\mathsf{{5}_{320}})j}
~-~\frac{i}{84\sqrt{435}}~\mathbf{H}^{(\mathsf{75_{560}})ij}_{kl} ~ \mathbf{H}^{(\mathsf{\overline{45}_{560}})k}_{ij}~ \chi^{(\mathsf{{5}_{320}})l}\\
&~~~+~{\frac{i}{84}}~\mathbf{H}^{(\mathsf{24_{560}})i}_k ~ \mathbf{H}^{(\mathsf{\overline{70}_{560}})k}_{ij}~ \chi^{(\mathsf{{5}_{320}})j}
\end{align*} 
\begin{align*}
M_{\bar c_i \bar c_j c_k}~t_{(c_ic_j)\bar c_k}&=\braket{\Theta_{\bar c_i\sigma}^{(\mathsf{560})*}|B~b_k|\Theta_{\bar c_j\sigma}^{(\mathsf{560})}}t_{(c_ic_j)\bar c_k}\\
&=~-~{i\sqrt{\frac{6}{721}}}~ \mathbf{H}^{(\mathsf{24_{560}})k}_j~ \mathbf{H}^{(\mathsf{\overline{5}_{560}})}_i ~\chi^{(\mathsf{{70}_{320}})ij}_k
~+~{2i\sqrt{\frac{2}{145}}}~ \mathbf{H}^{(\mathsf{24_{560}})j}_l~ \mathbf{H}^{(\mathsf{\overline{45}_{560}})i}_{jk} ~\chi^{(\mathsf{{70}_{320}})kl}_i\\
&~~~~~+~4i\sqrt{\frac{6}{145}}~ \mathbf{H}^{(\mathsf{75_{560}})jl}_{im}~ \mathbf{H}^{(\mathsf{\overline{45}_{560}})i}_{jk} ~\chi^{(\mathsf{{70}_{320}})km}_l
~+~\frac{i}{3\sqrt{15}}~ \mathbf{H}^{(\mathsf{1_{560}})}~ \mathbf{H}^{(\mathsf{\overline{70}_{560}})k}_{ij} ~\chi^{(\mathsf{{70}_{320}})ij}_k\\
&~~~~~+~{\frac{i}{3\sqrt{2}}}~\biggl(\mathbf{H}^{(\mathsf{24_{560}})l}_j~ \mathbf{H}^{(\mathsf{\overline{70}_{560}})k}_{li}+
\mathbf{H}^{(\mathsf{24_{560}})k}_l~ \mathbf{H}^{(\mathsf{\overline{70}_{560}})l}_{ij}\biggr)~\chi^{(\mathsf{{70}_{320}})ij}_k\\
&~~~~~+~i\sqrt{\frac{2}{3}}~\mathbf{H}^{(\mathsf{75_{560}})jl}_{km}~ \mathbf{H}^{(\mathsf{\overline{70}_{560}})k}_{ij}~\chi^{(\mathsf{{70}_{320}})im}_l
~+~\frac{16i}{7}\sqrt{\frac{2}{3605}}~ \mathbf{H}^{(\mathsf{1_{560}})}~ \mathbf{H}^{(\mathsf{\overline{5}_{560}})}_{i} ~\chi^{(\mathsf{{5}_{320}})i}\\
&~~~~~+~{\frac{2i}{7}\sqrt{\frac{3}{721}}}~ \mathbf{H}^{(\mathsf{24_{560}})i}_j~ \mathbf{H}^{(\mathsf{\overline{5}_{560}})}_{i} ~\chi^{(\mathsf{{5}_{320}})j}
~+~{\frac{4i}{7}\sqrt{\frac{5}{29}}}~ \mathbf{H}^{(\mathsf{24_{560}})i}_k~ \mathbf{H}^{(\mathsf{\overline{45}_{560}})k}_{ij} ~\chi^{(\mathsf{{5}_{320}})j}\\
&~~~~~-~\frac{8i}{7}\sqrt{\frac{3}{145}}~ \mathbf{H}^{(\mathsf{75_{560}})ij}_{kl}~ \mathbf{H}^{(\mathsf{\overline{45}_{560}})k}_{ij} ~\chi^{(\mathsf{{5}_{320}})l}
~+~{\frac{i}{7}}~ \mathbf{H}^{(\mathsf{24_{560}})k}_i~ \mathbf{H}^{(\mathsf{\overline{70}_{560}})k}_{ij} ~\chi^{(\mathsf{{5}_{320}})j}
\end{align*}
\begin{align*}
M_{c_i c_j \bar c_k}~t_{(\bar c_i\bar c_j)c_k}&=\braket{\Theta_{c_i\sigma}^{(\mathsf{560})*}|B~b_k^{\dagger}|\Theta_{c_j\sigma}^{(\mathsf{560})}}t_{(\bar c_i\bar c_j)c_k}\\
&=~-~{i}~ \mathbf{H}^{(\mathsf{24_{560}})i}_j~ \mathbf{H}^{(\mathsf{{45}_{560}})jk}_{l} ~\chi^{(\mathsf{\overline{70}_{320}})l}_{ik}
~+~{\frac{i\sqrt{2}}{7}}~\mathbf{H}^{(\mathsf{24_{560}})i}_j~ \mathbf{H}^{(\mathsf{{45}_{560}})jk}_{i} ~\chi^{(\mathsf{\overline{5}_{320}})}_k
\end{align*}
\begin{align*}
2M_{c_i \bar c_j \bar c_k}~t_{(\bar c_ic_j)c_k}&=2\braket{\Theta_{c_i\sigma}^{(\mathsf{560})*}|B~b_k^{\dagger}|\Theta_{\bar c_j\sigma}^{(\mathsf{560})}}t_{(\bar c_ic_j)c_k}\\
&=~~~\frac{83i}{7}\sqrt{\frac{2}{3605}}~ \mathbf{H}^{(\mathsf{1_{560}})}~ \mathbf{H}^{(\mathsf{\overline{5}_{560}})}_{i} ~\chi^{(\mathsf{{5}_{320}})i}
~-~{\frac{305i}{56}\sqrt{\frac{3}{721}}}~ \mathbf{H}^{(\mathsf{24_{560}})i}_j~ \mathbf{H}^{(\mathsf{\overline{5}_{560}})}_{i} ~\chi^{(\mathsf{{5}_{320}})j}\\
&~~~~~+~{\frac{671i}{84\sqrt{145}}} \mathbf{H}^{(\mathsf{24_{560}})i}_j~ \mathbf{H}^{(\mathsf{\overline{45}_{560}})j}_{ik} ~\chi^{(\mathsf{{5}_{320}})k}
~+~\frac{2i}{7\sqrt{435}}~ \mathbf{H}^{(\mathsf{75_{560}})ij}_{kl}~ \mathbf{H}^{(\mathsf{\overline{45}_{560}})k}_{ij} ~\chi^{(\mathsf{{5}_{320}})l}\\
&~~~~~+~{\frac{2i}{7}}~ \mathbf{H}^{(\mathsf{24_{560}})i}_{j}~ \mathbf{H}^{(\mathsf{\overline{70}_{560}})j}_{ik} ~\chi^{(\mathsf{{5}_{320}})k}
~+~{\frac{31i}{6}\sqrt{\frac{7}{206}}}~ \mathbf{H}^{(\mathsf{24_{560}})i}_j~ \mathbf{H}^{(\mathsf{\overline{5}_{560}})}_{k} ~\chi^{(\mathsf{{45}_{320}})jk}_i\\
&~~~~~-~\frac{i}{\sqrt{4326}}~ \mathbf{H}^{(\mathsf{75_{560}})ij}_{kl}~ \mathbf{H}^{(\mathsf{\overline{5}_{560}})}_{i} ~\chi^{(\mathsf{{45}_{320}})kl}_j
~+~\frac{49i}{15\sqrt{29}}~ \mathbf{H}^{(\mathsf{1_{560}})}~ \mathbf{H}^{(\mathsf{\overline{45}_{560}})i}_{jk} ~\chi^{(\mathsf{{45}_{320}})jk}_i\\
&~~~~~+~{\frac{i}{\sqrt{870}}}~ \mathbf{H}^{(\mathsf{24_{560}})i}_l~ \mathbf{H}^{(\mathsf{\overline{45}_{560}})k}_{ij} ~\chi^{(\mathsf{{45}_{320}})lj}_k
~+~{\frac{71i}{2\sqrt{870}}}~ \mathbf{H}^{(\mathsf{24_{560}})i}_l~ \mathbf{H}^{(\mathsf{\overline{45}_{560}})l}_{jk} ~\chi^{(\mathsf{{45}_{320}})jk}_i\\
&~~~~~+~\frac{i}{2\sqrt{290}}~ \mathbf{H}^{(\mathsf{75_{560}})ij}_{kl}~ \mathbf{H}^{(\mathsf{\overline{45}_{560}})m}_{ij} ~\chi^{(\mathsf{{45}_{320}})kl}_m
~-~\frac{i}{\sqrt{290}}~ \mathbf{H}^{(\mathsf{75_{560}})ij}_{kl}~ \mathbf{H}^{(\mathsf{\overline{45}_{560}})k}_{mi} ~\chi^{(\mathsf{{45}_{320}})ml}_j\\
&~~~~~+~{\frac{7i}{2}\sqrt{\frac{7}{618}}}~ \mathbf{H}^{(\mathsf{24_{560}})i}_j~ \mathbf{H}^{(\mathsf{\overline{5}_{560}})}_{k} ~\chi^{(\mathsf{{70}_{320}})jk}_i
~+~{\frac{i}{3\sqrt{290}}} \mathbf{H}^{(\mathsf{24_{560}})i}_j~ \mathbf{H}^{(\mathsf{\overline{45}_{560}})l}_{ik} ~\chi^{(\mathsf{{70}_{320}})jk}_l\\
&~~~~~-~\frac{i}{\sqrt{870}}~ \mathbf{H}^{(\mathsf{75_{560}})ij}_{kl}~ \mathbf{H}^{(\mathsf{\overline{45}_{560}})l}_{jm} ~\chi^{(\mathsf{{70}_{320}})km}_i
~-~\frac{4i}{3\sqrt{15}}~ \mathbf{H}^{(\mathsf{1_{560}})}~ \mathbf{H}^{(\mathsf{\overline{70}_{560}})i}_{jk} ~\chi^{(\mathsf{{70}_{320}})jk}_i\\
&~~~~~-~{\frac{i}{\sqrt{2}}}~ \mathbf{H}^{(\mathsf{24_{560}})i}_j~ \mathbf{H}^{(\mathsf{\overline{70}_{560}})j}_{kl} ~\chi^{(\mathsf{{70}_{320}})kl}_i
\end{align*}
\begin{align*}
2M_{\bar c_i c_j c_k}~t_{(c_i\bar c_j)\bar c_k}&=2 \braket{\Theta_{\bar c_i\sigma}^{(\mathsf{560})*}|B~b_k|\Theta_{c_j\sigma}^{(\mathsf{560})}} t_{(c_i\bar c_j)\bar c_k}\\
&=~~~{\frac{29i}{84\sqrt{2}}}~ \mathbf{H}^{(\mathsf{24_{560}})i}_{j}~ \mathbf{H}^{(\mathsf{{45}_{560}})jk}_{i} ~\chi^{(\mathsf{\overline{5}_{320}})}_k
~+~\frac{5i}{14\sqrt{6}}~ \mathbf{H}^{(\mathsf{75_{560}})ij}_{kl}~ \mathbf{H}^{(\mathsf{{45}_{560}})kl}_{i} ~\chi^{(\mathsf{\overline{5}_{320}})}_i\\
&~~~~~-~\frac{i}{3\sqrt{10}}~ \mathbf{H}^{(\mathsf{1_{560}})}~ \mathbf{H}^{(\mathsf{{45}_{560}})ij}_{k} ~\chi^{(\mathsf{\overline{45}_{320}})k}_{ij}
~-~{\frac{i}{2\sqrt{3}}}~ \mathbf{H}^{(\mathsf{24_{560}})i}_j~ \mathbf{H}^{(\mathsf{{45}_{560}})jk}_{l} ~\chi^{(\mathsf{\overline{45}_{320}})l}_{ik}\\
&~~~~~-~{\frac{i}{2\sqrt{3}}}~ \mathbf{H}^{(\mathsf{24_{560}})i}_j~ \mathbf{H}^{(\mathsf{{45}_{560}})kl}_{i} ~\chi^{(\mathsf{\overline{45}_{320}})j}_{kl}
~+~i~\mathbf{H}^{(\mathsf{75_{560}})ij}_{kl}~ \mathbf{H}^{(\mathsf{{45}_{560}})km}_{j} ~\chi^{(\mathsf{\overline{45}_{320}})l}_{im}\\
&~~~~~+~{\frac{i}{6}}~ \mathbf{H}^{(\mathsf{24_{560}})i}_j~ \mathbf{H}^{(\mathsf{{45}_{560}})jk}_{l} ~\chi^{(\mathsf{\overline{70}_{320}})l}_{ik}
~+~\frac{i}{\sqrt{3}}~ \mathbf{H}^{(\mathsf{75_{560}})ij}_{kl}~ \mathbf{H}^{(\mathsf{{45}_{560}})km}_{i} ~\chi^{(\mathsf{\overline{70}_{320}})l}_{jm}
\end{align*}
Using Eqs. (\ref{GUTsymmetryB5}, \ref{GUTsymmetryB6}, \ref{Extraction of 5+bar5 doublets} - \ref{Normalized doublets in 5+bar5 and 45+bar45}, \ref{Extraction of 70+bar70 doublets}, \ref{Normalized doublets in 70+bar70}) in each of the five expressions above, Eq. (\ref{intermevaldoubtrp}) gives
\begin{eqnarray}\label{560560320 coupling}
\beta\braket{\Theta_{\mu\sigma}^{(\mathsf{560})*}|B\Gamma_{\lambda}|\Theta_{\nu\sigma}^{(\mathsf{560})}}\Lambda^{(\mathsf{320})}
_{(\mu\nu)\lambda} &=&i\beta\Biggl\{{\frac{1}{144\sqrt{2}}\biggl(23\sqrt{5}~{{\mathcal V}}_{24}-16~{{\mathcal V}}_{75}\biggr)}~{}^{({\mathsf{{45}}}_{\mathsf{560}})}\!{{\mathbf D}}^{a}~{}^{({\mathsf{\overline{5}}}_{\mathsf{320}})}\!{{\mathbf D}}_{a}\nonumber\\
&&~~~~+{\frac{1}{504\sqrt{7210}}\biggl(13480~{{\mathcal V}}_{1}+7761~{{\mathcal V}}_{24}\biggr)}~{}^{({\mathsf{\overline{5}}}_{\mathsf{560}})}\!{{\mathbf D}}_{a}~{}^{({\mathsf{{5}}}_{\mathsf{320}})}\!{{\mathbf D}}^{a}\nonumber\\
&&~~~~+{\frac{1}{1008\sqrt{29}}\biggl(2695~{{\mathcal V}}_{24}+212\sqrt{5}~{{\mathcal V}}_{75}\biggr)}~{}^{({\mathsf{\overline{45}}}_{\mathsf{560}})}\!{{\mathbf D}}_{a}~{}^{({\mathsf{{5}}}_{\mathsf{320}})}\!{{\mathbf D}}^{a}\nonumber\\
&&~~~~+{\frac{1}{120\sqrt{174}}\biggl(65~{{\mathcal V}}_{24}+184\sqrt{5}~{{\mathcal V}}_{75}\biggr)}~{}^{({\mathsf{\overline{45}}}_{\mathsf{560}})}\!{{\mathbf D}}_{a}~{}^{({\mathsf{{70}}}_{\mathsf{320}})}\!{{\mathbf D}}^{a}\nonumber\\
&&~~~~+{\frac{1}{40\sqrt{3}}\biggl(-8\sqrt{5}~{{\mathcal V}}_{1}-\sqrt{5}~{{\mathcal V}}_{24}+20~{{\mathcal V}}_{75}\biggr)}~{}^{({\mathsf{\overline{70}}}_{\mathsf{560}})}\!{{\mathbf D}}_{a}~{}^{({\mathsf{{70}}}_{\mathsf{320}})}\!{{\mathbf D}}^{a}
\nonumber\\
&&~~~~+{\frac{1}{48\sqrt{3}}\biggl(-5\sqrt{5}~{{\mathcal V}}_{24}+16~{{\mathcal V}}_{75}\biggr)}~{}^{({\mathsf{{45}}}_{\mathsf{560}})}\!{{\mathbf D}}^{a}~{}^{({\mathsf{\overline{70}}}_{\mathsf{320}})}\!{{\mathbf D}}_{a}\nonumber\\
&&~~~~+{\frac{1}{60\sqrt{29}}\biggl(196~{{\mathcal V}}_{1}-129~{{\mathcal V}}_{24}+2\sqrt{5}~{{\mathcal V}}_{75}\biggr)}~{}^{({\mathsf{\overline{45}}}_{\mathsf{560}})}\!{{\mathbf D}}_{a}~{}^{({\mathsf{{45}}}_{\mathsf{320}})}\!{{\mathbf D}}^{a}
\nonumber\\
&&~~~~+{\frac{1}{24\sqrt{1442}}\biggl(217\sqrt{5}~{{\mathcal V}}_{24}-8~{{\mathcal V}}_{75}\biggr)}~{}^{({\mathsf{\overline{5}}}_{\mathsf{560}})}\!{{\mathbf D}}_{a}~{}^{({\mathsf{{45}}}_{\mathsf{320}})}\!{{\mathbf D}}^{a}\nonumber\\
&&~~~~+{\frac{1}{48\sqrt{10}}\biggl(-16~{{\mathcal V}}_{1}+33~{{\mathcal V}}_{24}\biggr)}~{}^{({\mathsf{{45}}}_{\mathsf{560}})}\!{{\mathbf D}}^{a}~{}^{({\mathsf{\overline{45}}}_{\mathsf{320}})}\!{{\mathbf D}}_{a}
\nonumber\\
&&~~~~+{\frac{37}{168}\sqrt{\frac{5}{2}}~{{\mathcal V}}_{24}}~{}^{({\mathsf{\overline{70}}}_{\mathsf{560}})}\!{{\mathbf D}}_{a}~{}^{({\mathsf{{5}}}_{\mathsf{320}})}\!{{\mathbf D}}^{a}\nonumber\\
&&~~~~+{\frac{37}{8}\sqrt{\frac{5}{2163}}~{{\mathcal V}}_{24}}~{}^{({\mathsf{\overline{5}}}_{\mathsf{560}})}\!{{\mathbf D}}_{a}~{}^{({\mathsf{{70}}}_{\mathsf{320}})}\!{{\mathbf D}}^{a}+\cdots\Biggr\}
\end{eqnarray}
The results for ${\mathsf{\overline{560}}\cdot\mathsf{\overline{560}}\cdot\mathsf{320}}$ can now be easily obtained from $\mathsf{560}\cdot\mathsf{560}\cdot\mathsf{320}$ by simply replacing barred with unbarred fields and vice versa:
\begin{eqnarray}\label{bar560bar560320 coupling}
\overline{\beta}\braket{\overline{\Theta}_{\mu\sigma}^{(\mathsf{\overline{560}})*}|B\Gamma_{\lambda}|\overline{\Theta}_{\nu\sigma}
^{(\mathsf{\overline{560}})}}\Lambda^{(\mathsf{320})}
_{(\mu\nu)\lambda} &=&i\overline{\beta}\Biggl\{{\frac{1}{144\sqrt{2}}\biggl(23\sqrt{5}~{\overline{\mathcal V}}_{24}-16~{\overline{\mathcal V}}_{75}\biggr)}~{}^{({\mathsf{\overline{45}}}_{\mathsf{\overline{560}}})}\!{{\mathbf D}}_{a}~{}^{({\mathsf{{5}}}_{\mathsf{320}})}\!{{\mathbf D}}^{a}\nonumber\\
&&~~~~+{\frac{1}{504\sqrt{7210}}\biggl(13480~{\overline{\mathcal V}}_{1}+7761~{\overline{\mathcal V}}_{24}\biggr)}~{}^{({\mathsf{{5}}}_{\mathsf{\overline{560}}})}\!{{\mathbf D}}^{a}~{}^{({\mathsf{\overline{5}}}_{\mathsf{320}})}\!{{\mathbf D}}_{a}\nonumber\\
&&~~~~+{\frac{1}{1008\sqrt{29}}\biggl(2965~{\overline{\mathcal V}}_{24}+212\sqrt{5}~{\overline{\mathcal V}}_{75}\biggr)}~{}^{({\mathsf{{45}}}_{\mathsf{\overline{560}}})}\!{{\mathbf D}}^{a}~{}^{({\mathsf{\overline{5}}}_{\mathsf{320}})}\!{{\mathbf D}}_{a}\nonumber\\
&&~~~~+{\frac{1}{120\sqrt{174}}\biggl(65~{\overline{\mathcal V}}_{24}+184\sqrt{5}~{\overline{\mathcal V}}_{75}\biggr)}~{}^{({\mathsf{{45}}}_{\mathsf{\overline{560}}})}\!{{\mathbf D}}^{a}~{}^{({\mathsf{\overline{70}}}_{\mathsf{320}})}\!{{\mathbf D}}_{a}\nonumber\\
&&~~~~+{\frac{1}{40\sqrt{3}}\biggl(-8\sqrt{5}~{\overline{\mathcal V}}_{1}-\sqrt{5}~{\overline{\mathcal V}}_{24}+20~{\overline{\mathcal V}}_{75}\biggr)}~{}^{({\mathsf{{70}}}_{\mathsf{\overline{560}}})}\!{{\mathbf D}}^{a}~{}^{({\mathsf{\overline{70}}}_{\mathsf{320}})}\!{{\mathbf D}}_{a}
\nonumber\\
&&~~~~+{\frac{1}{48\sqrt{3}}\biggl(-5\sqrt{5}~{\overline{\mathcal V}}_{24}+16~{\overline{\mathcal V}}_{75}\biggr)}~{}^{({\mathsf{\overline{45}}}_{\mathsf{\overline{560}}})}\!{{\mathbf D}}_{a}~{}^{({\mathsf{{70}}}_{\mathsf{320}})}\!{{\mathbf D}}^{a}\nonumber\\
&&~~~~+{\frac{1}{60\sqrt{29}}\biggl(196~{\overline{\mathcal V}}_{1}-129~{\overline{\mathcal V}}_{24}+2\sqrt{5}~{\overline{\mathcal V}}_{75}\biggr)}~{}^{({\mathsf{{45}}}_{\mathsf{\overline{560}}})}\!{{\mathbf D}}^{a}~{}^{({\mathsf{\overline{45}}}_{\mathsf{320}})}\!{{\mathbf D}}_{a}
\nonumber\\
&&~~~~+{\frac{1}{24\sqrt{1442}}\biggl(217\sqrt{5}~{\overline{\mathcal V}}_{24}-8~{\overline{\mathcal V}}_{75}\biggr)}~{}^{({\mathsf{{5}}}_{\mathsf{\overline{560}}})}\!{{\mathbf D}}^{a}~{}^{({\mathsf{\overline{45}}}_{\mathsf{320}})}\!{{\mathbf D}}_{a}\nonumber\\
&&~~~~+{\frac{1}{48\sqrt{10}}\biggl(-16~{\overline{\mathcal V}}_{1}+33~{\overline{\mathcal V}}_{24}\biggr)}~{}^{({\mathsf{\overline{45}}}_{\mathsf{\overline{560}}})}\!{{\mathbf D}}_{a}~{}^{({\mathsf{{45}}}_{\mathsf{320}})}\!{{\mathbf D}}^{a}
\nonumber\\
&&~~~~+{\frac{37}{168}\sqrt{\frac{5}{2}}~{\overline{\mathcal V}}_{24}}~{}^{({\mathsf{{70}}}_{\mathsf{\overline{560}}})}\!{{\mathbf D}}^{a}~{}^{({\mathsf{\overline{5}}}_{\mathsf{320}})}\!{{\mathbf D}}_{a}\nonumber\\
&&~~~~+{\frac{37}{8}\sqrt{\frac{5}{2163}}~{\overline{\mathcal V}}_{24}}~{}^{({\mathsf{{5}}}_{\mathsf{\overline{560}}})}\!{{\mathbf D}}^{a}~{}^{({\mathsf{\overline{70}}}_{\mathsf{320}})}\!{{\mathbf D}}_{a}+\cdots\Biggr\}
\end{eqnarray}
Again the ellipsis above represent terms containing fields which are not $\mathsf{SU(2)_L}$ doublets.

\subsection{Computation of $\boldsymbol{\mathsf{560\cdot560\cdot10_1}$, $\mathsf{\overline{560}\cdot\overline{560}\cdot\big(10_1+10_2\big)}}$ and $\boldsymbol{\mathsf{560\cdot\overline{560}}}$ interactions}
Using the Basic Theorem (\cite{Nath:2001uw}), we can expand the interaction $\alpha\braket{\Theta_{\mu\nu}^{(\mathsf{560})*}|B\Gamma_{\rho}|\Theta_{\mu\nu}^{(\mathsf{560})}}\Omega^{(\mathsf{10}_1)}_{\rho}\\
+\sum_{r=1}^2\overline{\alpha}_r\braket{\overline{\Theta}_{\mu\nu}^{(\overline{\mathsf{560}})*}|B\Gamma_{\rho}|
\overline{\Theta}_{\mu\nu}^{(\overline{\mathsf{560}})}}
\Omega^{(\mathsf{10}_r)}_{\rho}$ as follows:
\begin{eqnarray}
&&\alpha\braket{\Theta_{\mu\nu}^{(\mathsf{560})*}|B\Gamma_{\rho}|\Theta_{\mu\nu}^{(\mathsf{560})}}\Omega^{(\mathsf{10}_1)}_{\rho}
+\sum_{r=1}^2\overline{\alpha}_r\braket{\overline{\Theta}_{\mu\nu}^{(\overline{\mathsf{560}})*}|B\Gamma_{\rho}|
\overline{\Theta}_{\mu\nu}^{(\overline{\mathsf{560}})}}
\Omega^{(\mathsf{10}_r)}_{\rho}\nonumber\\
&&=2\alpha\Biggl[\biggl(\braket{\Theta_{c_ic_j}^{(\mathsf{560})*}|Bb_k|\Theta_{\overline{c}_i\overline{c}_j}^{(\mathsf{560})}}
-\braket{\Theta_{c_i\overline{c}_j}^{(\mathsf{560})*}|Bb_k|\Theta_{c_j\overline{c}_i}^{(\mathsf{560})}}\biggr)\Omega^{(\mathsf{10}_1)}_{\overline{c}_k}
+\biggl(\braket{\Theta_{c_ic_j}^{(\mathsf{560})*}|Bb_k^{\dagger}|\Theta_{\overline{c}_i\overline{c}_j}^{(\mathsf{560})}}
-\braket{\Theta_{c_i\overline{c}_j}^{(\mathsf{560})*}|Bb_k^{\dagger}|\Theta_{c_j\overline{c}_i}^{(\mathsf{560})}}\biggr)\Omega^{(\mathsf{10}_1)}_{c_k}\Biggr]
\nonumber\\
&&~~~~+ 2\sum_{r=1}^2 \overline{\alpha}_r\Biggl[\biggl(\braket{\overline{\Theta}_{c_ic_j}^{(\overline{\mathsf{560}})*}|Bb_k|
\overline{\Theta}_{\overline{c}_i\overline{c}_j}^{(\overline{\mathsf{560}})}}
-\braket{\overline{\Theta}_{c_i\overline{c}_j}^{(\overline{\mathsf{560}})*}|Bb_k|\overline{\Theta}_{c_j\overline{c}_i}^{(\overline{\mathsf{560}})}}\biggr)
\Omega^{(\mathsf{10}_r)}_{\overline{c}_k}
+\biggl(\braket{\overline{\Theta}_{c_ic_j}^{(\overline{\mathsf{560}})*}|Bb_k^{\dagger}|
\overline{\Theta}_{\overline{c}_i\overline{c}_j}^{(\overline{\mathsf{560}})}}
-\braket{\overline{\Theta}_{c_i\overline{c}_j}^{(\overline{\mathsf{560}})*}|Bb_k^{\dagger}|\overline{\Theta}
_{c_j\overline{c}_i}^{(\overline{\mathsf{560}})}}\biggl)
\Omega^{(\mathsf{10}_r)}_{c_k}\Biggr]\nonumber
\end{eqnarray}
Using Eqs. (\ref{tensor1} - \ref{tensor6}, \ref{Normalized SU(5) fields in 560}, \ref{Normalized 5+bar5 fields in 10plet}) above, we get
\begin{eqnarray}
&&\alpha\braket{\Theta_{\mu\nu}^{(\mathsf{560})*}|B\Gamma_{\rho}|\Theta_{\mu\nu}^{(\mathsf{560})}}\Omega^{(\mathsf{10}_1)}_{\rho}
+\sum_{r=1}^2\overline{\alpha}_r\braket{\overline{\Theta}_{\mu\nu}^{(\overline{\mathsf{560}})*}|B\Gamma_{\rho}|
\overline{\Theta}_{\mu\nu}^{(\overline{\mathsf{560}})}}
\Omega^{(\mathsf{10}_r)}_{\rho}\nonumber\\
&&=i\alpha\Biggl[\biggl(\sqrt{2}{\mathbf{H}}^{(\mathsf{45_{560}})ij}_k{\mathbf{H}}^{(\mathsf{{75}_{560}})kl}_{ij}-
{\sqrt{\frac{2}{3}}}{\mathbf{H}}^{(\mathsf{{45}_{560}})kl}_j
{\mathbf{H}}^{(\mathsf{{24}_{560}})j}_{k}\biggr)\mathbf{H}^{(\mathsf{\overline{5}_{10_1}})}_{l}
+\biggl(-{\frac{4}{\sqrt{721}}}{\mathbf{H}}^{(\mathsf{\overline{5}_{560}})}_k{\mathbf{H}}^{(\mathsf{{24}_{560}})k}_{l}\nonumber\\
&&~~~~~~+97\sqrt{\frac{2}{10815}}{\mathbf{H}}^{(\mathsf{\overline{5}_{560}})}_l
{\mathbf{H}}^{(\mathsf{{1}_{560}})}
+\frac{1}{\sqrt{145}}{\mathbf{H}}^{(\mathsf{\overline{45}_{560}})i}_{jk}{\mathbf{H}}^{(\mathsf{{75}_{560}})jk}_{il}+
{4\sqrt{\frac{5}{87}}}{\mathbf{H}}^{(\mathsf{\overline{45}_{560}})j}_{kl}
{\mathbf{H}}^{(\mathsf{{24}_{560}})k}_j\nonumber\\
&&~~~~~~-{\frac{\sqrt{3}}{2}}{\mathbf{H}}^{(\mathsf{\overline{70}_{560}})j}_{kl}{\mathbf{H}}^{(\mathsf{{24}_{560}})k}_j
\biggr)\mathbf{H}^{(\mathsf{{5}_{10_1}})l}+\dots\Biggr]\nonumber\\
&&+i\sum_{r=1}^2 \overline{\alpha}_r\Biggl[\biggl(\sqrt{2}{\mathbf{H}}^{(\mathsf{\overline{45}_{\overline{560}}})k}_{ij}{\mathbf{H}}^{(\mathsf{\overline{75}_{\overline{560}}})ij}_{kl}
-{\sqrt{\frac{2}{3}}}{\mathbf{H}}^{(\mathsf{\overline{45}_{\overline{560}}})j}_{kl}
{\mathbf{H}}^{(\overline{24}_{\mathsf{\overline{560}}})k}_{j}\biggr)\mathbf{H}^{({\mathsf{{5}_10}}_r)l}
+\biggl(-{\frac{4}{\sqrt{721}}}{\mathbf{H}}^{(\mathsf{{5}_{\overline{560}}})k}{\mathbf{H}}^{(\mathsf{\overline{24}_{\overline{560}}})l}_{k}\nonumber\\
&&~~~~~~~~~~~~~+97\sqrt{\frac{2}{10815}}{\mathbf{H}}^{(\mathsf{{5}_{\overline{560}}})l}
{\mathbf{H}}^{(\mathsf{\overline{1}_{\overline{560}}})}
+\frac{1}{\sqrt{145}}{\mathbf{H}}^{(\mathsf{{45}_{\overline{560}}})jk}_{i}{\mathbf{H}}^{(\mathsf{\overline{75}_{\overline{560}}})il}_{jk}
+{4\sqrt{\frac{5}{87}}}{\mathbf{H}}^{(\mathsf{{45}_{\overline{560}}})kl}_{j}
{\mathbf{H}}^{(\mathsf{\overline{24}_{\overline{560}}})j}_k\nonumber\\
&&~~~~~~~~~~~~~-{\frac{\sqrt{3}}{2}}{\mathbf{H}}^{(\mathsf{{70}_{\overline{560}}})kl}_{j}{\mathbf{H}}^{(\mathsf{\overline{24}_{\overline{560}}})j}_k
\biggr)\mathbf{H}^{(\mathsf{\overline{5}_{10}}_r)}_l+\dots\Biggr]\nonumber
\end{eqnarray}
Using Eqs. (\ref{GUTsymmetryB5}, \ref{GUTsymmetryB6}, \ref{Extraction of 5+bar5 doublets} - \ref{Normalized doublets in 5+bar5 and 45+bar45}, \ref{Extraction of 70+bar70 doublets}, \ref{Normalized doublets in 70+bar70}) in the expression above, we arrive 
\begin{eqnarray}
&&\alpha\braket{\Theta_{\mu\nu}^{(\mathsf{560})*}|B\Gamma_{\rho}|\Theta_{\mu\nu}^{(\mathsf{560})}}\Omega^{(\mathsf{10}_1)}_{\rho}
+\sum_{r=1}^2\overline{\alpha}_r\braket{\overline{\Theta}_{\mu\nu}^{(\overline{\mathsf{560}})*}|B\Gamma_{\rho}|
\overline{\Theta}_{\mu\nu}^{(\overline{\mathsf{560}})}}
\Omega^{(\mathsf{10}_r)}_{\rho}\nonumber\\
&&=i\alpha\Biggl[\biggl(-\sqrt{\frac{2}{3}}{{\mathcal V}}_{75}{}^{({\mathsf{{45}}}_{\mathsf{560}})}\!{{\mathbf D}}^{a}
- {\frac{1}{2}\sqrt{\frac{5}{6}}}{{\mathcal V}}_{24}{}^{({\mathsf{{45}}}_{\mathsf{560}})}\!{{\mathbf D}}^{a}
\biggr){}^{({\mathsf{\bar{5}}}_{\mathsf{10}_1})}\!{{\mathbf D}}_{a}
+\biggl({2\sqrt{\frac{6}{3605}}}{{\mathcal V}}_{24}{}^{({\mathsf{\bar{5}}}_{\mathsf{560}})}\!{{\mathbf D}}_{a}
+97\sqrt{\frac{2}{10815}}{{\mathcal V}}_{1}{}^{({\mathsf{\bar{5}}}_{\mathsf{560}})}\!{{\mathbf D}}_{a}\nonumber\\
&&~~~~~~~~
-\frac{1}{\sqrt{435}}{{\mathcal V}}_{75}{}^{({\mathsf{\overline{45}}}_{\mathsf{560}})}\!{{\mathbf D}}_{a}+{\frac{5}{\sqrt{87}}}{{\mathcal V}}_{24}{}^{({\mathsf{\overline{45}}}_{\mathsf{560}})}\!{{\mathbf D}}_{a}-{\frac{1}{4}\sqrt{\frac{15}{2}}}{{\mathcal V}}_{24}{}^{({\mathsf{\overline{70}}}_{\mathsf{560}})}\!{{\mathbf D}}_{a}
\biggr){}^{({\mathsf{{5}}}_{\mathsf{10}_1})}\!{{\mathbf D}}^a\Biggr]\nonumber\\
&&+i\sum_{r=1}^2 \overline{\alpha}_r\Biggl[\biggl(-\sqrt{\frac{2}{3}}{\overline{\mathcal V}_{75}}{}^{({\mathsf{\overline{45}}}_{\mathsf{\overline{560}}})}\!{{\mathbf D}}_{a}
- {\frac{1}{2}\sqrt{\frac{5}{6}}}{\overline{\mathcal V}_{24}}{}^{({\mathsf{\overline{45}}}_{\mathsf{\overline{560}}})}\!{{\mathbf D}}_{a}
\biggr){}^{({\mathsf{{5}}}_{\mathsf{10}_r})}\!{{\mathbf D}}^a
+\biggl({2\sqrt{\frac{6}{3605}}}{\overline{\mathcal V}_{24}}{}^{({\mathsf{{5}}}_{\mathsf{\overline{560}}})}\!{{\mathbf D}}^{a}
+97\sqrt{\frac{2}{10815}}{\overline{\mathcal V}_{1}}{}^{({\mathsf{{5}}}_{\mathsf{\overline{560}}})}\!{{\mathbf D}}^{a}\nonumber\\
&&~~~~~~~~~~~~~
-\frac{1}{\sqrt{435}}{\overline{\mathcal V}_{75}}{}^{({\mathsf{{45}}}_{\mathsf{\overline{560}}})}\!{{\mathbf D}}^{a}
+{\frac{5}{\sqrt{87}}}{\overline{\mathcal V}_{24}}{}^{({\mathsf{{45}}}_{\mathsf{\overline{560}}})}\!{{\mathbf D}}^{a}
-{\frac{1}{4}\sqrt{\frac{15}{2}}}{\overline{\mathcal V}_{24}}{}^{({\mathsf{{70}}}_{\mathsf{\overline{560}}})}\!{{\mathbf D}}^{a}
\biggr){}^{({\mathsf{\bar{5}}}_{\mathsf{10}_r})}\!{{\mathbf D}}_a\Biggl]
\end{eqnarray}
Finally, we compute the mass term for the $\mathsf{560+\overline{560}}$
\begin{eqnarray}
 M_{560}~\braket{\Theta_{\mu\nu}^{(\mathsf{560})*}|B|\overline{\Theta}_{\mu\nu}^{(\overline{\mathsf{560}})}}
&=&M_{560}\left[\braket{\Theta_{c_ic_j}^{(\mathsf{560})*}|B|\overline{\Theta}_{\overline{c}_i\overline{c}_j}^{(\overline{\mathsf{560}})}}
-2\braket{\Theta_{c_i\overline{c}_j}^{(\mathsf{560})*}|B|\overline{\Theta}_{c_j\overline{c}_i}^{(\overline{\mathsf{560}})}}+\dots\right]\nonumber\\
&=&iM_{560}\left[{\mathbf{H}}^{(\mathsf{\overline{5}_{560}})}_i{\mathbf{H}}^{(\mathsf{{5}_{\overline{560}}})i}-\frac{1}{2}
{\mathbf{H}}^{(\mathsf{{45}_{560}})ij}_k{\mathbf{H}}^{(\mathsf{\overline{45}_{\overline{560}}})k}_{ij}
+\frac{1}{2}{\mathbf{H}}^{(\mathsf{\overline{45}_{560}})i}_{jk}{\mathbf{H}}^{(\mathsf{{45}_{\overline{560}}})jk}_i+\frac{1}{2}{\mathbf{H}}^{(\overline{70}_{\mathsf{560}})i}_{jk}
{\mathbf{H}}^{(\mathsf{{70}_{\overline{560}}})jk}_{i}+\dots
\right]\nonumber\\
&=&iM_{560}\left[{}^{({\mathsf{\bar{5}}}_{\mathsf{560}})}\!{{\mathbf D}}_{a}{}^{({\mathsf{{5}}}_{\mathsf{\overline{560}}})}\!{{\mathbf D}}^{a}-\frac{1}{2}{}^{({\mathsf{{45}}}_{\mathsf{560}})}\!{{\mathbf D}}^{a}{}^{({\mathsf{\overline{45}}}_{\mathsf{\overline{560}}})}\!{{\mathbf D}}_{a}
+\frac{1}{2}{}^{({\mathsf{\overline{45}}}_{\mathsf{560}})}\!{{\mathbf D}}_{a}{}^{({\mathsf{{45}}}_{\mathsf{\overline{560}}})}\!{{\mathbf D}}^{a}
+\frac{1}{2}{}^{({\mathsf{\overline{70}}}_{\mathsf{560}})}\!{{\mathbf D}}_{a}{}^{({\mathsf{{70}}}_{\mathsf{\overline{560}}})}\!{{\mathbf D}}^{a}+\dots\right]\nonumber\\
\end{eqnarray}
}
}}

\end{appendices}

\end{document}